\documentclass[onecolumn,apj]{emulateapj}
\usepackage{graphicx}
\usepackage{subfigure}
\usepackage{bm}
\usepackage{dcolumn}
\usepackage{amsmath}
\usepackage{hyperref}
\usepackage{color}
\usepackage{mathrsfs}
\usepackage{bm}

\shorttitle{A Bayesian estimate}
\shortauthors{Moura-Santos et al.}

\begin{document}

\newcommand{\newc}{\newcommand}

\newc{\be}{\begin{equation}}
\newc{\ee}{\end{equation}}
\newc{\ba}{\begin{eqnarray}}
\newc{\ea}{\end{eqnarray}}
\newc{\bea}{\begin{eqnarray*}}
\newc{\eea}{\end{eqnarray*}}
\newc{\D}{\partial}
\newc{\ie}{{\it i.e.} }
\newc{\eg}{{\it e.g.} }
\newc{\etc}{{\it etc.} }
\newc{\etal}{{\it et al.}}
\newc{\lcdm }{$\Lambda$CDM }
\newcommand{\nn}{\nonumber}
\newc{\ra}{\rightarrow}
\newc{\lra}{\leftrightarrow}
\newc{\lsim}{\buildrel{<}\over{\sim}}
\newc{\gsim}{\buildrel{>}\over{\sim}}


\title{A Bayesian estimate of the CMB-large-scale structure
  cross-correlation}


%\author{E. Moura-Santos\altaffilmark{1}}
%\affil{Instituto de F\'isica, Universidade de S\~ao Paulo, Rua do Mat\~ao trav. R 187, 
%05508-090, S\~ao Paulo - SP, Brasil}
%\email{emoura@if.usp.br}
%
%\author{F. C. Carvalho\altaffilmark{2}}
%\affil{Departamento de F\'isica, Universidade do Estado do Rio Grande do Norte, 59610-210, Mossor\'o-RN, Brasil}
%\email{fabiocabral@uern.br}
%
%\author{M. Penna-Lima\altaffilmark{3,5}}
%\affil{APC, AstroParticule et Cosmologie, Universit\'e Paris Diderot, CNRS/IN2P3, CEA/Irfu, Observatoire de Paris, Sorbonne Paris
%  Cit\'e, 10, rue Alice Domon et L\'eonie Duquet, 75205 Paris Cedex 13, France}
%\email{pennal@apc.in2p3.fr}
%
%\author{C. P. Novaes\altaffilmark{4,5}}
%\affil{Observat\'orio Nacional, Rua General Jos\'e Cristino 77, S\~ao Crist\'ov\~ao, 20921-400, Rio de Janeiro, RJ, Brasil}
%\email{camilanovaes@on.br}
%
%\and
%
%\author{C. A. Wuensche\altaffilmark{5}}
%\affil{Divis\~ao de Astrof\'\i sica, Instituto Nacional de Pesquisas  Espaciais - INPE, S\~ao Jos\'e dos Campos-SP, Brasil}
%\email{cawuenschel@das.inpe.br}

\author{E. Moura-Santos\altaffilmark{1}, F. C. Carvalho\altaffilmark{2}, M. Penna-Lima\altaffilmark{3,4}, C. P. Novaes\altaffilmark{5,4} and C. A. Wuensche\altaffilmark{4}}
\altaffiltext{1}{Instituto de F\'isica, Universidade de S\~ao Paulo, Rua do Mat\~ao trav. R 187, 
05508-090, S\~ao Paulo - SP, Brasil}
\altaffiltext{2}{Departamento de F\'isica, Universidade do Estado do Rio Grande do Norte, 59610-210, Mossor\'o-RN, Brasil}
\altaffiltext{3}{APC, AstroParticule et Cosmologie, Universit\'e Paris Diderot, CNRS/IN2P3, CEA/Irfu, Observatoire de Paris, Sorbonne Paris Cit\'e, 10, rue Alice Domon et L\'eonie Duquet, 75205 Paris Cedex 13, France}
\altaffiltext{4}{Divis\~ao de Astrof\'\i sica, Instituto Nacional de
 Pesquisas  Espaciais - INPE, S\~ao Jos\'e dos Campos-SP, Brasil}
\altaffiltext{5}{Observat\'orio Nacional, Rua General Jos\'e Cristino 77, S\~ao Crist\'ov\~ao, 20921-400, Rio de Janeiro, RJ, Brasil}

\email{emoura@if.usp.br}
\email{fabiocabral@uern.br}
\email{pennal@apc.in2p3.fr}
\email{camilanovaes@on.br}
\email{cawuenschel@das.inpe.br}
\date{\today}

%\date{\today}


\begin{abstract}
Evidences for late-time acceleration of the Universe are provided by
multiple probes, such as Type Ia
supernovae, the cosmic microwave background (CMB) and large-scale structure (LSS). 
In this work, we focus on the integrated Sachs--Wolfe (ISW)
effect, i.e., secondary CMB fluctuations generated by evolving gravitational potentials 
due to the transition between, e.g., the matter and dark energy (DE) dominated phases.
Therefore, assuming a flat universe, DE properties can be inferred from ISW detections. We present a
Bayesian approach to compute the CMB--LSS cross-correlation signal. The method is based on the
estimate of the likelihood for measuring a combined set consisting of a CMB
temperature and a galaxy contrast maps, provided that we have some
information on the statistical properties of the fluctuations
affecting these maps. The likelihood is estimated by a
sampling algorithm, therefore avoiding the computationally demanding
techniques of direct evaluation in either pixel or harmonic space. As local
tracers of the matter distribution at large scales, we used the
Two Micron All Sky Survey (2MASS) galaxy catalog and, for the CMB
temperature fluctuations, the ninth-year data release of the Wilkinson
Microwave Anisotropy Probe ({\it WMAP9}). The results show a dominance of cosmic
variance over the weak recovered signal, due mainly to the shallowness
of the catalog used, with systematics associated with
the sampling algorithm playing a secondary role as sources of
uncertainty. When combined with other complementary probes, the method
presented in this paper is expected to be a useful tool to late-time
acceleration studies in cosmology.
\end{abstract}



%\linenumbers

%%%%%%%%%%%%%%%%%%%%%%%%%%%%%%%%%%%%%%%%%%%%%%%%%%%%%%%%%
\section{Introduction}
\label{sec:intro}
%%%%%%%%%%%%%%%%%%%%%%%%%%%%%%%%%%%%%%%%%%%%%%%%%%%%%%%%%

Understanding the origin of the late-time acceleration of the Universe
has been one of the biggest challenges in cosmology in the past almost 
20 yrs. A combined analysis of independent data sets, such as distance
measurements of Type Ia supernovae \citep{Betoule2014},
measurements of the cosmic microwave background (CMB) anisotropies \citep{Bennett:2012zja,Ade:2015xua} 
and observations of the large-scale structure (LSS) in the universe \citep{sdss}, 
indicate that the cosmos has the following energy budget: $\sim 5\%$ of
baryonic matter, $\sim 26\%$ of dark matter and $\sim 69\%$ of dark energy (DE).
From the theoretical point of view, a variety of DE models have been 
proposed, involving different DE equations of state \citep{Joyce2015}, 
canonical and noncanonical scalar fields \citep{Carvalho:2006fy,Copeland:2006wr}, 
and $f(R)$ gravity \citep{Nojiri2011}, among others, to explain the recent cosmic acceleration. 

The study of the statistical properties of fluctuations of the different 
constituents of the Universe is a powerful method to select among the various 
competing cosmological models. For example, useful probes for the late-time 
cosmic acceleration are the measurements of the cross-correlation between the CMB 
temperature fluctuations and LSS tracers.  
The CMB--LSS cross-correlation has been the subject of a number of papers since 
\citet{Crittenden:1995ak} proposed to use the method to detect the integrated Sachs--Wolfe
(ISW) effect independently of the intrinsic CMB fluctuations. In the case
where the gravitational potentials decay, a positive correlation is
expected \citep{Cooray:2001ab}, meaning that at large scales, hot spots
in the CMB will correspond to overdense regions in the galaxy
distribution. This positive correlation is also expected in open
universes \citep{Kamionkowski:1996ra,Kinkhabwala:1998zj}, whereas a
negative correlation will occur in closed ones. 

Several authors reported detections (sometimes in contradiction; see \citet{Dupe2011}) of the ISW 
effect by computing the cross-correlation between the first-year {\it Wilkinson Microwave Anisotropy
Probe} ({\it WMAP}) data with radio sources 
\citep{Boughn:2001zs, Nolta:2003uy, Boughn:2004zm, Boughn:2004yy}; the hard X-ray background 
\citep{Boughn:2004ah,Boughn:2003yz}; the Sloan Digital Sky Survey (SDSS) Data Realease 1 
\citep{Fosalba:2003ge,Padmanabhan:2004fy,Abazajian:2008wr}, the Two Micron All Sky Survey (2MASS; 
\citet{Afshordi:2003xu}), the APM Galaxy Survey \citep{Fosalba:2003iy,Hernandez-Monteagudo:2013vwa}, and a 
combination of the above \citep{Gaztanaga:2004sk}. 
The third-year {\it WMAP} data were correlated by \citet{Cabre:2006qm} with the fourth SDSS data 
release (DR4), showing a significant positive signal, while \citet{Giannantonio:2006du} compared 
it with high-redshift SDSS quasars, claiming a $2\sigma$ detection (see also \citet{Ho2008}). 
\citet{McEwen:2007ni} used a directional spherical wavelet analysis and found a positive detection 
at the $3.9\sigma$ level. ISW detections about $3\sigma$ were also obtained using the {\it Wide-field Infrared Survey Explorer} ({\it WISE}) and seventh-year \citep{Goto:2012} and ninth-year {\it WMAP} data \citep{Ferraro:2015}. 

More recently, the {\it Planck} satellite has released the results of its full data set, including the 
claim of an ISW effect detection at the $2\sigma-4\sigma$ level (\citet{Ade:2013dsi,2015arXiv150201595P}; see also \citet{Granett2015}).
They observed an exceeding amplitude of the ISW signal measured with superclusters and voids.
Such result confirms previous measurements by, e.g., \citet{Granett2008} and \citet{Nadathur2012}, also using 
localized measurements of superstructures, and \citet{Giannantonio2012}, cross-correlating {\it WMAP} CMB 
maps and SDSS observations, who found an ISW signal indicating a discrepancy with respect to what 
is expected by the $\Lambda \text{CDM}$ model.
In an attempt to find possible reasons for this tension, 
authors have dedicated efforts in better understanding the systematics and checking for possible contamination signals, e.g. point sources and Galactic foreground, besides using the recently released {\it Planck}
products, namely, CMB lensing potential maps and polarization data \citep{Goto:2012,Nadathur2012,Ferraro:2015,2015arXiv150201595P}.

Given the number of different data sets used, especially the ones taken
as LSS tracers, and the several methods employed, points 
such as (i) the statistical significance of the signal and (ii) the
systematics affecting both the data sets used, as well as (iii) the
procedure adopted, become particularly important in order to properly
interpret the results (see \citet{Giannantonio2012} and references therein). 
In this work we address some of these questions
by using Bayesian inference to study in detail the
correlation between the {\it WMAP}9 temperature maps \citep{Bennett:2012zja} and the catalog of luminous
sources of the 2MASS infrared survey \citep{Skrutskie:2006wh}.

Bayesian inference applied to CMB data has become very popular in the past two decades, 
particularly in situations in which the computation of a likelihood matrix is required. 
The particular case of obtaining the maximum likelihood estimate for the
CMB autocorrelation power spectrum, provided that some properties of the
noise and of the foregrounds are known, is an example where Bayesian
inference has shown its full power (see, e.g., \citet{Bond:1998zw}). However, the direct evaluation of this
likelihood for satellites with fine angular resolution, such as {\it WMAP} and
{\it Planck}, has also been revealed to be a hard computational problem, with
a number of operations scaling beyond the current capacity of computers
\citep{Hivon:2001jp}. To overcome this technical problem, cosmologists 
have developed techniques based on Markov chain algorithms in
order to reconstruct the likelihood for CMB measurements. Gibbs
sampling is one of these Monte Carlo (MC) based techniques and has been
successfully used to estimate the CMB (temperature and polarization) power
spectra \citep{Jewell:2002dz,Wandelt:2003uk,Eriksen:2004ss,Larson:2006ds,Dunkley:2008mk}. Here we apply the Gibbs sampling method to a combined CMB-galaxy
survey experiment whose measurements, as in the CMB-only case, contain
the primordial signal of interest, instrumental noise and, in
principle, residual foregrounds. Once some assumptions on the
statistical properties of the noise were made, we extracted a Bayesian estimate for the primordial
signal full covariance matrix, containing the autocorrelation power
spectra of the CMB and of the galaxies in its diagonal and, most important in this
work, the cross-correlation power spectrum as off-diagonal
elements. Here we focus on determining the potential of a galaxy
catalog with the widest possible sky coverage, such as 2MASS, to
show a cross-correlation signal at large angular scales with a CMB
temperature map. A study on the strength of such a signal, using deeper surveys (at the
cost of sky coverage), is left for a future work.

The paper is organized as follows. In Section~\ref{sec:xcross} we outline the
theoretical background in order to calculate the expected cross-correlation 
spectrum between the CMB and an LSS tracer when the ISW
effect is present. The effect of a cut-sky map on the estimates of cross-correlation
power spectrum is discussed in Section~\ref{sec:master} and the
usual mask deconvolution algorithm to retrieve the ensemble average
estimate for this spectrum is also discussed. Section~\ref{sec:gibbs}
is devoted to the detailed presentation of the Gibbs sampling
method applied here to a combined CMB--galaxy survey data set, the
validation of which is presented in Section~\ref{sec:valida}. The
application of the analysis pipeline, described in section \ref{sec:processing}, to the {\it WMAP}9--2MASS
combined data set is found in Section~\ref{sec:results}. Final remarks
are then presented in Section~\ref{sec:conclusions}.

%%%%%%%%%%%%%%%%%%%%%%%%%%%%%%%%%%%%%%%%%%%%%%%%%%%%%%%%%
\section{The CMB--LSS cross-correlation}
\label{sec:xcross}
%%%%%%%%%%%%%%%%%%%%%%%%%%%%%%%%%%%%%%%%%%%%%%%%%%%%%%%%%

In this section we briefly present the theoretical background to compute the ISW and LSS 
auto- and cross-correlation functions. The ISW effect is a secondary anisotropy in the CMB 
temperature field due to a variation of the gravitational potential along the line of sight 
in a direction $\hat{\mathbf{n}}$, i.e., 
\begin{equation}
\Delta_t^{\text{ISW}}(\hat{\mathbf{n}})\equiv\left(\frac{\Delta T(\hat{\mathbf{n}})}{T}\right)=-2\int dz \, e^{-\tau(z)}
~{d\Phi\over{dz}}(\hat{\mathbf{n}},z),
\label{eq:isw} 
\end{equation}
where $\tau$ is the optical depth of CMB photons and $\Phi (\hat{\mathbf{n}},z)$ is the 
Newtonian gravitational potential at redshift $z$. Using the Poisson equation in the Fourier 
space, $\Phi (\mathbf{k} ,z)$ is given by
\begin{equation}
\Phi (\mathbf{k}, z) = \frac{3}{2} \Omega_m
\left(\frac{H_0}{c\,k}\right)^2 (1 + z) \delta (\mathbf{k}, z),
\label{eq:phi_delta}
\end{equation}
where $\Omega_m$ is the dimensionless matter density today, $c$ is the
speed of light, $H_0$ is the Hubble parameter, and $\delta(\mathbf{k},
z)$ is the density contrast of matter.

Similarly, the galaxy fluctuations can be written as
\begin{equation}
\Delta_g (\hat{\mathbf{n}}) \equiv \left(\frac{n_g (\hat{\mathbf{n}})
    - \bar{n}_g}{\bar{n}_g}\right) = \int dz \, b_g(z)
\frac{dN}{dz}(z) \delta (\mathbf{k}, z), 
\label{eq:gal}
\end{equation}
where $\bar{n}_g$ is the mean number density of galaxies, $b_g(z)$ is the linear bias factor [$\delta_g(z) = b_g(z) \delta (z)$] and $dN/dz$ is the normalized galaxy redshift distribution \citep{Afshordi:2003xu, Rassat:2006kq,Giannantonio:2013uqa}.

From Equations~\eqref{eq:isw} -- \eqref{eq:gal}, the two-point correlation function $\langle \Delta_x (\hat{\mathbf{n}}) \Delta_y (\hat{\mathbf{n}}^\prime)\rangle$ in harmonic space is
\begin{equation}
C_l^{xy} = \frac{2}{\pi}\int dk \, k^2 W_l^{x}(k)W_l^{y}(k) P(k),
\label{eq:cl}
\end{equation}
where $x, y = (t, g)$, and $P(k)$ is the matter power spectrum. The ISW and galaxy autocorrelation functions, $C_l^{tt}$ and $C_l^{gg}$, as well as the CMB(ISW)--galaxy cross-correlation $C_l^{tg}$ are computed substituting into Equation~\eqref{eq:cl} the respective kernels, namely,
\begin{eqnarray}
W^{t}(k,z)&=&-3 \Omega_m \left(\frac{H_0}{k\,c}\right)^2 \int dz \, \frac{d\left[(1+z)D(z)\right]}{dz}\, j_l[k \chi(z)],
\nonumber\\
W^{g}(k,z)&=& \int dz \, b_g(z) \frac{dN}{dz} D(z) j_l[k \chi(z)],
\label{eq:Cls2}
\end{eqnarray} 
where $D(z)$ is the linear growth function (normalized to one at $z=0$), $j_l[k \chi(z)]$ is the spherical Bessel function of the first kind, and $\chi(z)$ is the comoving distance. 

%%%%%%%%%%%%%%%%%%%%%%%%%%%%%%%%%%%%%%%%%%%%%%%%%%%%%%%%%
\section{Cross-correlation with partial sky coverage}
\label{sec:master}
%%%%%%%%%%%%%%%%%%%%%%%%%%%%%%%%%%%%%%%%%%%%%%%%%%%%%%%%%

Partial and/or nonuniform sky coverage is a typical situation for CMB
and galaxy survey experiments due to a limited field of view, galactic
foregrounds, dust extinction, etc. For a CMB experiment, the strong
galactic emission at radio and microwave wavelengths, for example,
forces the masking of large regions of the sky maps close to the
galactic plane, in order to get a reliable estimate of the CMB temperature
power spectrum. At small angular scales, the positions of identified
extragalactic sources also have to be removed. The same galactic
emissions affect the detection of galaxies close to the Milky Way
plane, and even at high galactic latitudes, dust extinction has to be
taken into account when calculating magnitudes.
\begin{figure*}[ht]
\begin{center}
\subfigure[]{
  \includegraphics[width=.42\textwidth]{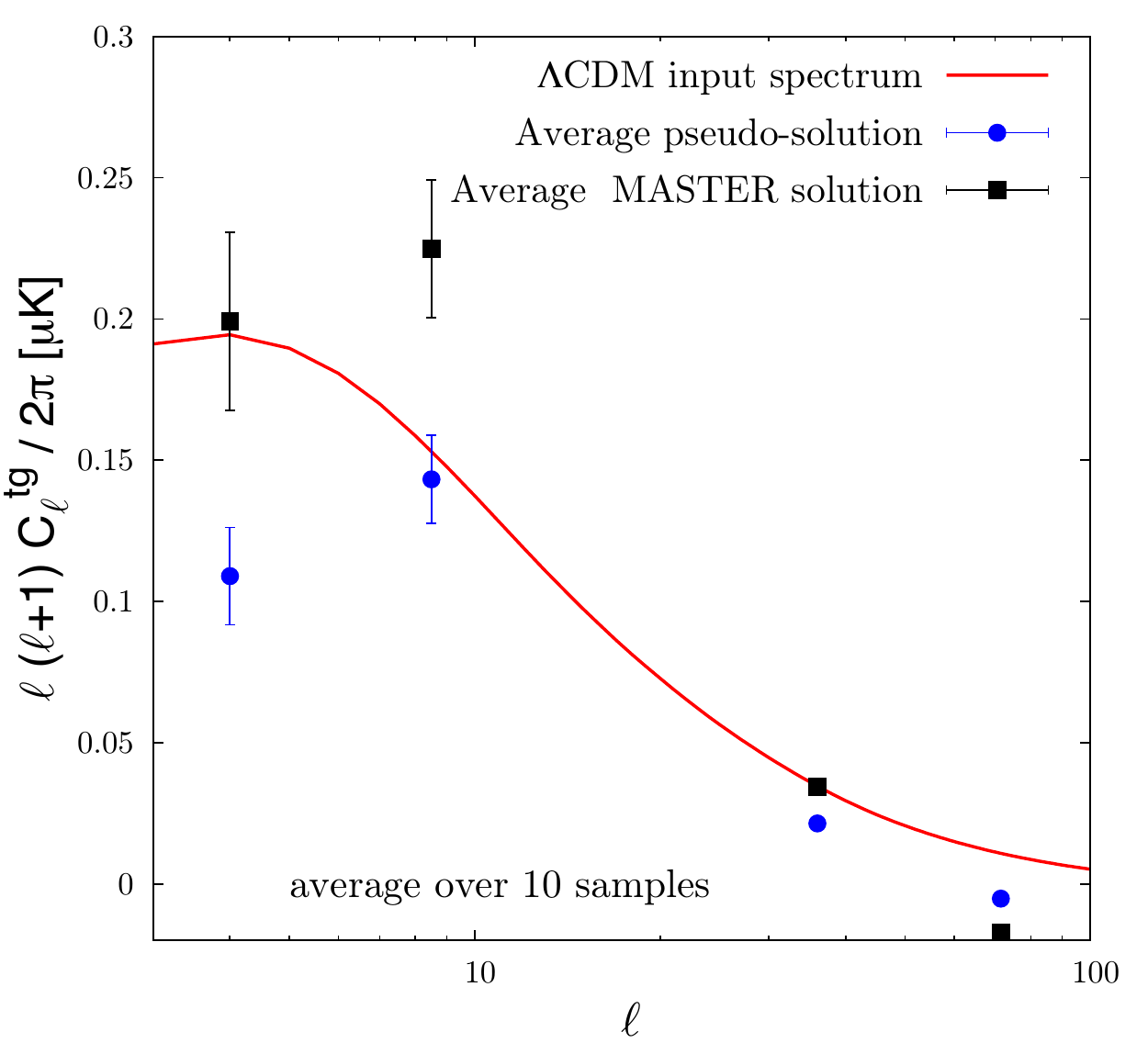}
}%
\subfigure[]{
  \includegraphics[width=.42\textwidth]{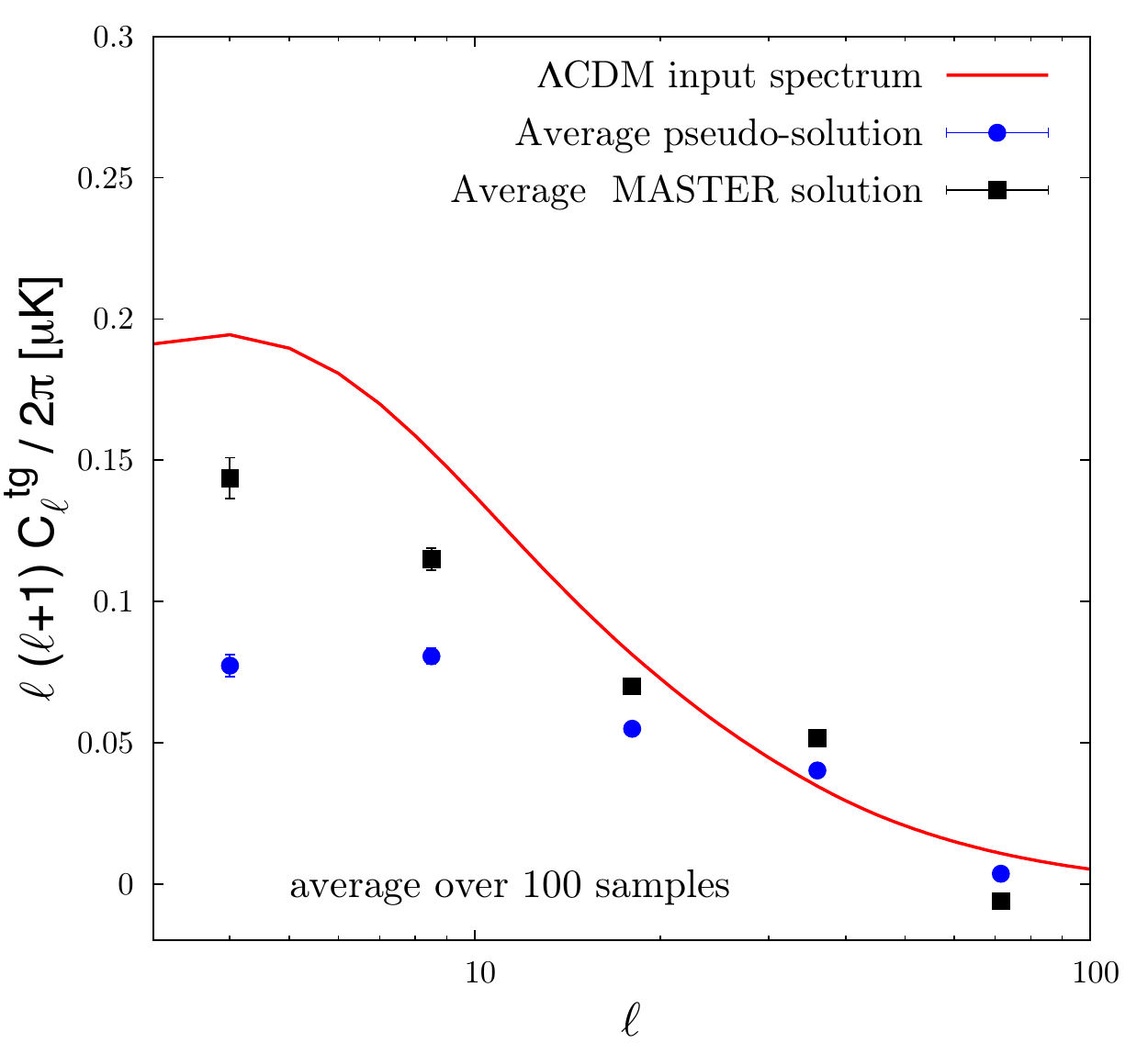}
} \\ % end of first row
\subfigure[]{
  \includegraphics[width=.42\textwidth]{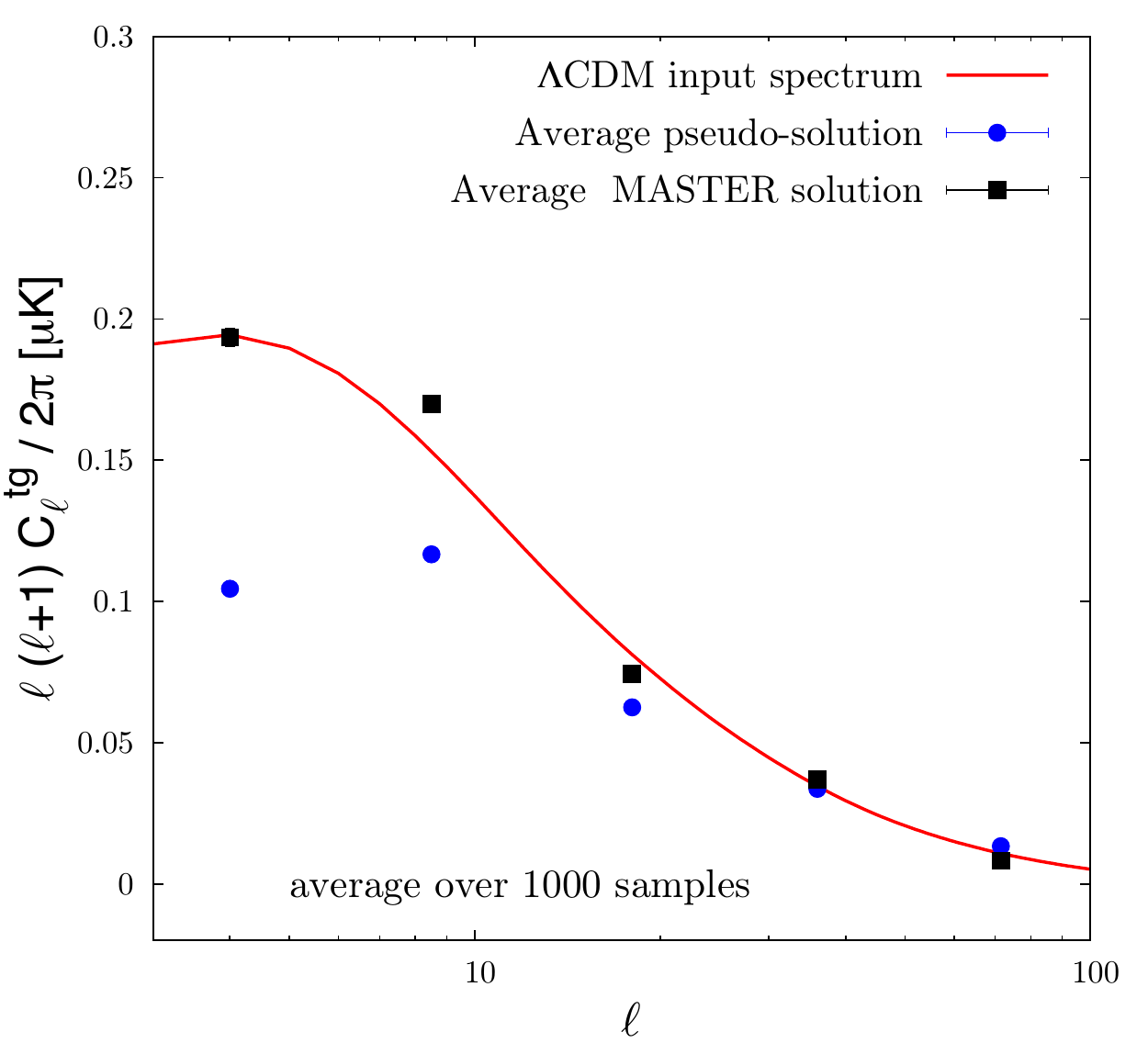}
}%
\subfigure[]{
  \includegraphics[width=.42\textwidth]{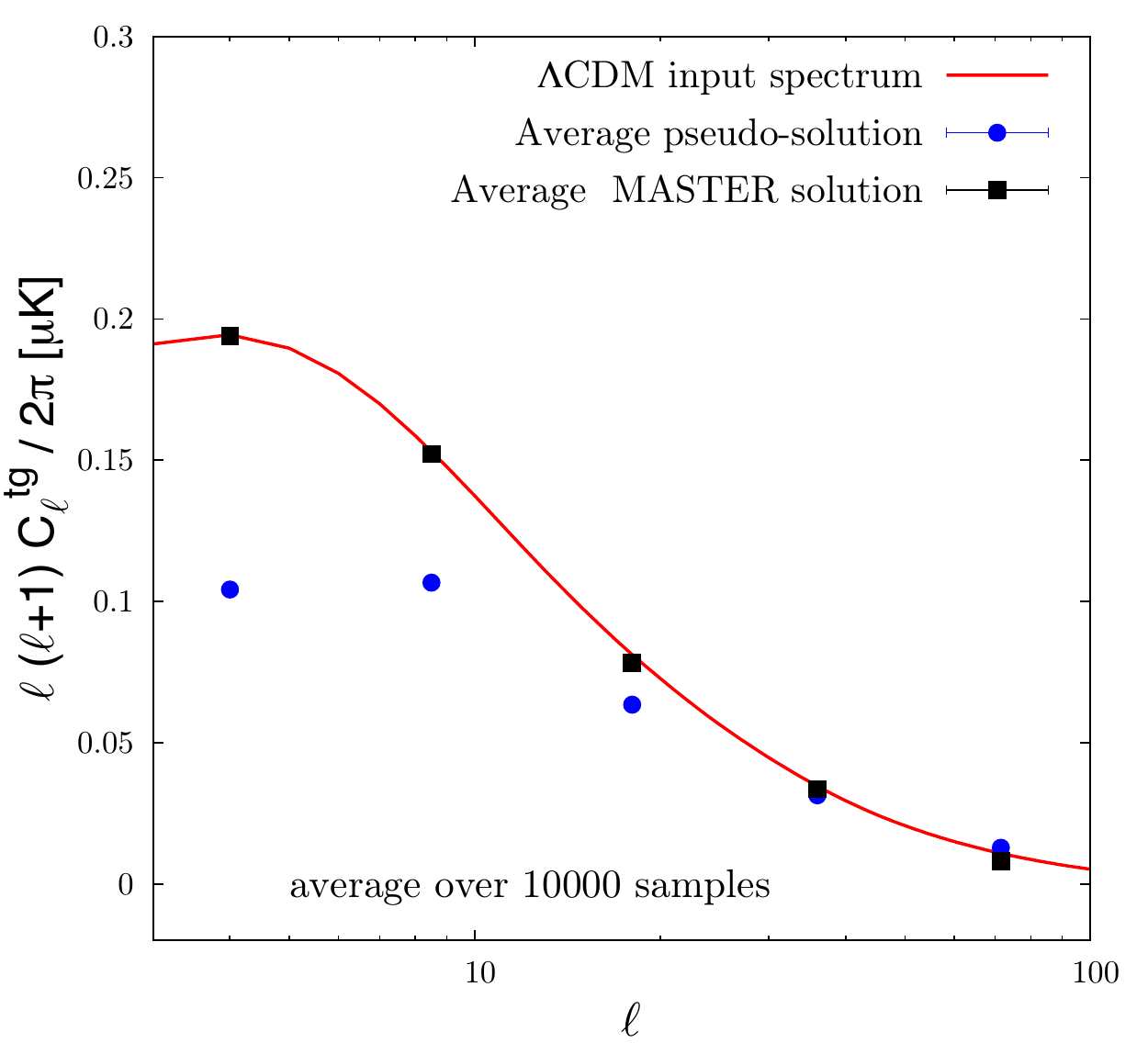}
}%
\end{center}
\caption{Convergence of the MASTER binned (a total of six bins
  approximately logarithmically spaced
  over $3\le\ell\le 192$) solution for the CMB--LSS cross-power
  spectrum for different numbers of sky realizations. The mask used was
  {\it WMAP}9 KQ85 with $f_{sky}=0.77$; the fiducial cosmological model is
  $\Lambda$CDM (only linear theory), with parameters given in Table~\ref{table:inputpars}
  and with a galaxy bias factor $b_g=1.4$ and the selection suited for
  band 4 of the 2MASS catalog. The pseudo-$C_\ell$ estimate 
  is also shown, and its power is depleted at all scales, but the damping is
  particularly strong at large angular scales ($\ell<10$),
  irrespective of the number of samples. Uncertainties are the errors
  on the average. For the cosmic variance contribution to $C_{\ell}^{tg}$ see Figures
  \ref{fig:bestfit_mc} and \ref{fig:systematics_allbands}.} 
\label{fig:masterconv}
\end{figure*}

When retrieving spherical harmonics coefficients from a cut-sky map,
the effect of the mask, taken here as a weight function
$W(\hat{\mathbf{n}})$ in the direction of the unit vector $\hat{\mathbf{n}}$, is to mix different multipoles
\citep{Hivon:2001jp}, leading to the so-called pseudo-power spectra
$\langle\tilde{C}_{\ell}\rangle$. If the mask power spectrum ${\cal
  W}_\ell$, the $\ell$-space window function of the beam $B_{\ell}$, and an estimate of
the instrumental noise power $N_{\ell}$ are available, as shown in
\citet{Hivon:2001jp}, one can retrieve an estimate for the ensemble
average of the true power spectrum $\langle C_{\ell}\rangle$, by solving the following linear system:
\begin{equation}
\langle\tilde{C}_\ell\rangle=\sum_{\ell'}M_{\ell\ell'}B_{\ell'}^2\langle
C_{\ell'}\rangle+\langle N_{\ell}\rangle,
\label{eq:masterlinsys}
\end{equation}
where $M_{\ell\ell'}$ is the multipole mixing matrix that can be written in
terms of the Wigner-$3j$ symbols
\begin{equation}
M_{\ell\ell^\prime}=\frac{2\ell^{\prime}+1}{4\pi}\sum_{\ell^{\prime\prime}}(2\ell^{\prime\prime}+1){\cal W}_{\ell^{\prime\prime}}\left(
\begin{array}{ccc}
\ell & \ell^\prime & \ell^{\prime\prime} \\
0 & 0 & 0
\end{array}
\right)^2.
\label{eq:mixkernel}
\end{equation}

The mixing pattern is strongly dependent on the
shape of the mask, but for the particular case of a constant power
spectrum $\langle C_{\ell} \rangle=C=cte$, like in the case of white noise, it leads to
a simple power damping at all scales given by the sky fraction $f_{\text{sky}}$
covered by the unmasked directions $\langle \tilde{C}
\rangle=Cf_{sky}$ \footnote{This is valid for the case of a mask $W(\hat{\mathbf{n}})$
  that is equal to 0 in the masked region and equal to 1 in the rest of
  the sky. For a different weighting function, the correction factor
  is $f_{sky}w_2\equiv (1/4\pi)\int_{4\pi} W^2(\hat{\mathbf{n}})d\Omega$.}.

Cross-correlation analyses of the CMB and galaxy survey maps have been
performed in the literature using pseudo-$C_\ell$'s estimators
\citep{Rassat:2006kq,Ho2008,Xia:2011hj,Giannantonio2012}. As an ensemble average, 
the solution of Equation \eqref{eq:masterlinsys}
will eventually converge to the true power spectrum in the limit of a 
large number of sky realizations. Having a single universe to retrieve the sky
map multipoles, such an estimator is severely affected by cosmic
variance at low multipoles. We have quantified the impact of cosmic
variance in the MASTER (Monte Carlo Apodized Spherical Transform
Estimator; \citet{Hivon:2001jp}) 
solution for the CMB--LSS cross-power spectrum
using MC simulations and $\Lambda$CDM as the fiducial cosmological model (see Table~\ref{table:inputpars}). Figure~\ref{fig:masterconv} shows the convergence of a binned MASTER solution toward the true
value for different sets of sky realizations. Auto- and cross-correlation power spectra have been generated
from the fiducial model with a bias factor $b_g=1.4$ and a selection
function appropriate for band 4 of the 2MASS catalog (see eq. \ref{eq:window}
and section \ref{sec:gibbs} ahead), from which the respective sky maps
for the CMB and galaxy survey were synthesized using the HEALPix synfast
program \footnote{http://healpix.jpl.nasa.gov/html/facilitiesnode14.htm}.  The {\it WMAP}9 KQ85 
analysis mask was then applied to these maps,
and the MASTER solution was retrieved. No instrumental noise has been
added to the maps, since for the particular case of cross-correlation
between maps measured independently, the average noise cross-power spectrum
appearing in Equation~\eqref{eq:masterlinsys} is zero. As one can see, the solution will
be a reasonable approximation to the true underlying power spectrum for
about 1000 sky samples. A single sky-based estimate can be very
far from the true value. One can also see the effect of power depletion
at large scales due to the masked region, which for the KQ85 mask is
about 50\% for the quadrupole and octopole.

\begin{table*}[ht]
\caption{Fiducial Cosmological Model: $\Lambda$CDM, flat Universe}
\begin{center}
\begin{tabular}{lcc}
\hline
\hline
Parameter & Symbol & Value \\
\hline
\hline
Baryon density & $\Omega_b h^2$ & 0.0226  \\
Cold dark matter density & $\Omega_c h^2$ & 0.112  \\
%Dark energy density & $\Omega_\Lambda$ & ***  
Curvature perturbation ($k_0=0.002$ Mpc$^{-1}$) & $\Delta_{\cal R}^2$ & $2.4\times 10^{-9}$  \\
Scalar spectral index & $n_s$ & 0.96 \\
Reionization optical depth & $\tau$ & 0.09 \\
Hubble constant (100 km s$^{-1}$Mpc$^{-1}$) & $h$ & 0.7 \\
\hline 
\hline 
\end{tabular}
\end{center}
\label{table:inputpars}
\end{table*}

%%%%%%%%%%%%%%%%%%%%%%%%%%%%%%%%%%%%%%%%%%%%%%%%%%%%%%%%%
\section{Gibbs sampling approach}
\label{sec:gibbs}
%%%%%%%%%%%%%%%%%%%%%%%%%%%%%%%%%%%%%%%%%%%%%%%%%%%%%%%%%

Bayesian estimates of the CMB temperature and polarization power
spectra have been successfully used by several experiments in recent
years. The fluctuations observed today in the photon field temperature 
distribution over the sky are assumed to come from three main sources:
primordial fluctuations generated in the early universe, instrumental
noise, and foregrounds. The likelihood for observing a
given temperature angular distribution can be calculated once the galactic emission and extragalactic
point sources have been removed from the sky maps, and also if appropriate
assumptions are made on the statistical properties of the primordial
signal and the instrumental noise
\citep{Bond:1998zw}. If performed in pixel space, these 
calculations were shown to scale as  $\mathcal{O}(N_p^3)$, where $N_p$ is the total number of
pixels, becoming prohibitively expensive in
terms of computing power for satellite experiments like {\it WMAP} ($N_p\sim
10^6$) and {\it Planck} ($N_p\sim 10^7$) \citep{Hivon:2001jp}. In order to
avoid such brute-force approach of direct evaluation of the
likelihood, MC techniques have been implemented in which
sample maps of the primordial signal are drawn from the {\it a posteriori}
probability distribution taking into account cosmic variance, 
instrumental noise, and residual foregrounds 
\citep{Eriksen:2004ss,Jewell:2002dz,Wandelt:2003uk,Dunkley:2008mk}. The method can be
  applied to either Time Order Data (TOD) or pixelized maps and
  can take into account filtering and beaming effects.

Let $\mathbf{d}$ be a pixelized temperature map consisting of
primordial signal $\mathbf{s}$ and instrumental noise
$\mathbf{n}$. Assuming that both $\mathbf{s}$ and $\mathbf{n}$
follow Gaussian distributions with covariance matrices
$\mathbf{S}$ and $\mathbf{N}$, respectively, the likelihood for
observing such a map, after marginalizing over the unknown true signal
$\mathbf{s}$, is given by
\begin{equation}
\mathscr{L}=P(\mathbf{d}|\mathbf{C})=\frac{1}{(2\pi)^{n_{\textrm{dim}}/2} |\mathbf{C}|^{1/2}}\textrm{exp}\left(-\frac{1}{2}\mathbf{d}^\textrm{T}\mathbf{C}^{-1}\mathbf{d}\right),
\label{eq:likelihood}
\end{equation}
where $n_{\textrm{dim}} = N_p^2$ and $\mathbf{C}$ represents the full covariance matrix
($\mathbf{C}=\mathbf{S}+\mathbf{N}$). The difficulties arising from a
direct evaluation of the Equation~\eqref{eq:likelihood} become evident, since it
requires the calculation of the determinant of $\mathbf{C}$ and, in a
typical CMB experiment, $\mathbf{S}$ is diagonal in harmonic space, whereas $\mathbf{N}$ is usually close to diagonal in pixel space.

Using the Bayes theorem, we write the likelihood as 
$P(\mathbf{C}|\mathbf{d})\propto \pi(\mathbf{S}) P(\mathbf{d}|\mathbf{C})$, where $P(\mathbf{C}|\mathbf{d})$ 
is the {\it a posteriori} probability density and $\pi(\mathbf{S})$ is the prior on the power spectrum.
In \citet{Jewell:2002dz} and \citet{Wandelt:2003uk}, an MC
method was devised in which $P(\mathbf{C}|\mathbf{d})$ is estimated by
building a Markov chain whose stationary state is the joint probability
distribution $P(\mathbf{C,s,d}) = \pi(\mathbf{S}) P(\mathbf{d} | \mathbf{s})P(\mathbf{s} |
 \mathbf{C})$, and assuming that $\pi(\mathbf{S}) = const$.
In such a state, samples drawn along the chain can be used to recover $\mathscr{L}$ by marginalizing over
the unknown true temperature signal $\mathbf{s}$ with a Blackwell--Rao
estimator \citep{Chu:2004zp,Dunkley:2008mk}. The transition
probabilities between two chain states are taken to be the conditional
probabilities $P(\mathbf{s}|\mathbf{C,d})$ and
$P(\mathbf{C}|\mathbf{s,d})$. Thus, at a certain state in the chain,
the current value of $\mathbf{S}$ \text{and} the measured sky
temperature map $\mathbf{d}$ are used to extract
a new signal map $\mathbf{s}$, leading to a
new signal covariance matrix $\mathbf{S}$. The method has been successfully
applied to MC simulations using high-resolution sky maps in
\citet{Eriksen:2004ss}, where procedures to deal with the effects of
foregrounds and to properly treat the presence of noncosmological
monopoles and dipoles have also been presented. Later, the method was
naturally extended beyond the $tt$ covariance matrix estimation,
allowing for polarization measurements to be included in the vector 
$\mathbf{d}$ \citep{Larson:2006ds}. Recently this approach was applied 
to characterize localized secondary anisotropies \citep{Bull2015}.

In this paper, we apply the extended Gibbs sampling method of
\citet{Larson:2006ds} to a combined CMB--galaxy survey experiment whose
measurements can be represented by a vector $\mathbf{d}$ which, in our
case, is the image of a mapping from the 2-sphere $S^2$ into $\mathbb{R}^2$:
\begin{eqnarray*}
f:&S^2& \longrightarrow \mathbb{R}^2 \\
&(\theta,\phi) & \mapsto (\Delta_t,\Delta_g) . 
\end{eqnarray*}
As in the CMB case, these measurements contain the primordial signal, instrumental 
noise, and, in principle,
irreducible foregrounds. The signal is also subject to the effect of the
experiment beam, represented here by a matrix $\mathbf{B}$. Neglecting
the foregrounds for a while, one can write $\mathbf{d}=\mathbf{Bs+n}$ 
\footnote{See ahead how one can deal with residual foregrounds.}. We are interested in
extracting from $\mathbf{d}$ a Bayesian estimate of the primordial
signal covariance matrix $\mathbf{S}$, provided that the full noise
covariance matrix $\mathbf{N}$ is known. In harmonic space, we can
write $\mathbf{s}$ as
\begin{equation}
\mathbf{s}^{\textrm{T}}=(\mathbf{s}^{tg}_{00},\mathbf{s}^{tg}_{10},\mathbf{s}^{tg}_{11},\cdots,\mathbf{s}^{tg}_{\ell_{max}0}\cdots,\mathbf{s}^{tg}_{\ell_{max}\ell_{max}})
\end{equation}
and matrix $\mathbf{S}$ as
\begin{equation}
\mathbf{S}=\textrm{diag}(\mathbf{S}_0^{tg},\mathbf{S}_1^{tg},\mathbf{S}_1^{tg},\cdots,\mathbf{S}_{\ell_{max}}^{tg},\cdots,
\mathbf{S}_{\ell_{max}}^{tg})
\end{equation}
with repeated elements along its diagonal associated with $m=0,\cdots,\ell$. We will assume that the noise
covariance is diagonal in pixel space, that is,
$\mathbf{N}=\textrm{diag}(\mathbf{N}_{11}^{tg},\cdots,\mathbf{N}_{ii}^{tg}, \cdots, \mathbf{N}_{pp}^{tg})$, with
\begin{equation}
\mathbf{s}^{tg}_{\ell m}=\left(
\begin{array}{c}
a_{\ell m}^t \\
a_{\ell m}^g 
\end{array}
\right), 
\end{equation}
and
\begin{equation}
\mathbf{S}^{tg}_{\ell}=\left(
\begin{array}{cc}
C_{\ell}^{tt} & C_{\ell}^{tg}\\
C_{\ell}^{tg} & C_{\ell}^{gg} 
\end{array}
\right),
\quad
\mathbf{N}^{tg}_{ii}=\left(
\begin{array}{cc}
N_{ii}^{tt} & 0\\
0 & N_{ii}^{gg} 
\end{array}
\right),
\label{eq:matrices}
\end{equation}
since the noises affecting $\Delta_t$ and $\Delta_g$ are
uncorrelated. One also sees that $\mathbf{S}$ is in fact block-diagonal.

In order to draw samples from the joint probability distribution
$P(\mathbf{C,s,d})$, we have to set up all the Markov chain machinery described in
\citet{Jewell:2002dz,Wandelt:2003uk,Eriksen:2004ss,Larson:2006ds}, and \citet{Dunkley:2008mk}. In particular, the conditional probability
$P(\mathbf{S}|\mathbf{s,d})=P(\mathbf{S}|\mathbf{s})$ was shown in \citet{Larson:2006ds} to be
proportional to an inverse-Wishart distribution \citep{gupta1999matrix}:
\begin{equation}
P(\mathbf{S}|\mathbf{s}) \propto
\pi(\mathbf{S})\prod_{\ell}\frac{1}{\sqrt{|\mathbf{S}_{\ell}|^{2\ell+1}}}\textrm{exp}\left[-\frac{1}{2}\textrm{Tr}\left(\bm{\sigma}_{\ell}\mathbf{S}_{\ell}^{-1}
  \right)\right],
\label{eq:iwishart}
\end{equation}
where $\bm{\sigma}_{\ell}$ is a $2 \times 2$ matrix
containing the total primordial signal power at scale $\ell$,
\begin{equation}
\bm{\sigma}_{\ell}=\sum_{m=-\ell}^{\ell}\mathbf{s}_{\ell m}^{tg}\mathbf{s}_{\ell m}^{tg\dagger}.
\end{equation}

The signal map samples are drawn according to
$P(\mathbf{s}|\mathbf{S,d})$, which is the {\it a posteriori}
probability density of the Wiener-filtered data, given the power
spectrum of the primordial signal and the observed data. This posterior is a multivariate
Gaussian with mean $\bm{\mu}=\mathbf{SB}^{\textrm{T}}(\mathbf{N}+\mathbf{BSB}^{\textrm{T}})^{-1}\mathbf{d}$ and
covariance $\mathbf{C}=(\mathbf{S}^{-1}+\mathbf{B}^{\textrm{T}}\mathbf{N}^{-1}\mathbf{B})^{-1}$ \citep{hobson2010bayesian}.

To produce these samples, we have used the 
transformed white-noise sampling technique, where a linear transformation 
is applied over initially
independent Gaussian random variables in order to induce the
appropriate covariance among them. In \citet{Eriksen:2004ss}, it was shown that an
elegant way to do that is to write the primordial signal as a sum of
a Wiener-filtered map $\boldsymbol{x}$ and a fluctuation field
$\boldsymbol{y}$, i.e., $\mathbf{s}=\boldsymbol{x+y}$. If $\bm{\chi}$ and $\bm{\xi}$ are Gaussian random
variables with zero mean and unit variance, then $\boldsymbol{x}$ and $\boldsymbol{y}$
can be obtained by solving the following $n_{dim}=(2\times
(\ell_{max}+1)^2)^2$-dimensional linear systems:\footnote{In fact, to
  save computer time, the system solved is the sum of these two
  equations, that is, we solve for the combined field $\boldsymbol{x}+\boldsymbol{y}$.}
\begin{eqnarray}
\left[\mathbf{1+ S^{\textrm{\tiny{1/2}}}B}^{\textrm{\tiny{T}}}\mathbf{N^{\textrm{\tiny{-1}}}BS^{\textrm{\tiny{1/2}}}}\right]\mathbf{S^{\textrm{\tiny{-1/2}}}}\boldsymbol{x}&=&\mathbf{S^{\textrm{\tiny{1/2}}}B}^{\textrm{\textrm{\tiny{T}}}}\mathbf{N^{\textrm{\tiny{-1}}}d},\label{eq:linsys0} \\ 
\left[\mathbf{1+ S^{\textrm{\tiny{1/2}}}B}^{\textrm{\tiny{T}}}\mathbf{N^{\textrm{\tiny{-1}}}BS^{\textrm{\tiny{1/2}}}}\right]\mathbf{S^{\textrm{\tiny{-1/2}}}}\boldsymbol{y}&=&
\bm{\xi}+\mathbf{S^{\textrm{\tiny{1/2}}}B^{\textrm{\tiny{T}}}N^{\textrm{\tiny{-1/2}}}}\bm{\chi}.
\label{eq:linsys}
\end{eqnarray}

To speed up the process, this system is solved iteratively by using a Conjugate
Gradient (CG) method, provided that a good preconditioner matrix is chosen that 
approximates the inverse of the matrix $\mathbf{1+
  S^{1/2}B^{T}N^{-1}BS^{1/2}}$ appearing on the left-hand side of Equation
\eqref{eq:linsys}. Here we have used a block-diagonal preconditioner
$\mathbf{M}=\textrm{diag}(\mathbf{M}_{00},...,\mathbf{M}_{\ell m},...,\mathbf{M}_{\ell_{max}\ell_{max}})$, whose
elements in harmonic space are simply
\begin{equation}
\mathbf{M}_{\ell
  m}=\left[\mathbf{1}+\mathbf{S}_{\ell}^{1/2}\mathbf{B}^{T}_{\ell}\mathbf{N}^{-1}_{\ell
    m}\mathbf{B}_{\ell}\mathbf{S}^{1/2}\right]^{-1},
\label{eq:preconditioner}
\end{equation}
with the harmonic space beam matrix given by
$\mathbf{B}_{\ell}=\textrm{diag}(B_{\ell}^t,B_{\ell}^g)$. The
inverse noise covariance matrix $\mathbf{N}^{-1}_{\ell
  m}=\textrm{diag}(1/N^{tt}_{\ell m},1/N^{gg}_{\ell m})$ is obtained
by decomposing the corresponding pixel-space inverse
noise RMS maps into spherical harmonics $N^{-1}(\theta,\phi)=\sum_{\ell
m}a_{\ell m}^{(n)}Y_{\ell m}(\theta,\phi)$. In the case of partial and/or
nonuniform sky coverage, $N^{-1}(\theta,\phi)$ is set to zero at
unobserved pixels and proportional to the number of observations in
all the other pixels. This, in turn, provides an elegant
way to deal with foregrounds, where the statistical significance of
contaminated pixels can be set to zero by taking
$N^{-1}(\theta,\phi)=0$ for these pixels. This is the so-called COMMANDER 
method \citep{Eriksen:2004ss}. Thus, one can write \citep{Hivon:2001jp}
\begin{equation}
N^{-1}_{\ell m}=(-1)^{m}(2\ell+1)\sum_{\ell'=0}^{\ell_{max}}a_{\ell'  0}^{(n)}\left(\frac{2\ell'+1}{4\pi}\right)^{1/2}
\alpha_{\ell\ell'm}
\label{eq:noisestruct}
\end{equation}
with
\begin{equation}
\alpha_{\ell\ell'm}=
\left(
\begin{array}{ccc}
\ell & \ell & \ell' \\
0 & 0 & 0
\end{array}
\right)
\left(
\begin{array}{ccc}
\ell & \ell & \ell' \\
m & -m & 0
\end{array}
\right).
\end{equation}

%%%%%%%%%%%%%%%%%%%%%%%%%%%%%%%%%%%%%%%%%%%%%%%%%%%%%%%%%
\section{Data sets, Processing, and analysis pipeline}
\label{sec:processing}
%%%%%%%%%%%%%%%%%%%%%%%%%%%%%%%%%%%%%%%%%%%%%%%%%%%%%%%%%
\begin{figure*}[ht]
\begin{center}
\subfigure[]{
  \includegraphics[width=.48\textwidth]{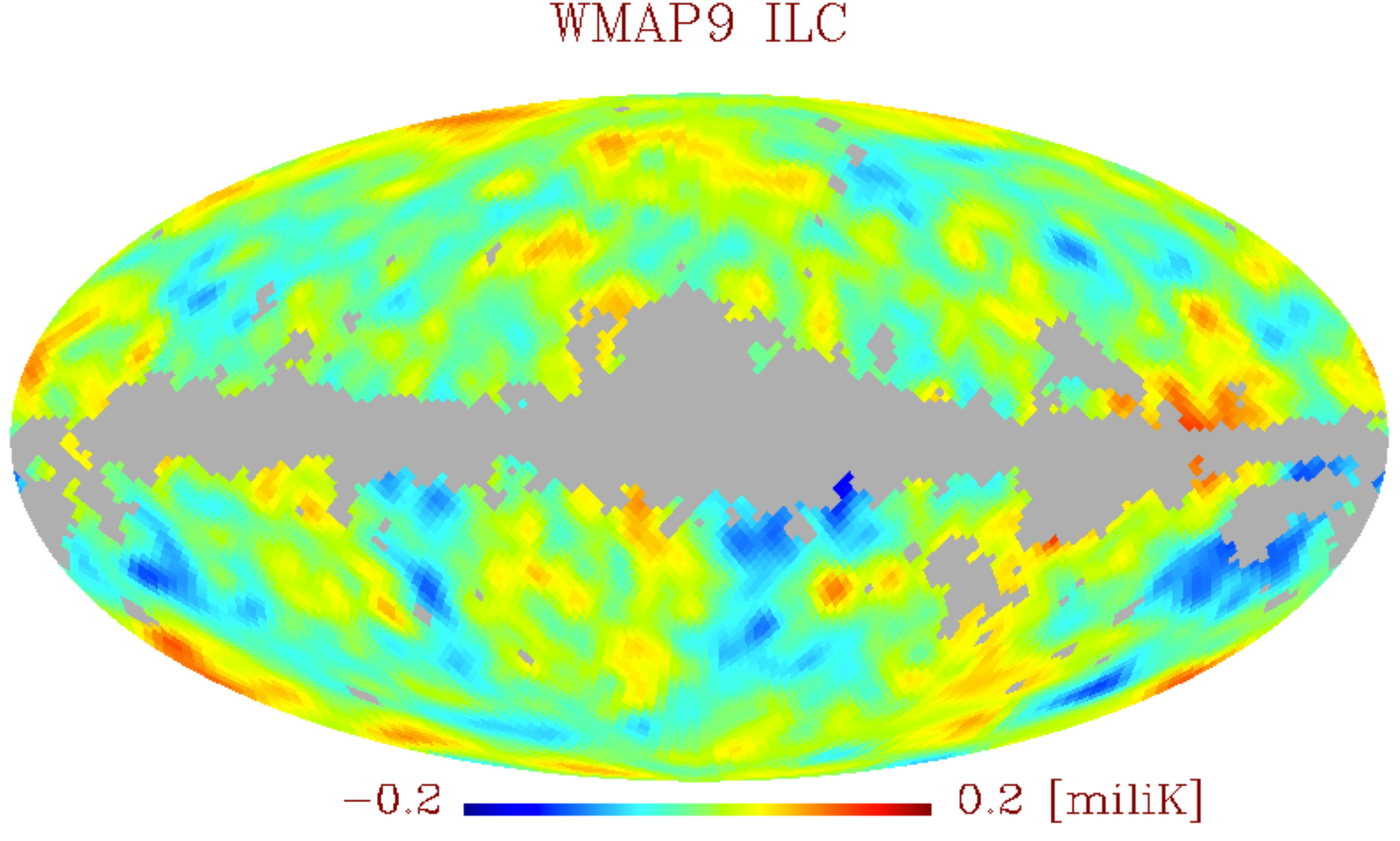}
}%
\subfigure[]{
  \includegraphics[width=.48\textwidth]{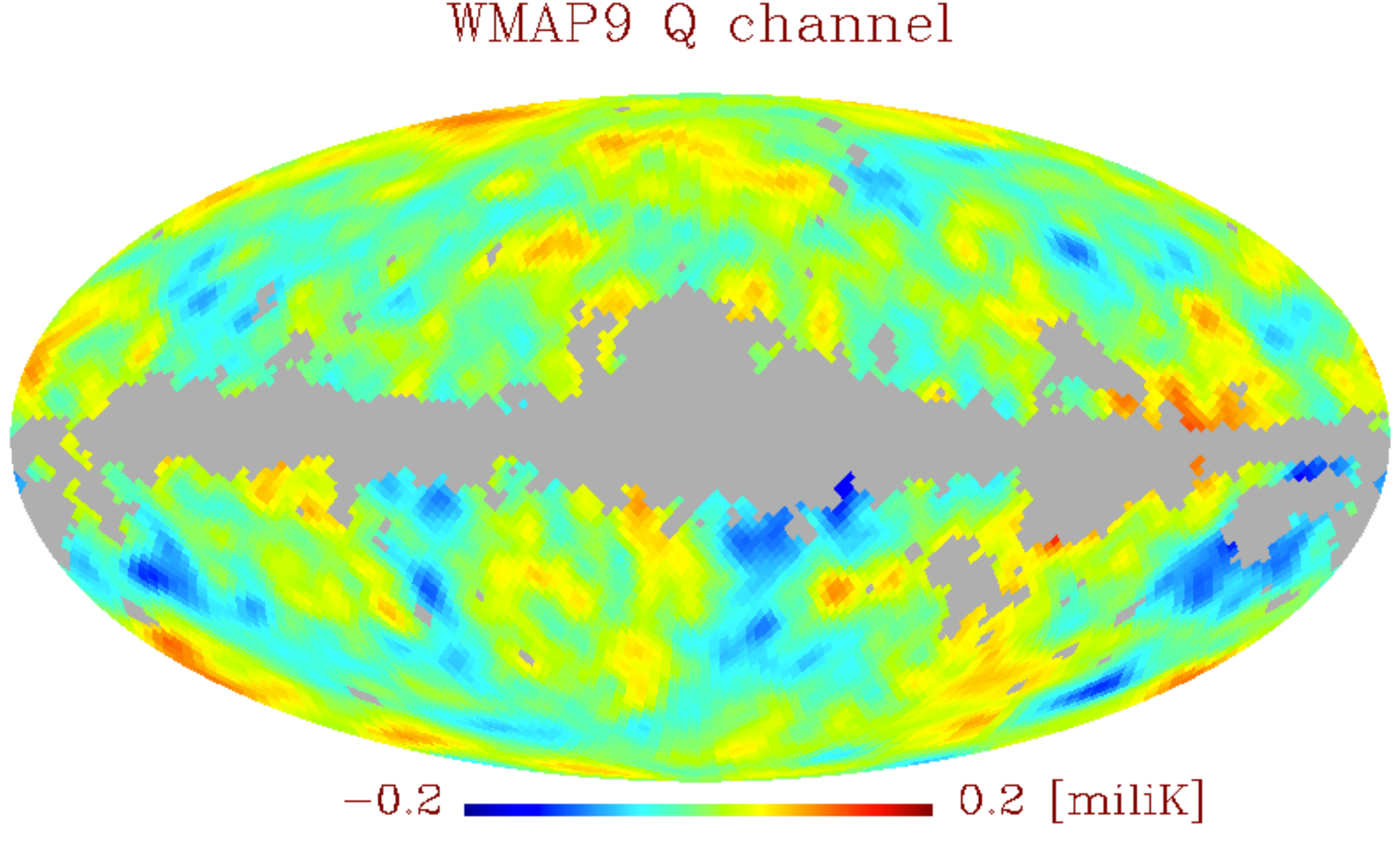}
} \\ % end of first row
\subfigure[]{
  \includegraphics[width=.48\textwidth]{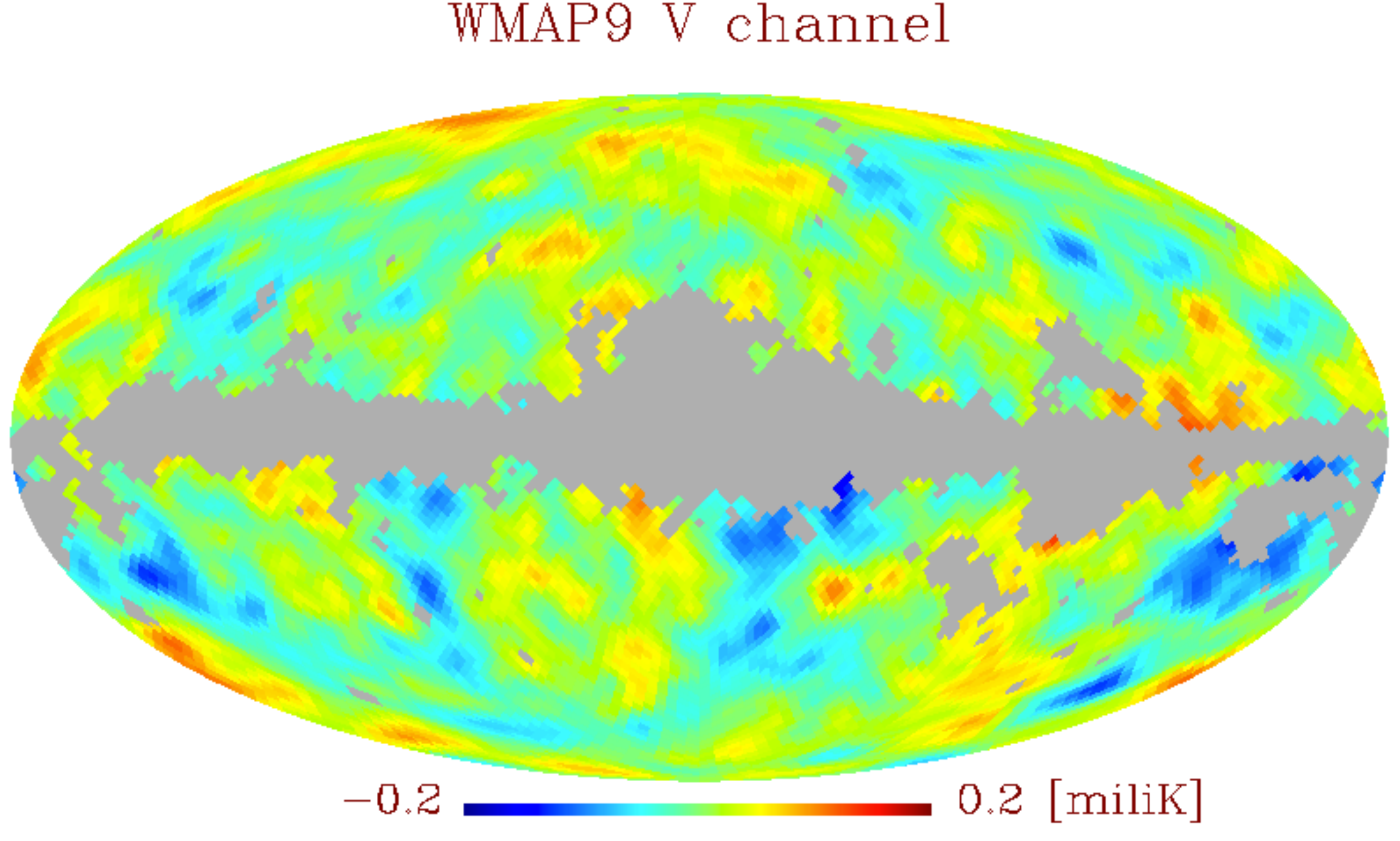}
}%
\subfigure[]{
  \includegraphics[width=.48\textwidth]{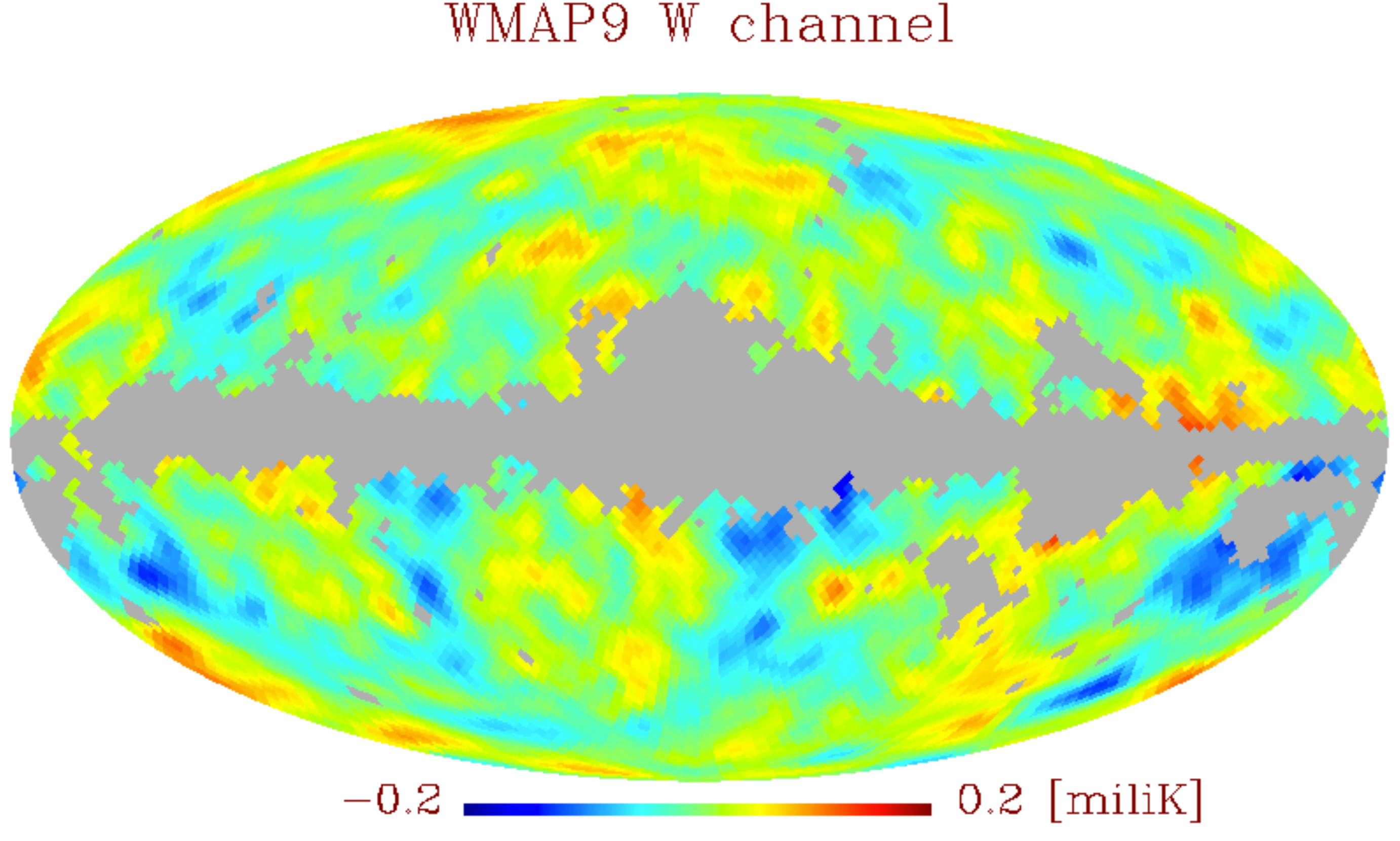}
}%
\end{center}
\caption{Mollweide projections in galactic coordinates of the four {\it WMAP}9 temperature maps (ILC,
  {\it Q}, {\it V} and {\it W} channels) in HEALPix
  pixelization ($nside=32$). A Gaussian beam of
  $4^\circ .9$ (see text) has been applied to the ILC map and $5^\circ$
  to the other three maps, before the primary
  temperature analysis mask KQ85 was used.}
\label{fig:wmap9maps}
\end{figure*}

As explained in Section \ref{sec:gibbs}, the Bayesian method
underlying the Gibbs sampling relies on the fact that the 
statistical properties of the instrumental noise are known through the covariance matrix ${\bf
  N}$. Moreover, the conditional probabilities presented there assume that the fluctuations in the signal and in the noise are Gaussian. Finally, to ensure that the algorithm converges, the signal has to be sampled to sufficiently large multipoles, that is, up to
regions of low signal-to-noise ratio (S/N).

In this work, we have used the data from {\it WMAP}9 {\it Q} (40 GHz), {\it V} (60 GHz) and {\it W} (90
GHz) channels, as well as the Internal Linear Combination (ILC)
CMB temperature map \footnote{http://lambda.gsfc.nasa.gov}
\citep{Bennett:2012zja,Hinshaw:2012aka}, formed from the
linear combination of five smoothed intensity maps (taken from the three
previous channels) by minimizing the temperature variance. 
We have verified that the noise power observed in the high-resolution
temperature intensity ({\it I}) maps of {\it WMAP}9 is fairly well modeled by
uncorrelated and Gaussian fluctuations with variances given by 
$\sigma_0^2/N_{obs}$, and $\sigma_0$ for each channel taken from \cite{Greason:2012}. To
avoid sampling the signal at very high multipoles, where the CMB power
spectrum can be alternatively estimated by the MASTER algorithm, they
have applied an effective $5^\circ$ beam to the ILC map, followed by
the addition of uncorrelated white noise of appropriate RMS strength
($\sigma_0= 2 \ \mu$K). The noise amplitude is chosen so that (a) it
overcomes the correlation introduced in the noise at small scales by the beam; (b) the added 
noise dominates over the primordial signal for
multipoles $\ell \gsim 60$, while keeping the likelihood essentially
unchanged at large scales; and (c) the statistical properties of
the added noise are completely known.
\begin{figure*}[ht]
\begin{center}
\subfigure[]{
  \includegraphics[width=.48\textwidth]{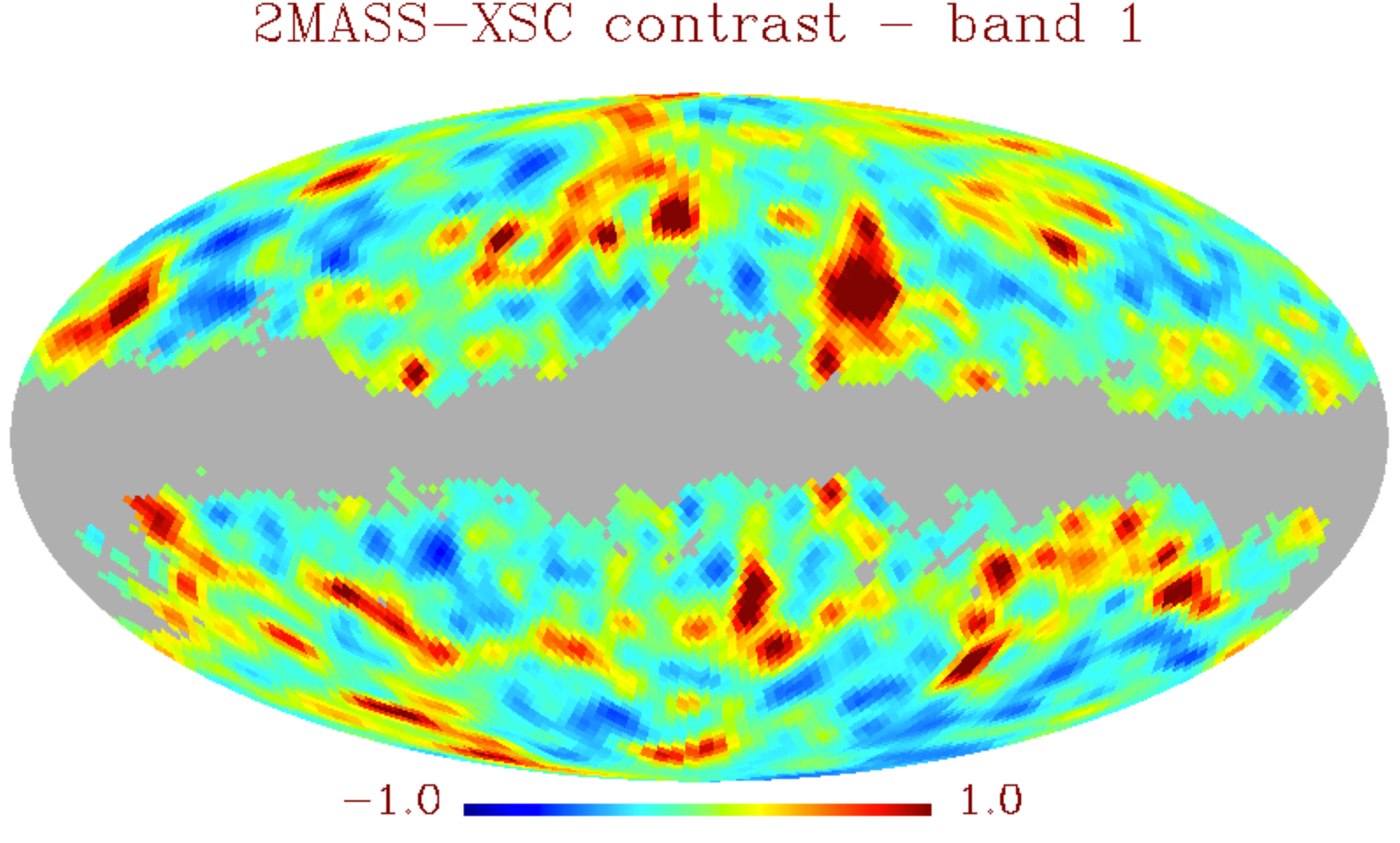}
}%
\subfigure[]{
  \includegraphics[width=.48\textwidth]{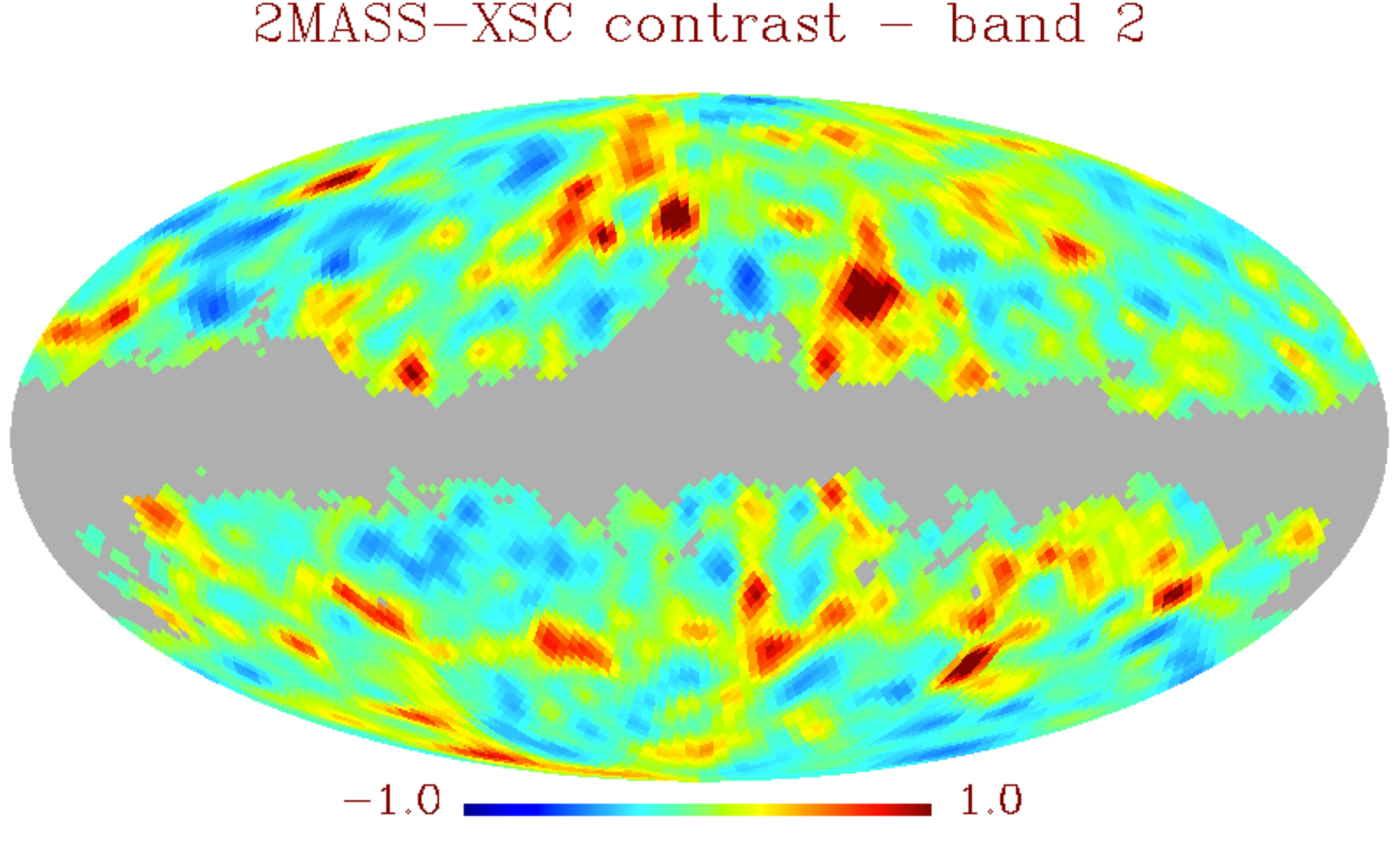}
} \\ % end of first row
\subfigure[]{
  \includegraphics[width=.48\textwidth]{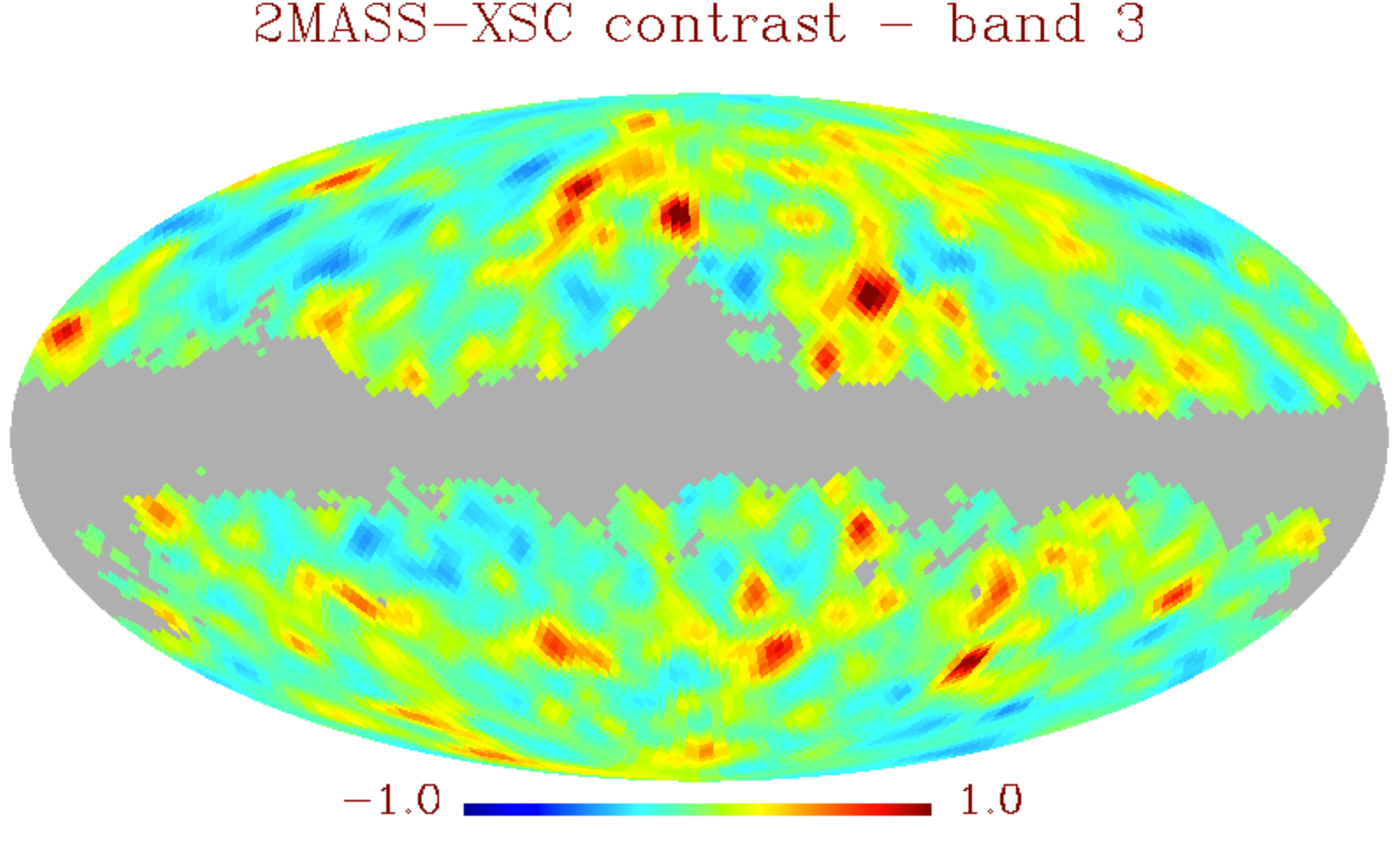}
}%
\subfigure[]{
  \includegraphics[width=.48\textwidth]{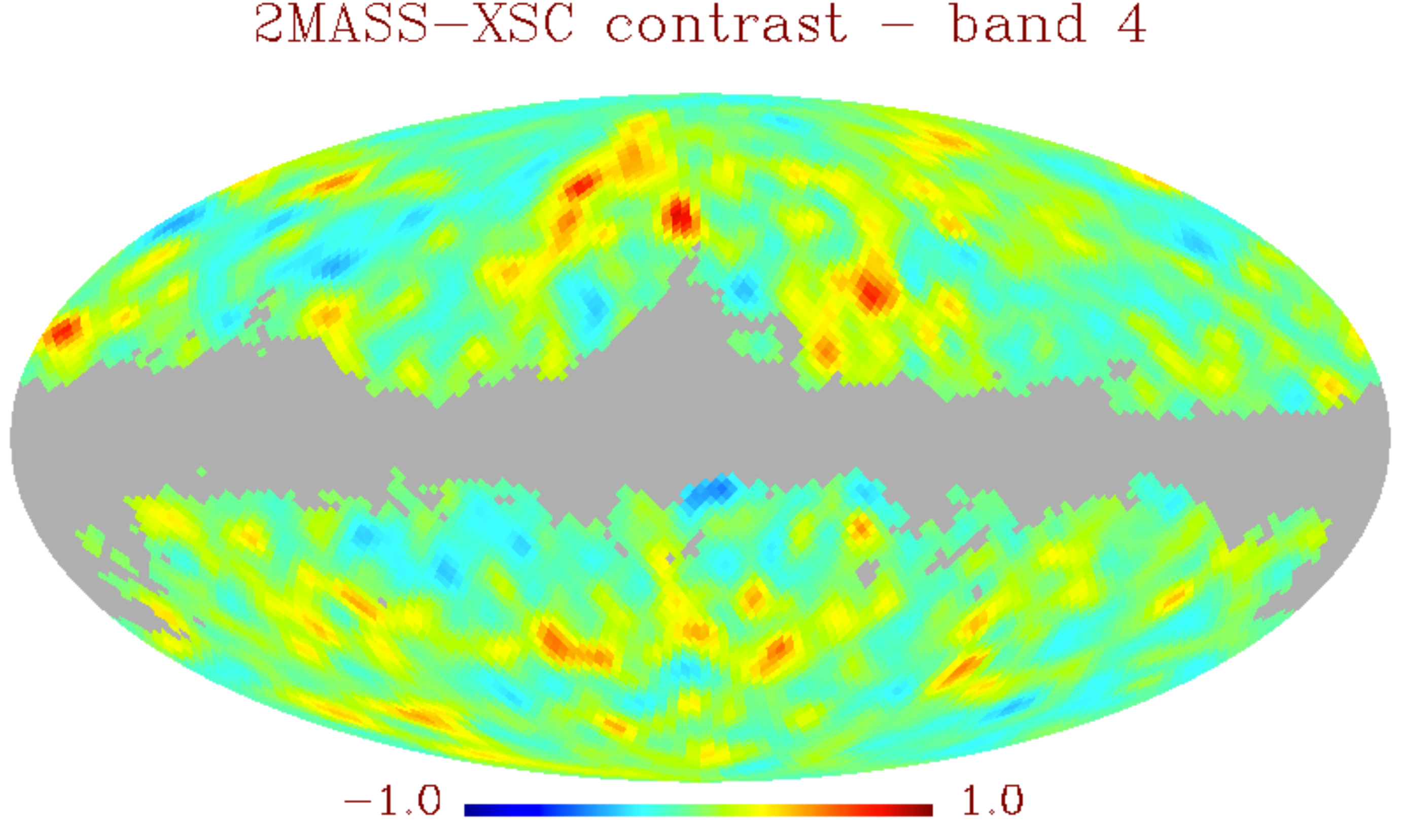}
}%
\end{center}
\caption{Mollweide projections in galactic coordinates of the 2MASS--XSC contrast maps in HEALPix
  pixelization ($nside=32$) for the four corrected
  $K_{20}$ magnitude bands. A Gaussian beam of $5^\circ$ FWHM has been
  applied. Pixels were masked to avoid the intense
  emission and dust extinction close to the galactic plane.}
\label{fig:2massmaps}
\end{figure*}

The high-resolution
maps provided by the LAMBDA website ($nside=512$) were degraded to
$nside=32$. The primary temperature analysis mask KQ85 was also degraded
using the method described in \citet{Bennett:2012zja}, where a pixel at the
new resolution is unmasked if at least 50\% of the higher-resolution
pixels are unmasked. In particular, this corresponds to 9496 out of
12,288 pixels, or 77.3\% of the sky. Figure \ref{fig:wmap9maps} shows the final cut-sky 
temperature maps, with a $5^\circ$ Gaussian beam applied in the case of {\it Q-}, {\it V-}, and {\it W-}band maps 
(original beams can be neglected) and an additional $4^\circ .9$ on the ILC map (since it 
is already smoothed at $1^\circ$), in order to get an effective beam of $5^\circ$.
The resolution degradation is
performed after the beaming.
\begin{figure}[ht]
\begin{center}
  \includegraphics[width=.48\textwidth]{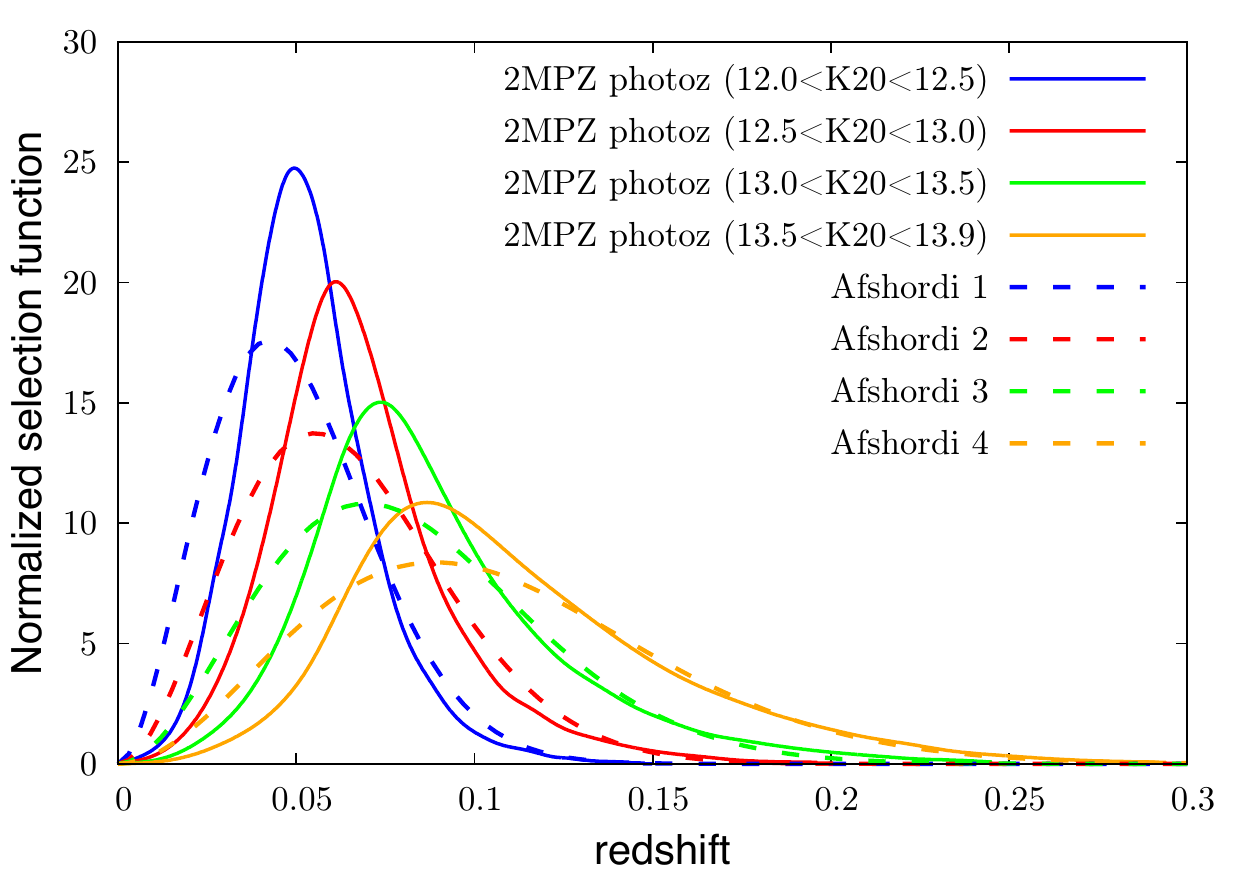}
\caption{Normalized selection functions for the 2MASS catalog. The
  dashed lines show the parameterizations of \cite{Afshordi:2003xu} based on a fit to the
  $K$-band magnitudes distributions, and the solid lines the
  photometric redshift distributions of 2MPZ.}
\label{fig:xsc2mpzsels}
\end{center}
\end{figure}

\begin{figure*}[ht]
\begin{center}
\subfigure[]{
  \includegraphics[width=.48\textwidth]{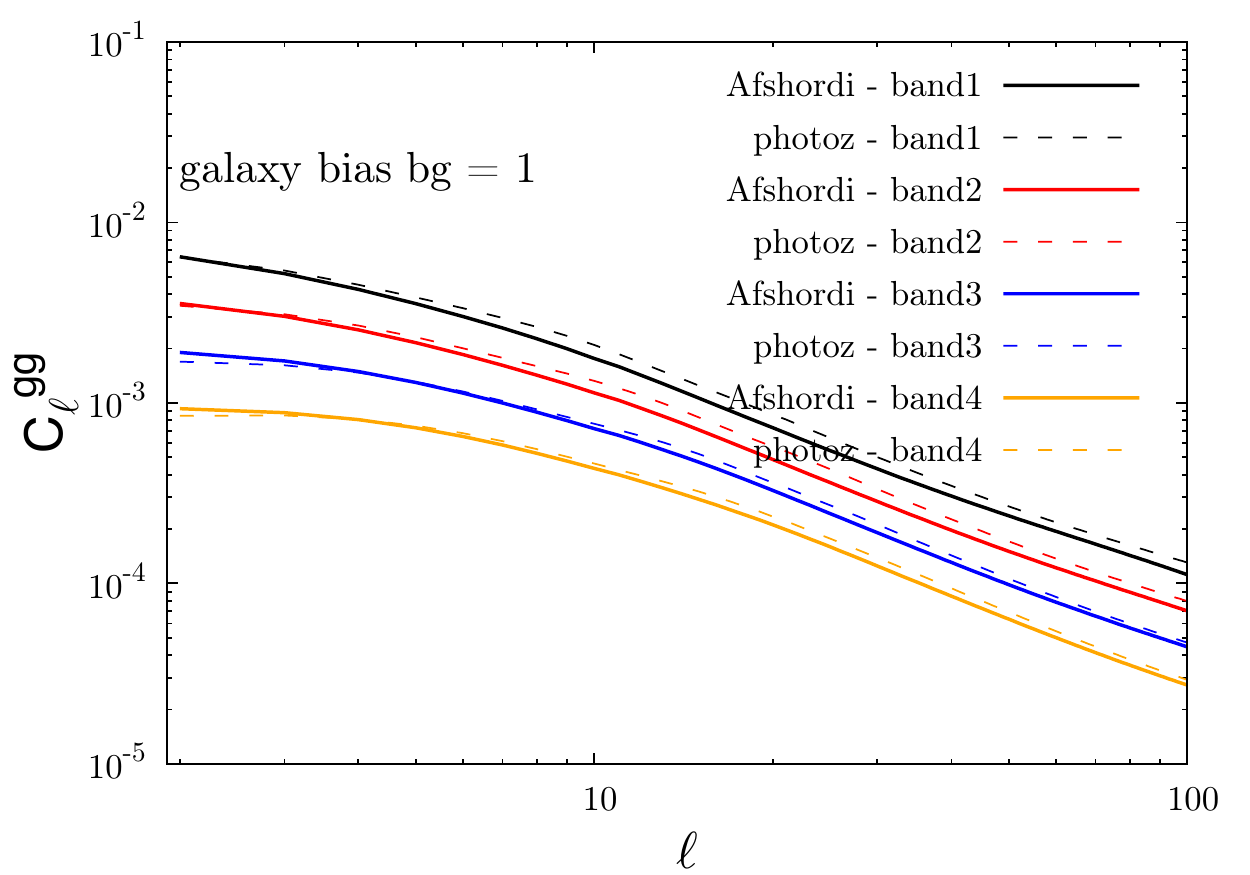}
}%
\subfigure[]{
  \includegraphics[width=.48\textwidth]{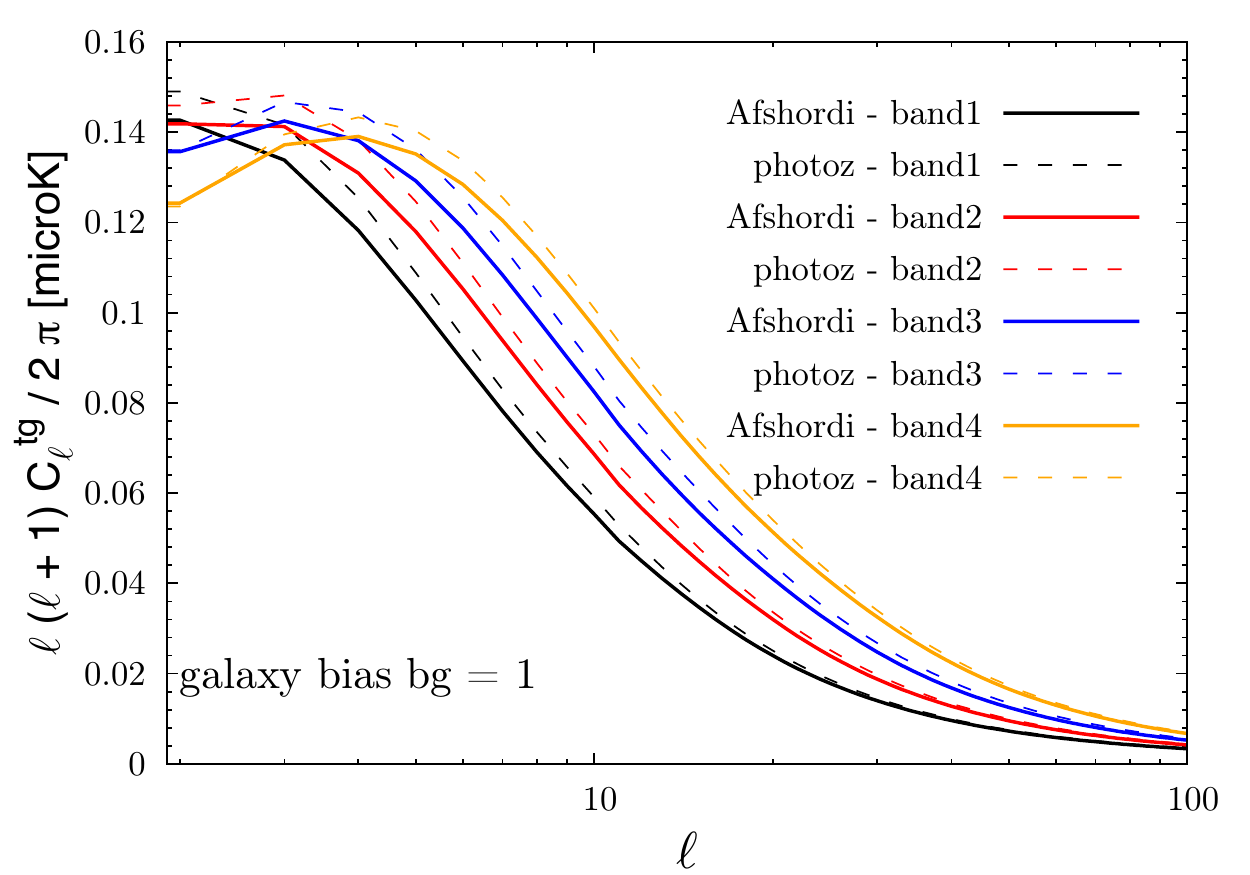}
} \\ % end of first row
\caption{Expected signals for different
  2MASS galaxy selection functions (see eqs. \ref{eq:cl} and \ref{eq:Cls2}). (a) Galaxy autocorrelation signal. (b) CMB--galaxy cross-correlation signal.}
\label{fig:xsc2mpzcompara}
\end{center}
\end{figure*}

As a tracer of the dark matter distribution at large scales, we have used
the XSC (eXtended Source Catalog) of the near-IR 2MASS
\citep{Skrutskie:2006wh}.\footnote{The catalog is publicly
  available at ftp://ftp.ipac.caltech.edu/pub/2mass/allsky/} This
catalog contains positions, photometry, and basic shape information of
1,647,599 resolved sources, most of which are galaxies ($\sim$
97\%). In order to build the galaxy count map, the following
procedure was adopted:

\begin{enumerate}

\item The 2MASS $K_S$-band (2.16 $\mu$m) 20 mag arcsec$^{-2}$
  isophotal circular aperture magnitudes (\texttt{‘k\_m\_i20c’} called
  here simply $K_{20}$) were corrected for galactic extinction using
  the reddening maps \footnote{These reddening maps are available at http://lambda.gsfc.nasa.gov/product/foreground/fg\_sfd\_get.cfm.} $(E(B-V))$
  at 100 $\mu$m of \citet{Schlegel:1997yv}, according to the expression 
  $K_{20} \rightarrow K_{20}' = K_{20} - A_k$, where $A_k = 0.367 E(B-V)$.

\item A HEALPix resolution parameter $nside=32$ was adopted, and pixels
  were masked whenever $A_k>0.05$, which leaves 71.6\% of the sky with
  unmasked pixels; only objects with uniform detection were kept
  (\texttt{cc\_flag} != ``a" and ``z" ) and artifacts have been removed (flag \texttt{use\_src} = 1).

\item The final galaxy counting map has 801,476 objects with
  $12<K'_{20}<14$, which were further divided into four magnitude
  bands. Table~\ref{table:2massbands} summarizes the number of objects 
  in each band, as well as the parameters of the redshift distribution 
  of these sources [see Equation~\eqref{eq:gal}]. Following \citet{Afshordi:2003xu}, 
  this distribution can be parameterized by a generalized gamma function, namely,
\begin{equation}
\frac{dN}{dz}(z|\{\lambda,\beta,z_0\}) dz =
\frac{\beta}{\Gamma(\lambda)}\left(\frac{z}{z_0}\right)^{\beta\lambda-1}
\exp\left[-\left(\frac{z}{z_0}\right)^\beta\right] \frac{dz}{z_0},
\label{eq:window}
\end{equation}
where $\lambda > 0$, $\beta > 0$ and $z_0 >0$. Figure \ref{fig:2massmaps}
shows the four 2MASS final contrast maps, already masked and smoothed
by a $5^\circ$ FWHM Gaussian beam. The input map has $nside=64$, and then
the beam is applied (excluding masked pixels) and
the resolution is degraded to $nside=32$. 

\end{enumerate}

\begin{table*}[ht]
\setlength{\tabcolsep}{8pt}
\caption{Summary of 2MASS Galaxy Counts in Each of the Four $K_{20}'$
  Magnitude Bands and the Parameters Describing the Redshift Distribution of These 
  Objects.}
\begin{center}
\begin{tabular}{ccccc} %{p{4cm}cccc}
\hline
\hline
Band & $z_0$ & $\beta$ & $\lambda$ & $N_{\text{total}}$ (Unmasked) \\
\hline
\hline
$12.0<K_{20}'<12.5$ & 0.043 & 1.825 & 1.524 & 51263 \\
$12.5<K_{20}'<13.0$ & 0.054 & 1.800 & 1.600 & 105930 \\
$13.0<K_{20}'<13.5$ & 0.067 & 1.765 & 1.636 & 224345 \\
$13.5<K_{20}'<14.0$ & 0.084 & 1.723 & 1.684 & 469782 \\
\hline 
\hline 
Note: Magnitudes were corrected by reddening.
\end{tabular}
\end{center}
\label{table:2massbands}
\end{table*}

It is worth mentioning that the 2MASS-XSC catalog has
  been recently improved with the addition of photometric redshifts to a large 
  fraction of its sources \citep{2mpzpaper}. This new sample is called the 2MPZ
  (2MASS Photometric Redshift) catalog.\footnote{Available at
    http://ssa.roe.ac.uk/TWOMPZ.html} By cross-correlating XSC with {\it WISE} and
  SuperCOSMOS, the authors were able to obtain photo-$z$'s with errors
  essentially independent of distance, determined with
  a spectroscopic redshift subsample, being the median of relative error of about 12\%. A hard magnitude cut at
  $K_{20}'<13.9$, the completeness limit of 2MASS, was applied over a slightly different magnitude (calculated
  using the isophotal elliptical aperture magnitude
  \texttt{k\_m\_k20fe}, instead of the circular aperture \texttt{‘k\_m\_i20c’} adopted by us). In terms of galaxy counts, this cut 
  introduces differences that are maximum in band 4, reducing the
  counts by about 30\% with respect to the full XSC. However, we have checked
  that the galaxy reduction is mostly uniform over the sky and that their
  partial-sky autocorrelation power spectra, after an $f_{\text{sky}}$
  correction and shot-noise subtraction, are completely consistent,
  within statistical uncertainties, in all four corrected magnitude
  bands.
\begin{figure*}[ht]
\begin{center}
\subfigure[]{
  \includegraphics[width=.47\textwidth]{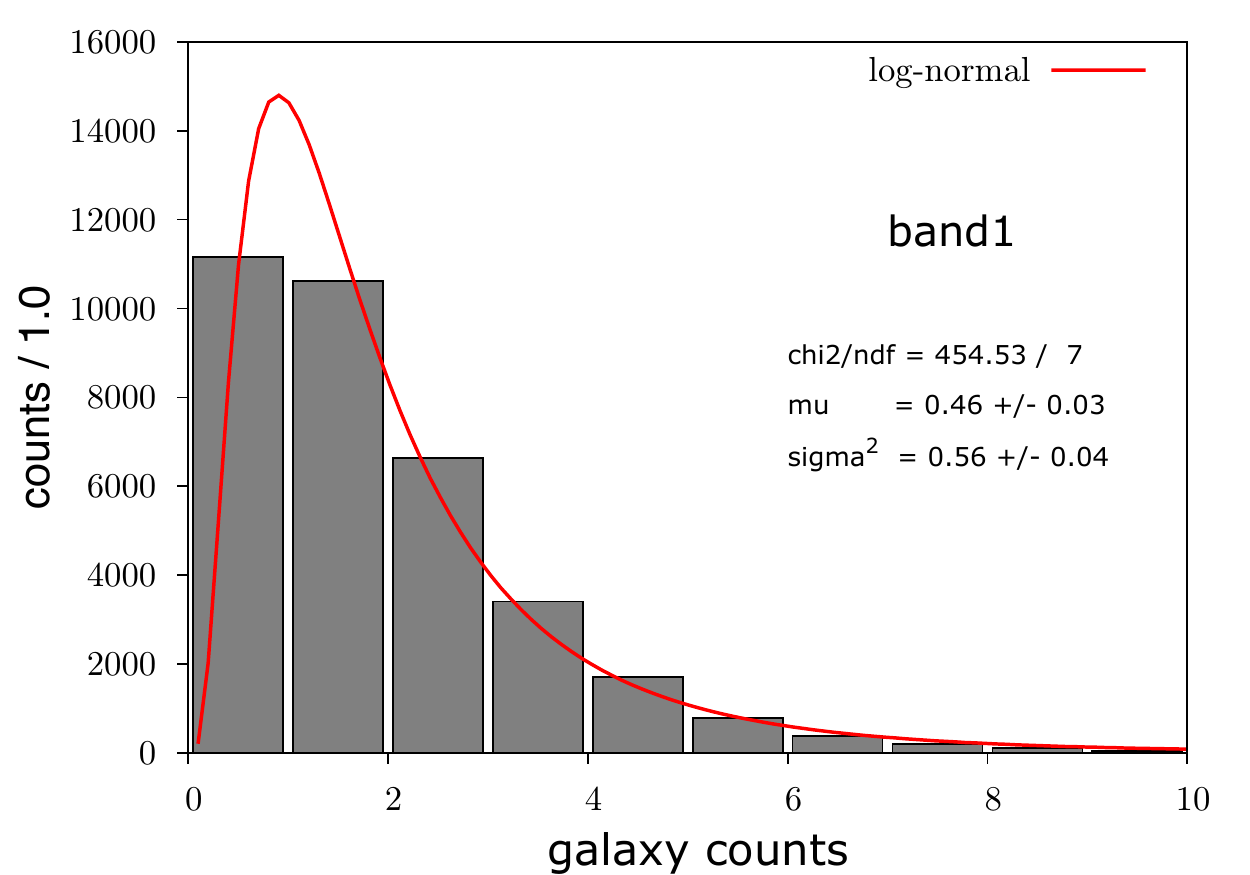}
}%
\subfigure[]{
  \includegraphics[width=.47\textwidth]{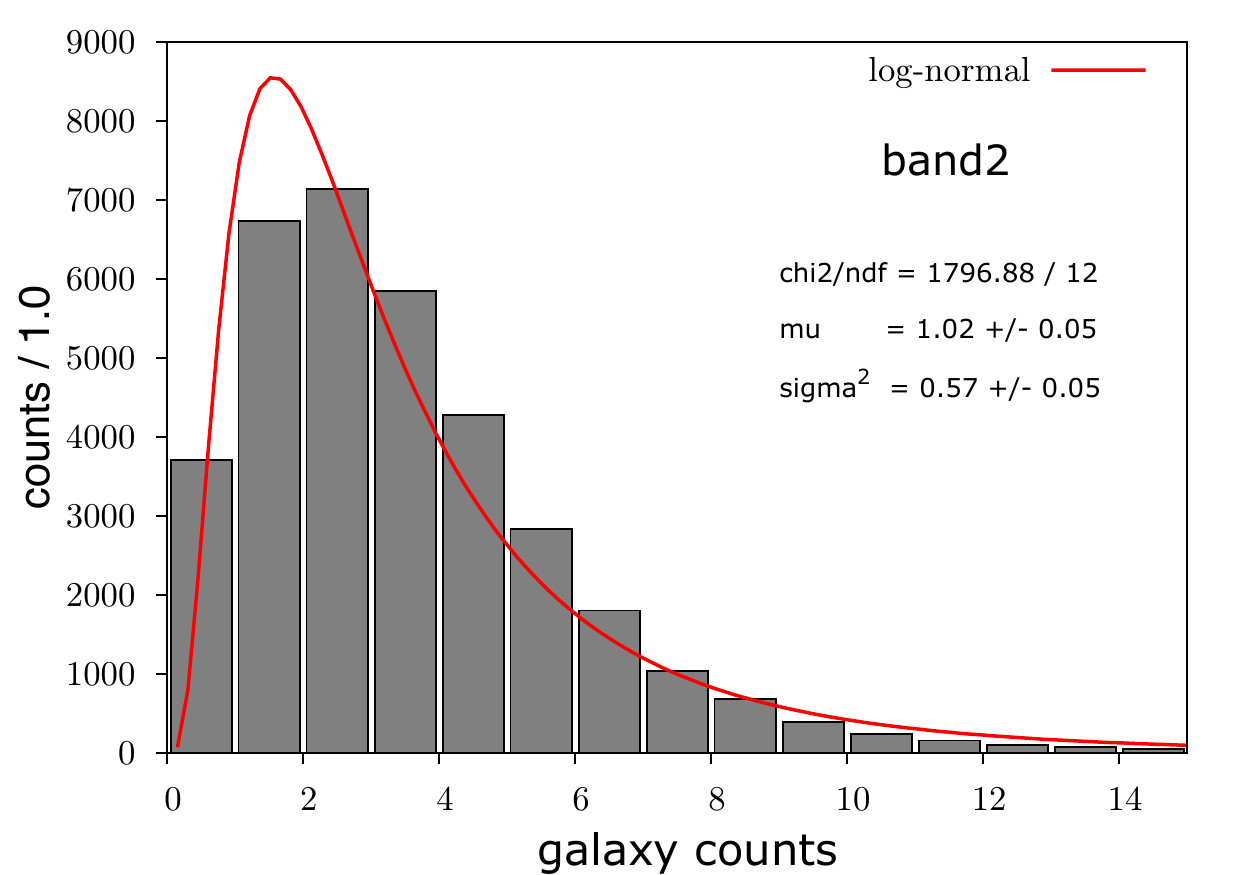}
} \\ % end of first row
\subfigure[]{
  \includegraphics[width=.47\textwidth]{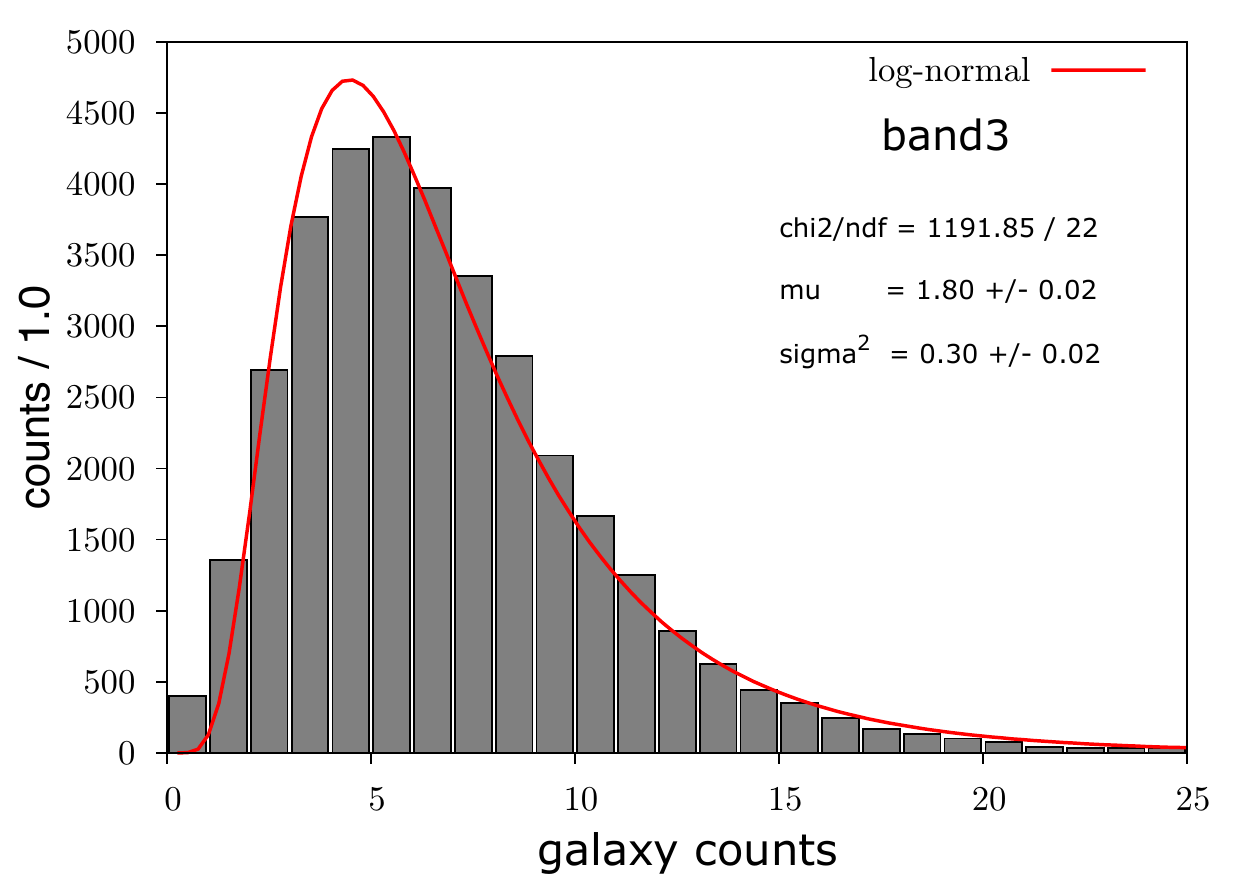}
}%
\subfigure[]{
  \includegraphics[width=.47\textwidth]{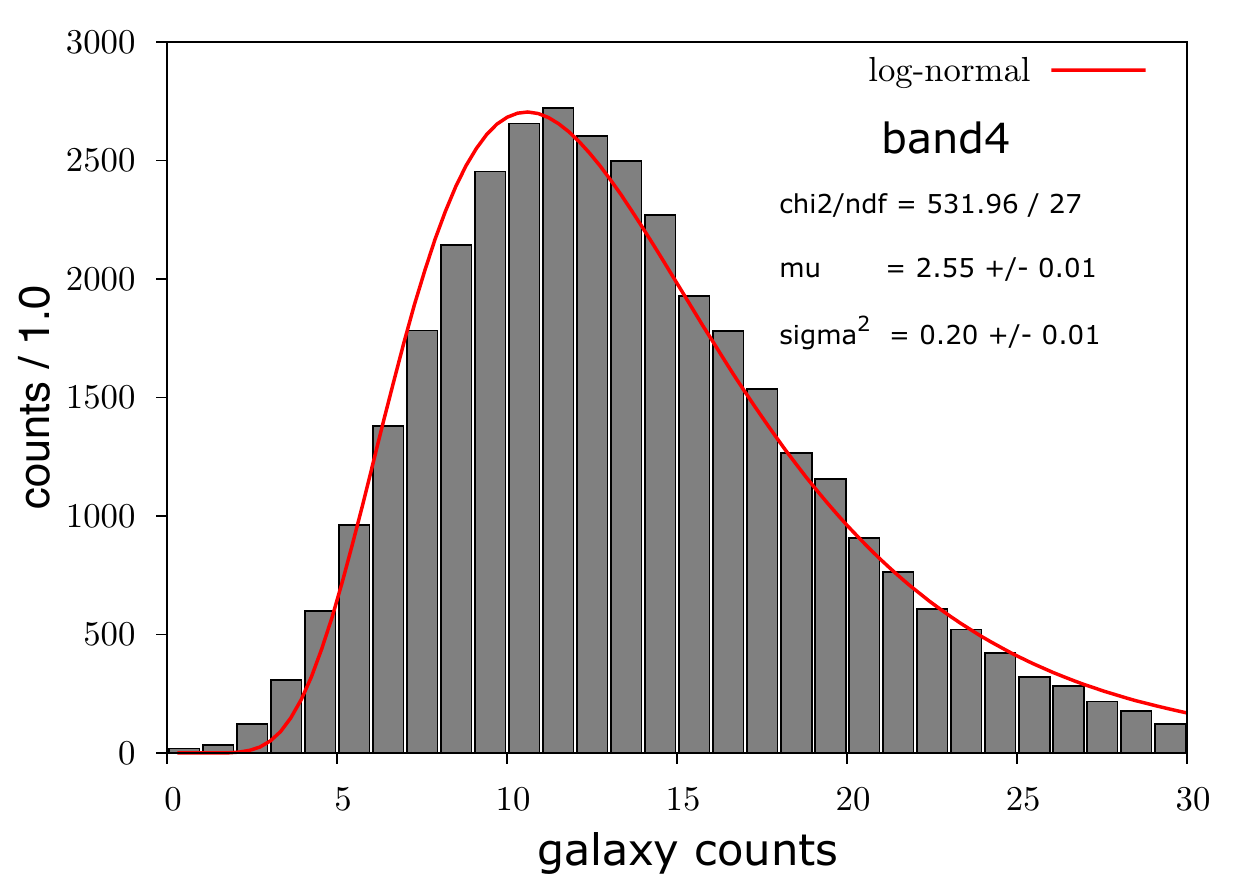}
}%
\caption{Distribution of 2MASS galaxy counts per pixel with resolution
  parameter $nside=64$. A lognormal fit was performed, and the result
  is shown. The average $\mu$ and variance $\sigma^2$ are the ones of
  $\ln(N_{gal}(i))$ at pixel $i$.}
\label{fig:2masscounts}
\end{center}
\end{figure*}

  The expected auto- and cross-correlation signals for both catalogs
  depend on their selection functions. We have compared the Afshordi's
  parameterizations [Equation~\eqref{eq:window}] for XSC with the corresponding
  2MPZ selection functions taken as the photometric redshift
  distributions of sources in this catalog. We see in Figure~\ref{fig:xsc2mpzsels} that
  the parameterized selection functions systematically predict more
  galaxies at low redshifts in comparison to the 2MPZ photo-$z$
  distributions. Nonetheless, for the real data analysis presented in this
  paper, the selection functions are used only to fix the auto-
  and cross-correlation power spectra in the low-S/N ratio
  region of the harmonic space (the large-$\ell$ region) during the covariance matrix sampling 
  along the Gibbs chain. Therefore, here, the important ingredient
  to the analysis pipeline is the result of the convolution presented in
  Equations~\eqref{eq:cl} and \eqref{eq:Cls2} for the region $51< \ell \le 96$
  (see sections \ref{sec:valida} and \ref{sec:results} ahead). Figure
  \ref{fig:xsc2mpzcompara} shows that the expected spectra are very similar in this
  multipole region, with differences much smaller than the expected
  noise. Thus, we will keep using the  XSC catalog in this work, together with the parameterized selection
  functions given by Equation~\eqref{eq:window} and Table~\ref{table:2massbands}. We finally decided 
to add to the 2MASS maps uncorrelated white noise of 5\% pixel$^{-1}$ (RMS amplitude).

\begin{figure*}[ht]
\begin{center}
\subfigure[]{
  \includegraphics[width=.47\textwidth]{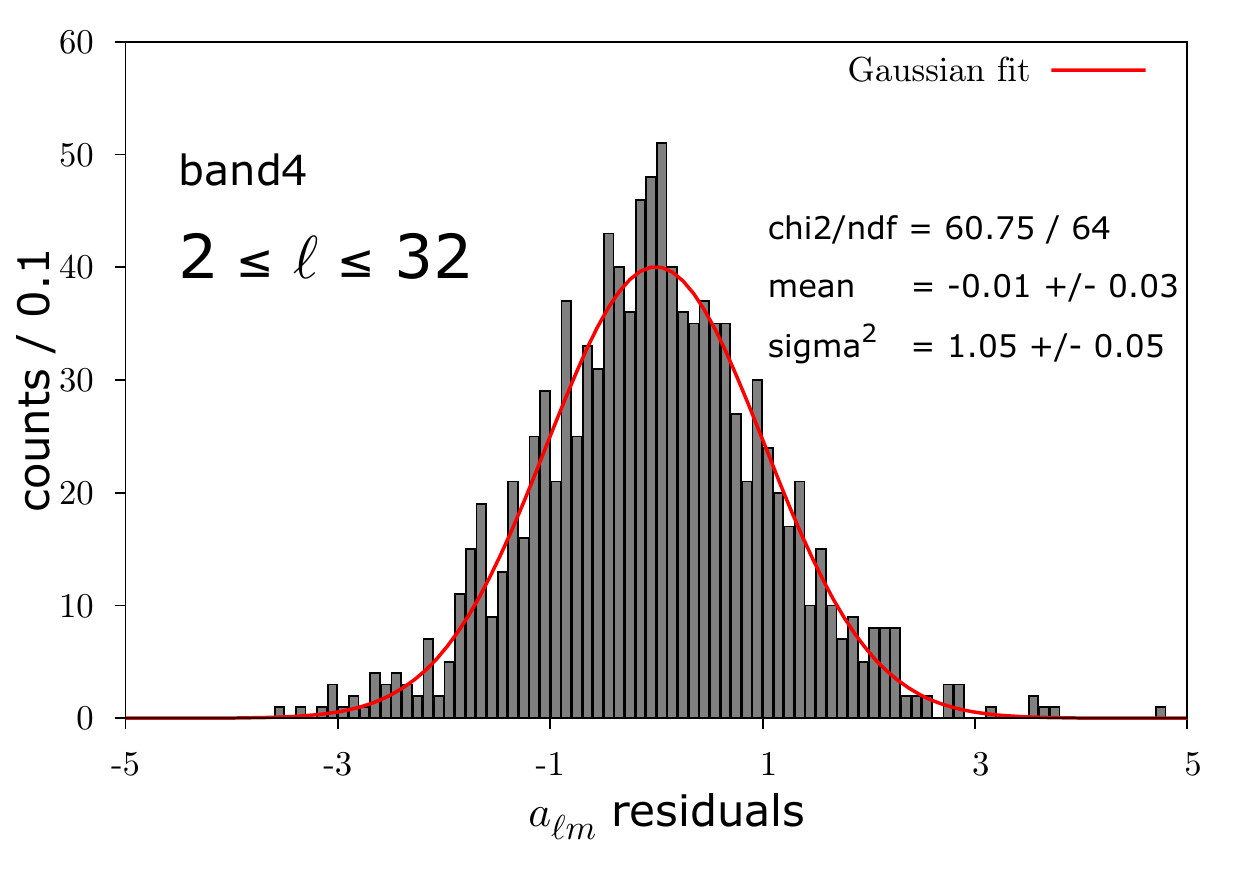}
}%
\subfigure[]{
  \includegraphics[width=.47\textwidth]{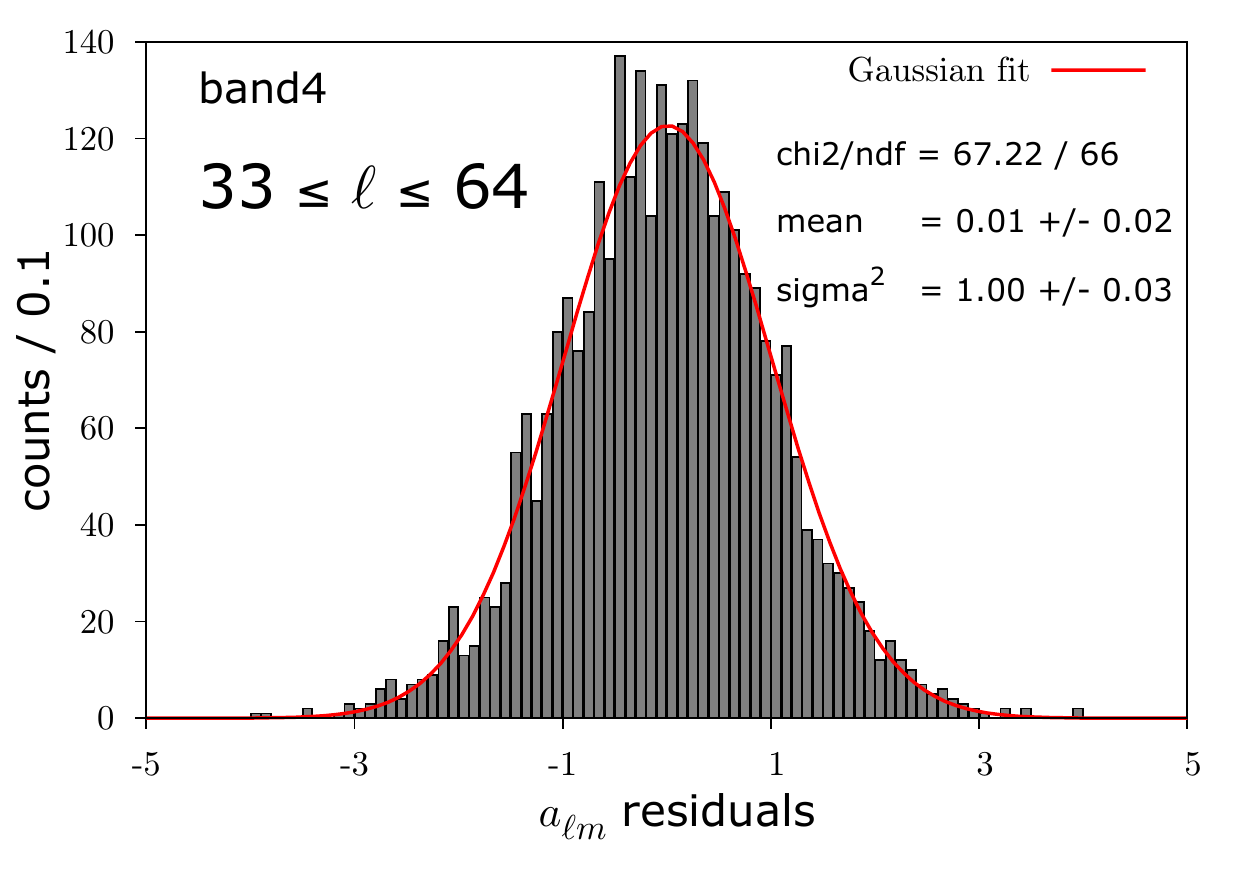}
} \\ % end of first row
\subfigure[]{
  \includegraphics[width=.47\textwidth]{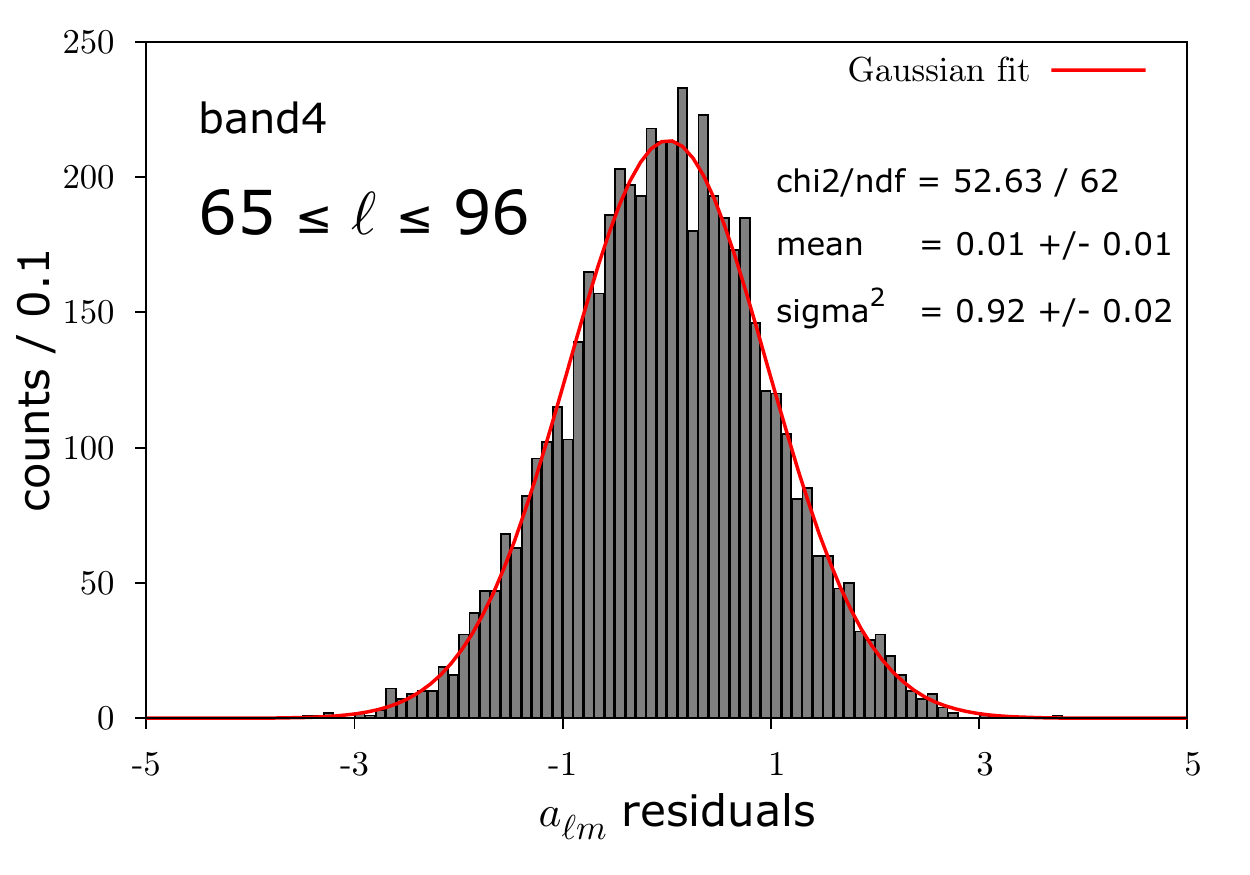}
}%
\subfigure[]{
  \includegraphics[width=.47\textwidth]{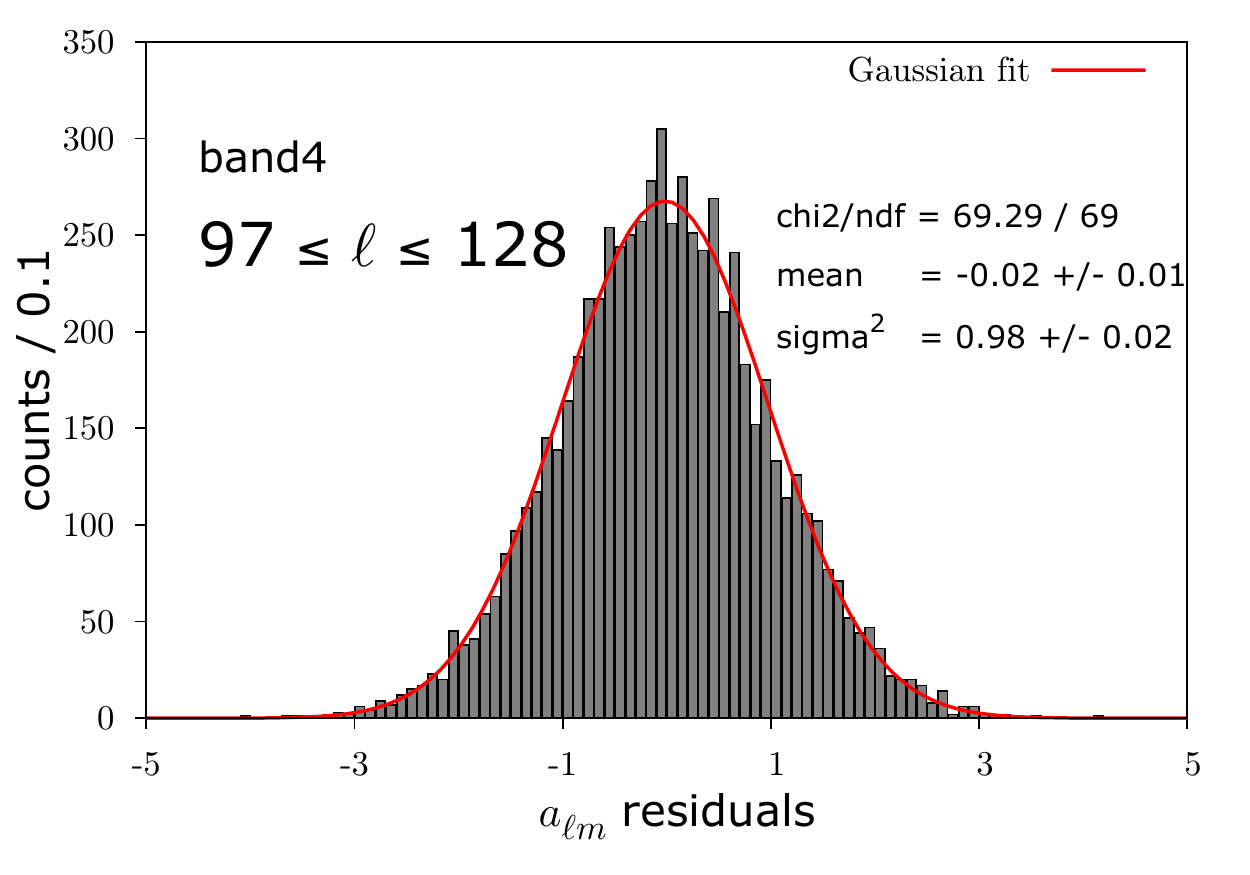}
}%
\caption{Distribution of 2MASS band 4 map residuals in harmonic space,
  according to Equation~\eqref{eq:residuals}, for different ranges of multipoles
  $\ell$ and a variance model given by $\sigma^2=f_{sky}B_{\ell}^2C_\ell + \eta$. The variance 
  $C_{\ell}$ is taken from the fiducial $\Lambda$CDM model, with nonlinearities in the 
  LSS given by CAMB's 
halofit implementation (see Section \ref{sec:valida} for details). The
effect of the mask is partially corrected by rescaling the
$C_{\ell}$ values by an $f_{\text{sky}}$ factor, and the pixelization is also taken
into account. Shot noise is included in the total variance, and the bias factor $b_g$ was taken by fitting
this model to the the survey power spectrum in the region
$2\le\ell\le 50$ (see Table~\ref{table:biasvalues}). The 
decomposed map has $nside = 64$, and no beaming,
except for the pixelization, has been applied. Gaussian fits are shown on
top of each histogram.}
\label{fig:2massflucs}
\end{center}
\end{figure*}

In order to apply the method of Section \ref{sec:gibbs}
to the CMB--galaxy cross-correlation, we first need to test whether
the hypothesis of Gaussianity is, at least approximately, respected in
harmonic space by the distribution of $a_{\ell m}$
coefficients. Especially at small scales, it is hard to predict the
nature of the fluctuations of these coefficients in the nonlinear regime. In
pixel space, lognormality as an approximate property of the galaxy
count distributions in 2-dimensional
cells was first observed by \cite{lognormal1}. This was
proposed as a model when \cite{lognormal2} showed that such a
distribution could be obtained from a purely Gaussian field through
gravitational evolution. Figure~\ref{fig:2masscounts} displays the
distributions of the 2MASS galaxy counts per pixel and evinces that, as the
density of galaxies per pixel increases, the distributions tend to
lognormal. However, one can see that for 2MASS statistics, even for
band 4 with about 13 galaxies pixels$^{-1}$ on average, the $\chi^2/n$ dof is still bad. 
Figure~\ref{fig:2massflucs} shows the distribution of the harmonic space residuals
\begin{equation}
a_{\ell m} \textrm{ residuals} = 
\left\{
\begin{array}{ll}
\sqrt{2}\textrm{Re}(a_{\ell m})/\sigma & \quad (m\ne 0)\\
\sqrt{2}\textrm{Im}(a_{\ell m})/\sigma & \quad (m\ne 0)\\
\textrm{Re}(a_{\ell m})/\sigma             & \quad (m= 0),
\end{array}
\right.
\label{eq:residuals}
\end{equation}
for 2MASS band 4, considering as a model 
for the total variance $\sigma^2=f_{sky}B_{\ell}^2C_\ell+\eta$, with shot noise
$\eta$ taken into account (see Table~\ref{table:biasvalues}). The $C_{\ell}$ values have been taken
from a $\Lambda$CDM cosmology and a suitable galaxy
bias parameter to describe the pseudo-power spectrum (corrected by $f_{\text{sky}}$) for this
band (see Section~\ref{sec:valida} for details). The $a_{\ell m}$ values
are the partial-sky pseudo-coefficients in harmonic
space. The panels present the $a_{\ell m}$'s histograms in four different multipole ranges. 
We can see that, except for the range $65\le \ell < 96$, fluctuations in harmonic space are fairly well described by
zero-mean, unit-variance Gaussians at the $1\sigma$ level, even though the
(co) variance model used is simple, not including the mixture induced
by the mask, but a simple scale-independent $f_{\text{sky}}$ correction. The
distributions for the other three bands are similar to those for band 4. To avoid aliasing for multipoles as large as $\ell_{max}=128$ in these
plots, the resolution parameter for the contrast map 
was taken to be a bit larger ($nside=64$) than the final map
resolution used in the Gibbs chains ($nside=32$). We stress that
the Gaussianity of fluctuations seen in harmonic space is not at all inconsistent with
the histograms of galaxy counts in pixel space of Figure~\ref{fig:2masscounts}. The statistical
properties of the $a_{\ell m}$ depend on the $n$-point correlation
functions. For example, their variances ($C_{\ell}^{gg}$) are related to the two-point
angular autocorrelation function $w(\theta)$ by
\begin{equation}
w(\theta)=\sum_{\ell}\frac{2\ell+1}{4\pi}C_{\ell}^{gg}P_{\ell}(\cos\theta),
\end{equation}
and therefore are not determined simply by the distribution of
galaxy counts, but by how these counts are correlated at a given angular scale. At each bin of the
histograms in Figure~\ref{fig:2masscounts}, several different angular scales are mixed, and the
information on the correlation at each scale is lost. It is also important to
say that the aspect of the histograms of Figure~\ref{fig:2massflucs}
does not prove complete Gaussianity of the spherical harmonics of 2MASS
contrast maps, since small non-Gaussianities should be probed with
more sensitive estimators like the bi-spectrum, for example.

%%%%%%%%%%%%%%%%%%%%%%%%%%%%%%%%%%%%%%%%%%%%%%%%%%%%%%%%%
\section{Validating the Bayesian method}
\label{sec:valida}
%%%%%%%%%%%%%%%%%%%%%%%%%%%%%%%%%%%%%%%%%%%%%%%%%%%%%%%%%

We validated the methodology described so far via the MC approach. For this, 
we adopted $\Lambda$CDM as the fiducial model, with the cosmological parameter 
values listed in Table~\ref{table:inputpars}. We have obtained the power spectrum $C_{\ell}^{tt}$ 
for the CMB temperature fluctuations and the 3D matter power spectrum $P(k)$ 
at large scales, taking into account nonlinear effects in the structure 
formation, using the CAMB \citep{camb2,camb3,camb1,camb4} + HALOFIT \citep{halofit2012} code. 

\begin{figure*}[ht]
\begin{center}
\subfigure[]{
  \includegraphics[width=.43\textwidth]{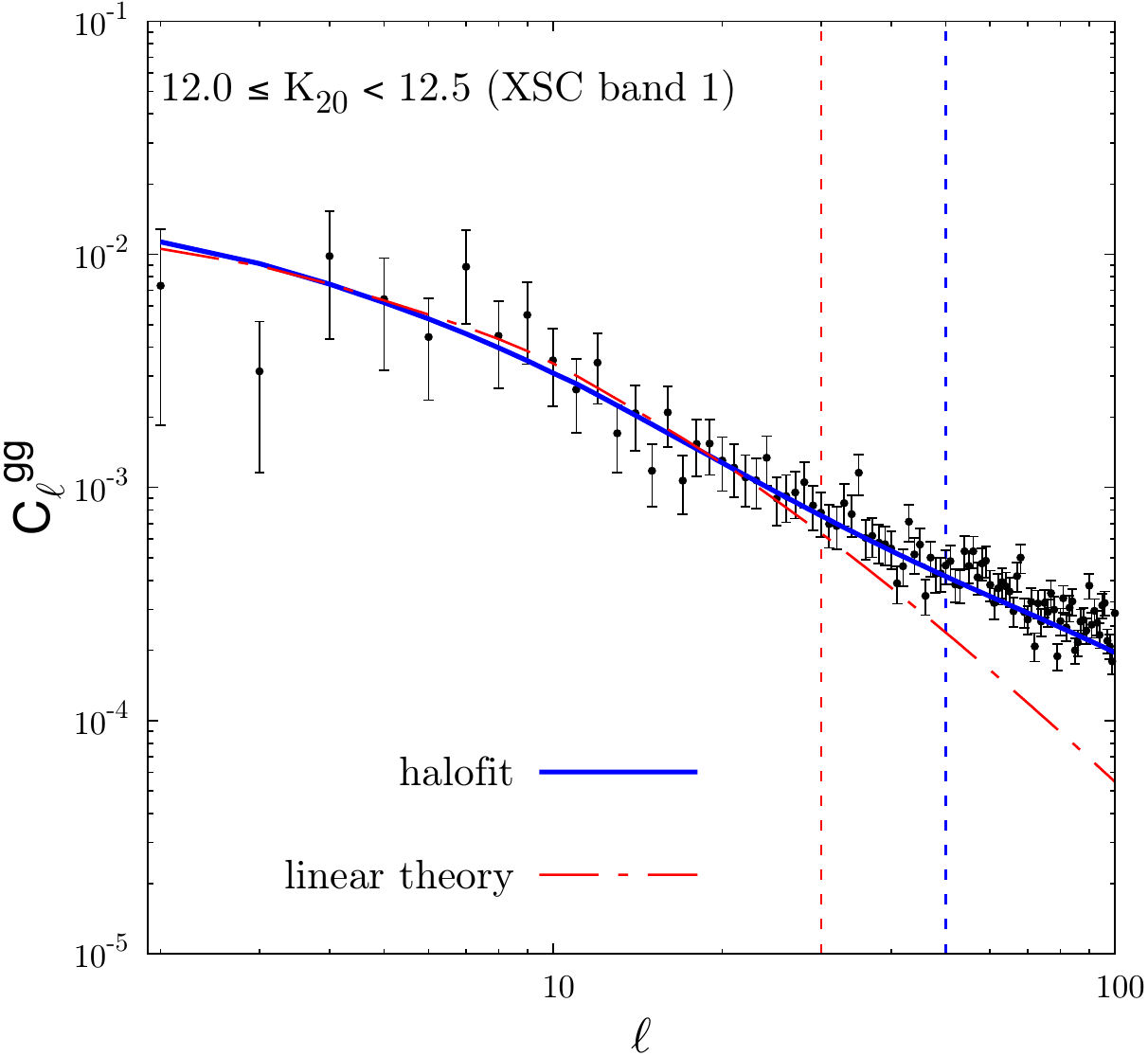}
}%
\subfigure[]{
  \includegraphics[width=.43\textwidth]{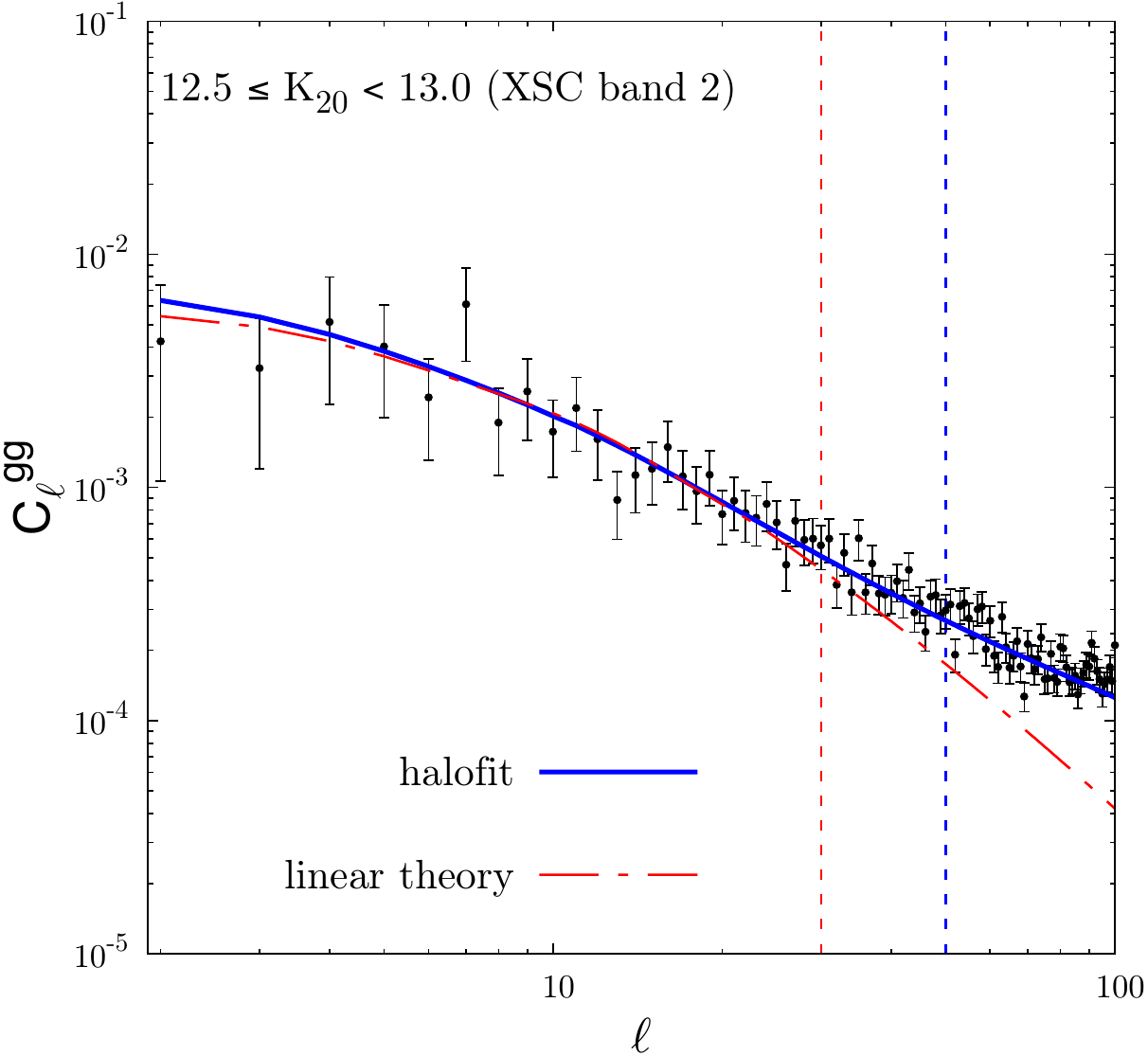}
} \\ % end of first row
\subfigure[]{
  \includegraphics[width=.43\textwidth]{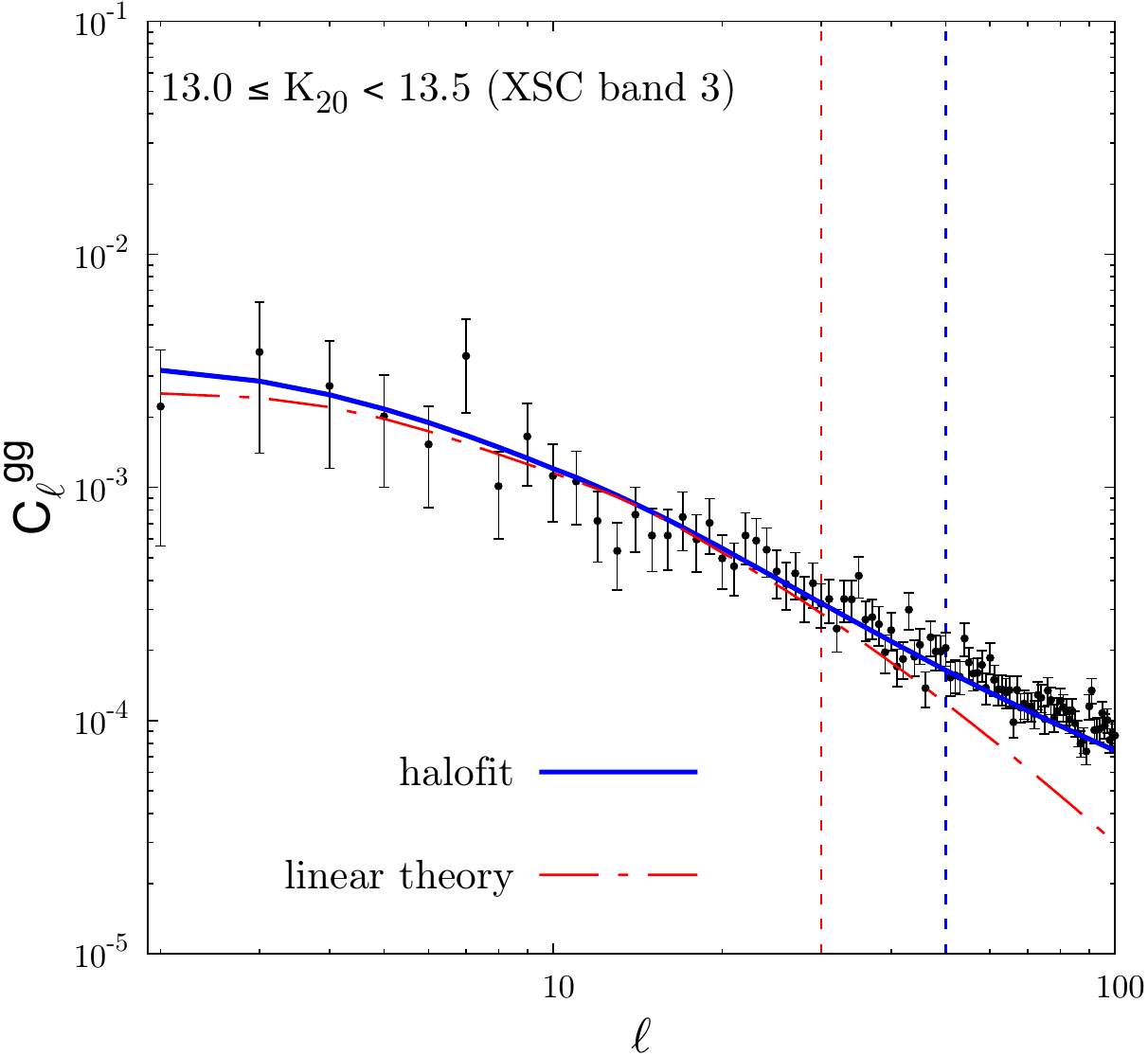}
}%
\subfigure[]{
  \includegraphics[width=.43\textwidth]{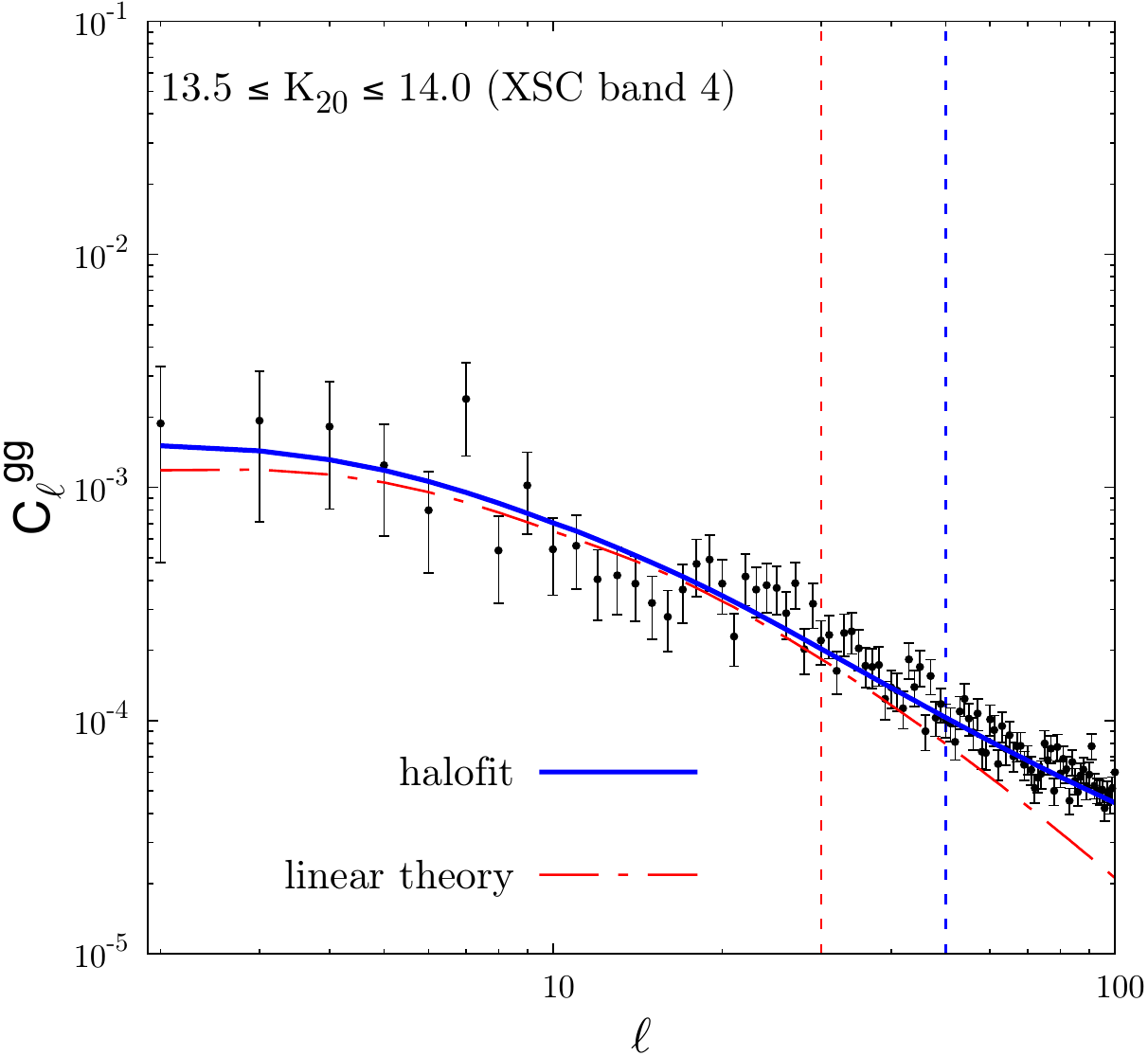}
}%
\caption{Data vs. theory comparisons for the galaxy autocorrelation power
spectrum $C_{\ell}^{gg}$. The dark matter power spectrum was scaled 
to data by fitting the galaxy bias factor $b_g$ (see Table~\ref{table:biasvalues} for the
best-fitting values and errors for each of the 2MASS magnitude bands). The vertical lines
mark the maximum multipole $\ell$ used in the fit. Both a pure linear
model (red) and a model with nonlinearities given by CAMB's
HALOFIT implementation (blue) are shown. Shot noise has been
subtracted from the data.}
\label{fig:biasfit}
\end{center}
\end{figure*}

We have then used $P(k)$ to build the corresponding galaxy 
autocorrelation spectrum $C_{\ell}^{gg}$ and the temperature--galaxy
cross-correlation spectrum $C_{\ell}^{tg}$ by assuming a linear
galaxy bias $b_g$. In order to mitigate the bias to be introduced by such a
linearity hypothesis, even at scales below about $10\ h^{-1}$ Mpc where
collapsed halos start to dominate, we fitted $b_g$ to the 2MASS autocorrelation angular
power spectrum considering just multipoles no larger than
$\ell=50$ (intermediate scales). The corresponding selection
functions for each of the four 2MASS magnitude bands were also included (see Table~\ref{table:2massbands}). Figure~\ref{fig:biasfit}
shows data versus theory comparisons, accounting for the $b_g$
best-fitting values (displayed on Table~\ref{table:biasvalues}) in the
2MASS linear (red line) and nonlinear (blue line) spectra. The data point central values were
obtained by decomposing the masked galaxy contrast map into spherical
harmonics with HEALPix routines, and then rescaled by a scale factor $1/f_{\text{sky}}$, 
to partially correct for the power loss introduced by the
mask, and also by the power dump introduced by the pixelization at
small scales. Finally, shot noise (given by $1/\overline{N}$, where
$\overline{N}$ is the average number of galaxies per steradian) was
also subtracted from the data. Figure~\ref{fig:biasfit} also evinces 
  another reason for using a broad $5^\circ$ beam, even
  for the galaxy contrast map: both the CMB and
  galaxy original maps are not bandwidth limited in the region $\ell
  \lsim 96$, so that spherical harmonics decompositions using low-resolution ($nside=32$)
  unbeamed maps would be severely affected by aliasing, especially in
  the multipole range $64\lsim\ell\lsim 96$. Moreover, to keep integration errors under
  control in the high-$\ell$ region, all decompositions have been performed with at least four 
  iterations with HEALPix routines.

In linear theory, the autocorrelation matter power spectrum normalization depends on the product
$\sigma_8 b_g$. Here we have chosen to fit $b_g$ fixing $\sigma_8$ at
0.78. For comparison, fitting all four 2MASS-XSC bands together
  (using $\Omega_m=0.30$, $\Omega_b=0.05$, $\sigma_8=0.75$ and
  $h=0.7$) in the range $\ell=1-50$, \cite{Rassat:2006kq} found
  $b_g=1.40\pm 0.03$. Binning multipoles, \cite{Afshordi:2003xu} fitted each band separately
and found that the biases for the different magnitude bands lie within
$b_{g}=1.18\pm 0.08$ for $\ell \lsim 70$.

\begin{table*}[ht]
\setlength{\tabcolsep}{8pt}
\caption{Results of the fit for the four 2MASS bands of the bias
  parameter $b_g$.}
\begin{center}
\begin{tabular}{lcccccccc}
\hline
\hline
& & \multicolumn{2}{ c }{Linear ($\ell_{max}=30$)} & \multicolumn{2}{
                                                     c }{Halofit
                                                     ($\ell_{max}=50$)}
& \multicolumn{2}{ c }{Halofit ($\ell_{max}=96$)}\\
\hline
Band & Shot Noise & $b_g\pm \sigma_b$ & $\chi^2$/ndof & $b_g\pm \sigma_b$ & $\chi^2$/ndof  & $b_g\pm \sigma_b$ & $\chi^2$/ndof \\
\hline
1 & $1.8\times 10^{-4}$ & $1.27 \pm 0.04$ & 19.1/28 & $1.32 \pm 0.02$ & 33.5/48 & $1.37 \pm 0.01$ & 127.3/94\\
2 & $8.5\times 10^{-5}$ & $1.25 \pm 0.03$ & 17.3/28 & $1.34 \pm 0.03$ & 29.2/48 & $1.35 \pm 0.01$ & 102.1/94\\
3 & $4.0\times 10^{-5}$ & $1.22 \pm 0.03$ & 16.9/28 & $1.29 \pm 0.02$ & 37.3/48 & $1.34 \pm 0.01$ & 86.2/94\\
4 & $1.9\times 10^{-5}$ & $1.18 \pm 0.03$ & 32.2/28 & $1.28 \pm 0.02$ & 52.5/48 & $1.29 \pm 0.01$ & 105.9/94\\
\hline 
\hline 
\end{tabular}
\end{center}
Note: Both a pure linear model and one with nonlinearities (HALOFIT) are
  shown, as well as $\ell_{max}$ used in each fit.
\label{table:biasvalues}
\end{table*}

\begin{figure*}[ht]
\begin{center}
\includegraphics[width=0.6\textwidth]{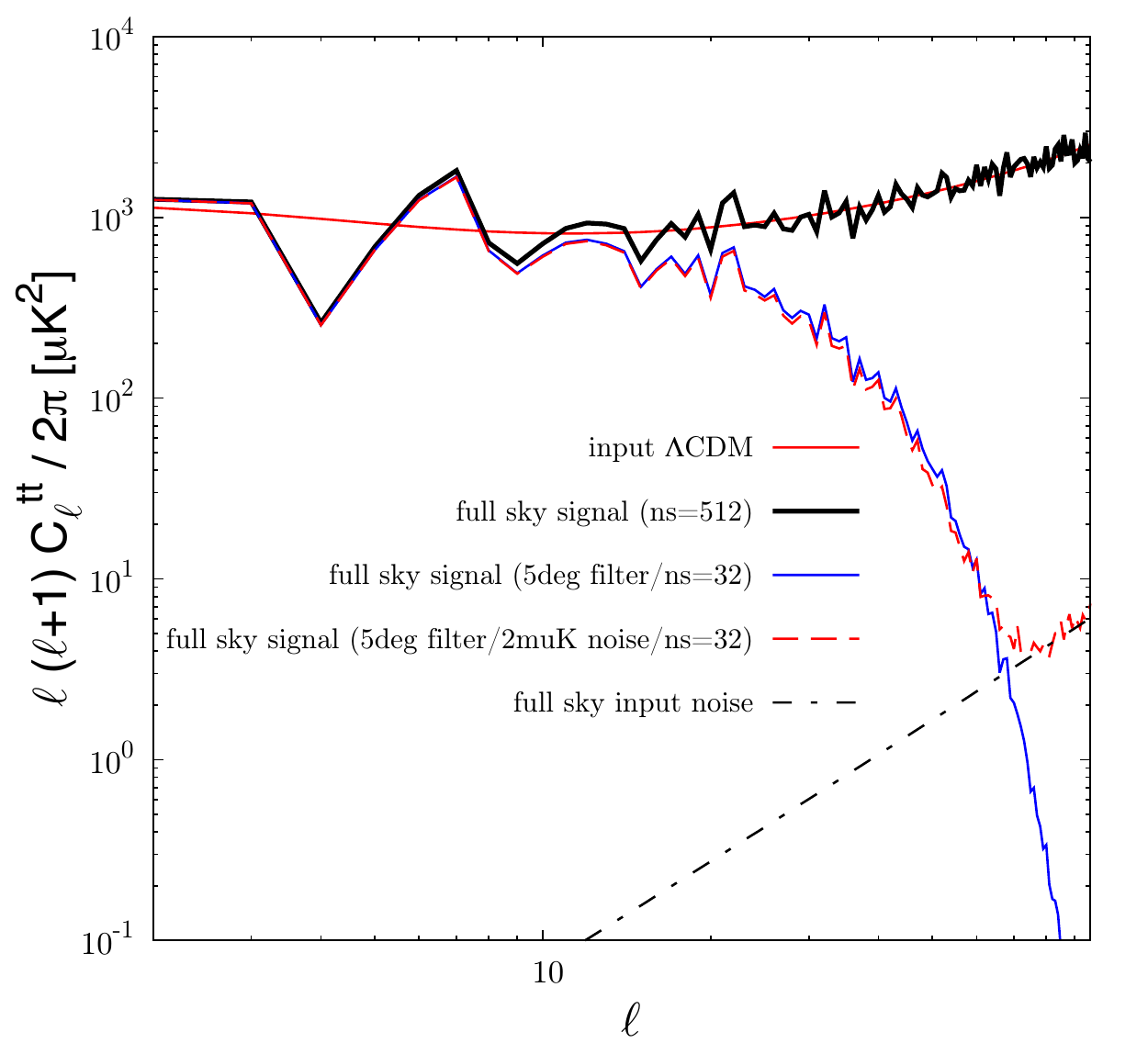}
\end{center}
%\vspace*{-1.0cm}
\caption{CMB autocorrelation power spectra: input $\Lambda$CDM (red solid line),
 full-sky spectrum for a given CMB map realization at $nside=512$ (black
 solid line), spectrum convoluted with a $5^\circ$ Gaussian beam and
 sampled at $nside=32$ (blue
 solid line), spectrum after the addition of 2$\mu$K pixel$^{-1}$ RMS white
 noise (dashed red line), and added noise power (dashed black line).}
\label{fig:inputmapspowercmb}
\end{figure*}

\begin{figure*}[ht]
\begin{center}
\subfigure[]{
  \includegraphics[width=.45\textwidth]{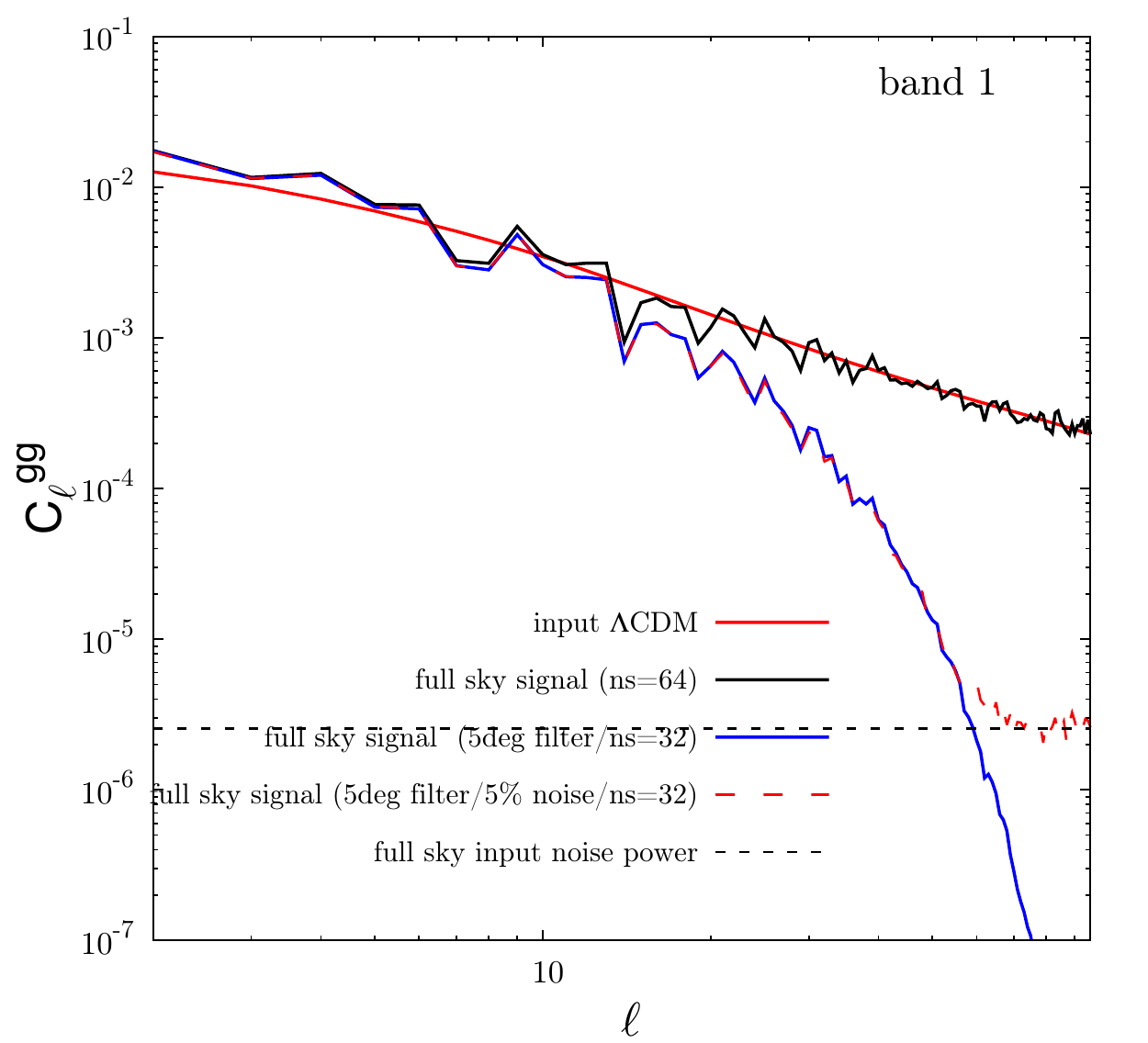}
}%
\subfigure[]{
  \includegraphics[width=.45\textwidth]{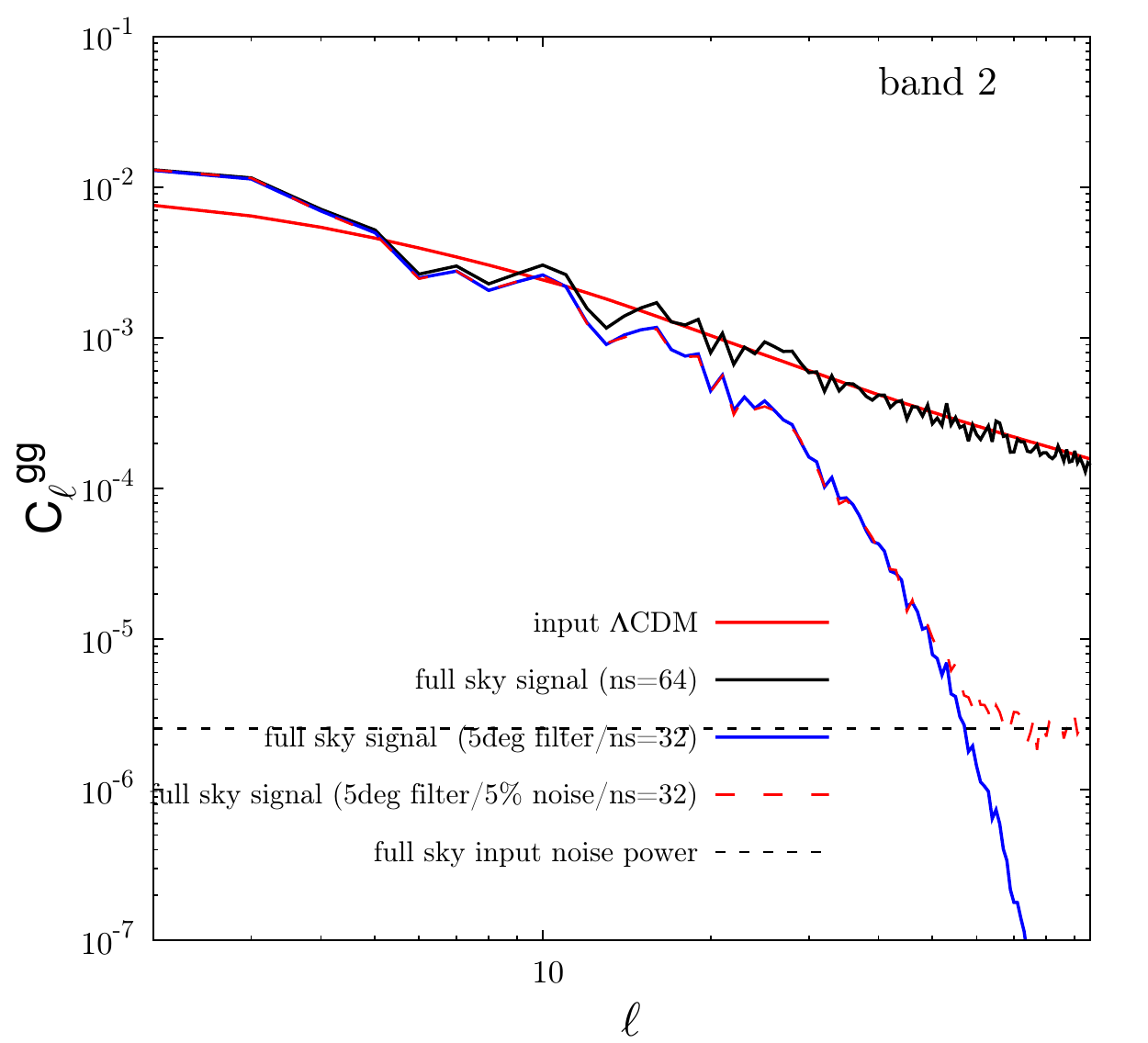}
} \\ % end of first row
\subfigure[]{
  \includegraphics[width=.45\textwidth]{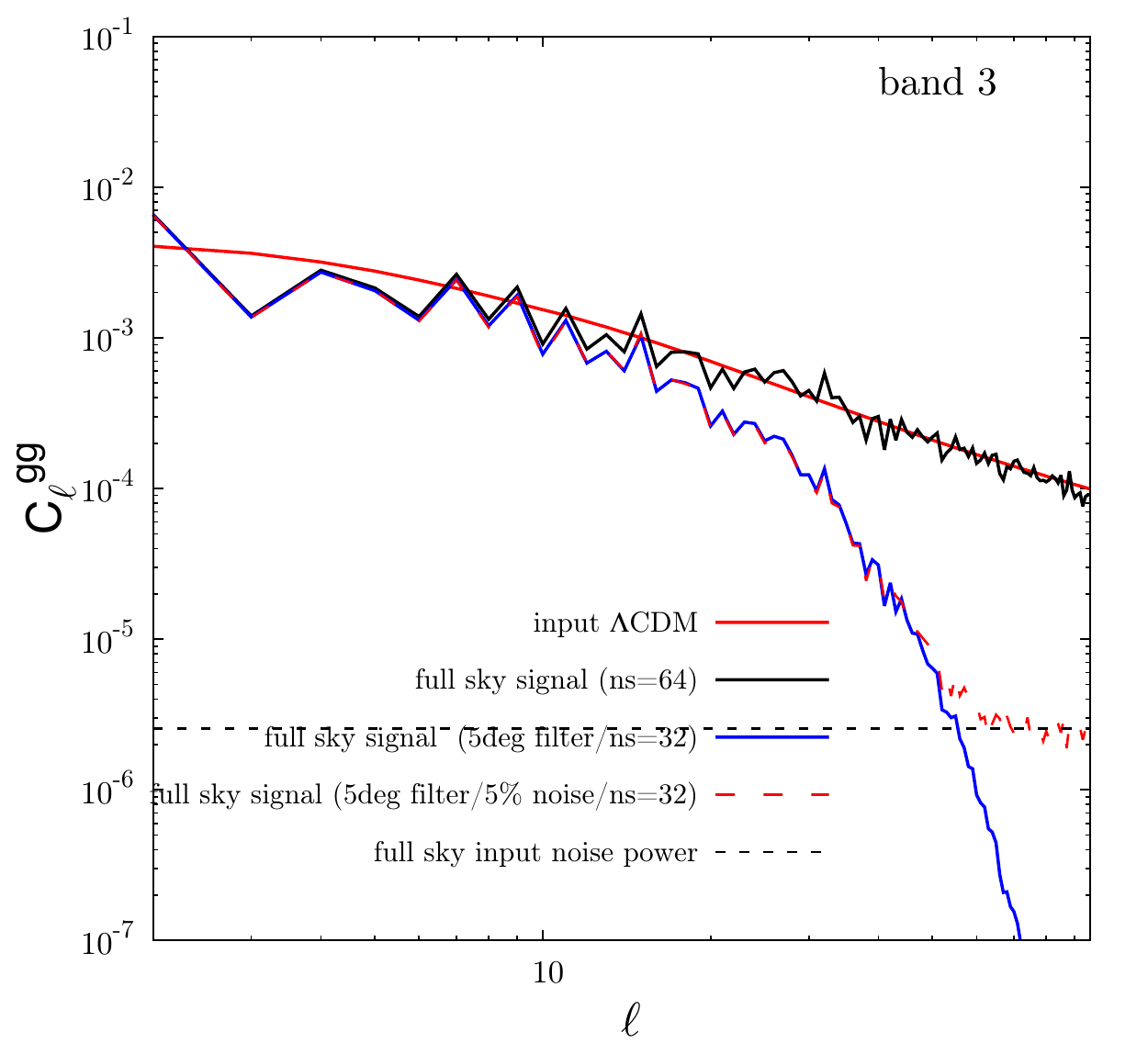}
}%
\subfigure[]{
  \includegraphics[width=.45\textwidth]{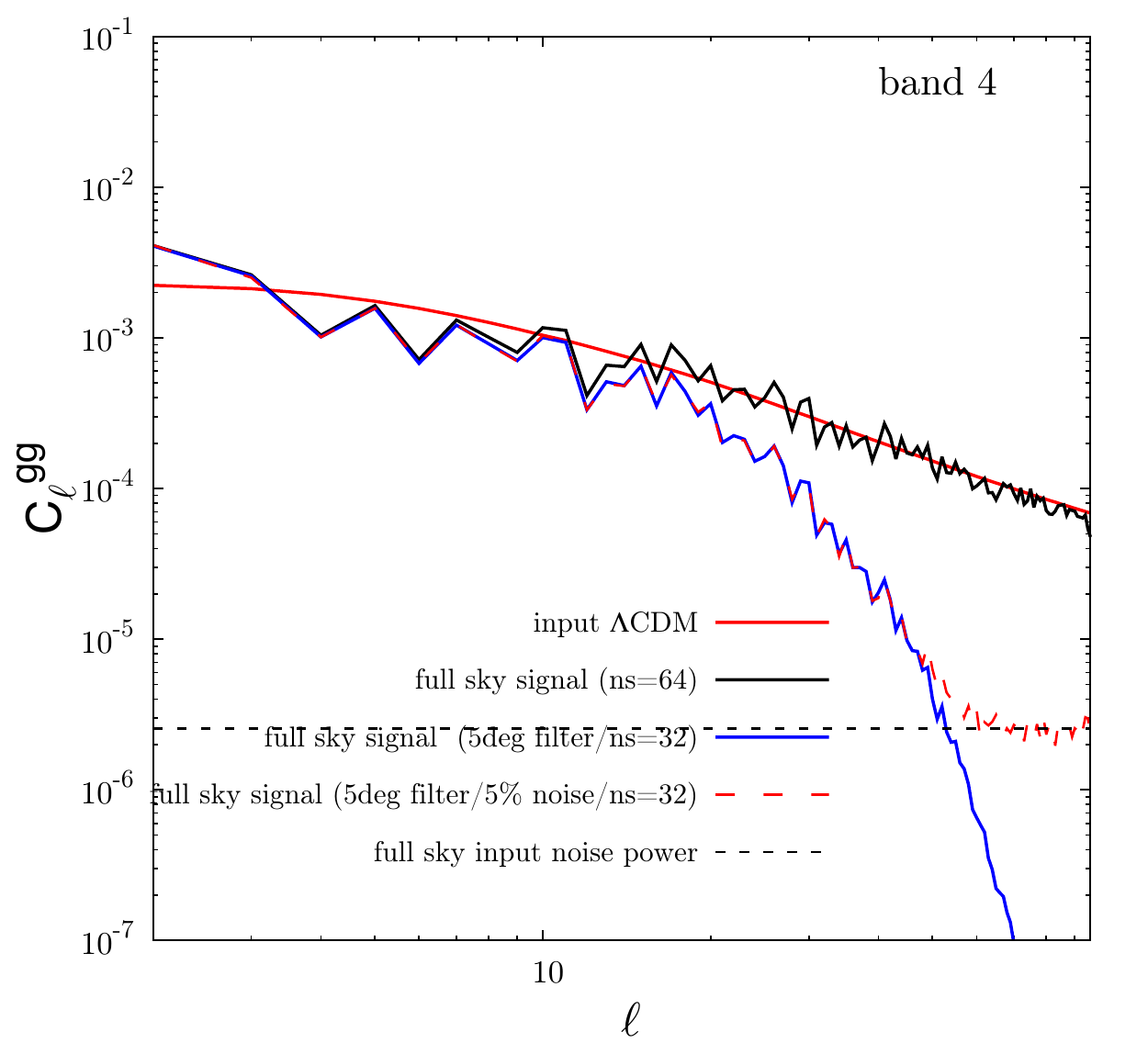}
}%
\end{center}
%\vspace*{-1.0cm}
\caption{Galaxy contrast autocorrelation power spectra for the four 2MASS
  selection functions: input
  $\Lambda$CDM with biases given in Table~\ref{table:biasvalues} (red solid line),
 full-sky spectrum for a given galaxy contrast map realization at $nside=64$ (black
 solid line), spectrum convoluted with a $5^\circ$ Gaussian beam and
 sampled at $nside=32$ (blue
 solid line), spectrum after the addition of 5\% pixel$^{-1}$ RMS white
 noise (dashed red line), and added noise power (dashed black line).}
\label{fig:inputmapspowerlss}
\end{figure*}

Once $b_g$ has been estimated, a combined realization of a CMB temperature map
($nside=512$) and 2MASS galaxy contrast map ($nside=64$) was
produced, including their cross-correlation signal. A $5^\circ$
Gaussian beam was applied to both maps, and their resolutions were
degraded to $nside=32$. Finally, white noise of RMS strength
$\sigma_0=2 \ \mu$K pixel$^{-1}$ was added to the CMB low-resolution map and
of $5\%$ pixel$^{-1}$ to the galaxy contrast map. Such noise levels dominate
the power in both maps for scales above $\ell=60$, as can be seen in
Figures~\ref{fig:inputmapspowercmb} and \ref{fig:inputmapspowerlss},
where the full-sky $tt$ and $gg$ power
spectra were retrieved from the input maps, respectively.

\begin{figure*}[ht]
%\captionsetup[subfigure]{aboveskip=-1pt,belowskip=-1pt}
\begin{center}
\subfigure[]{
  \includegraphics[width=.48\textwidth]{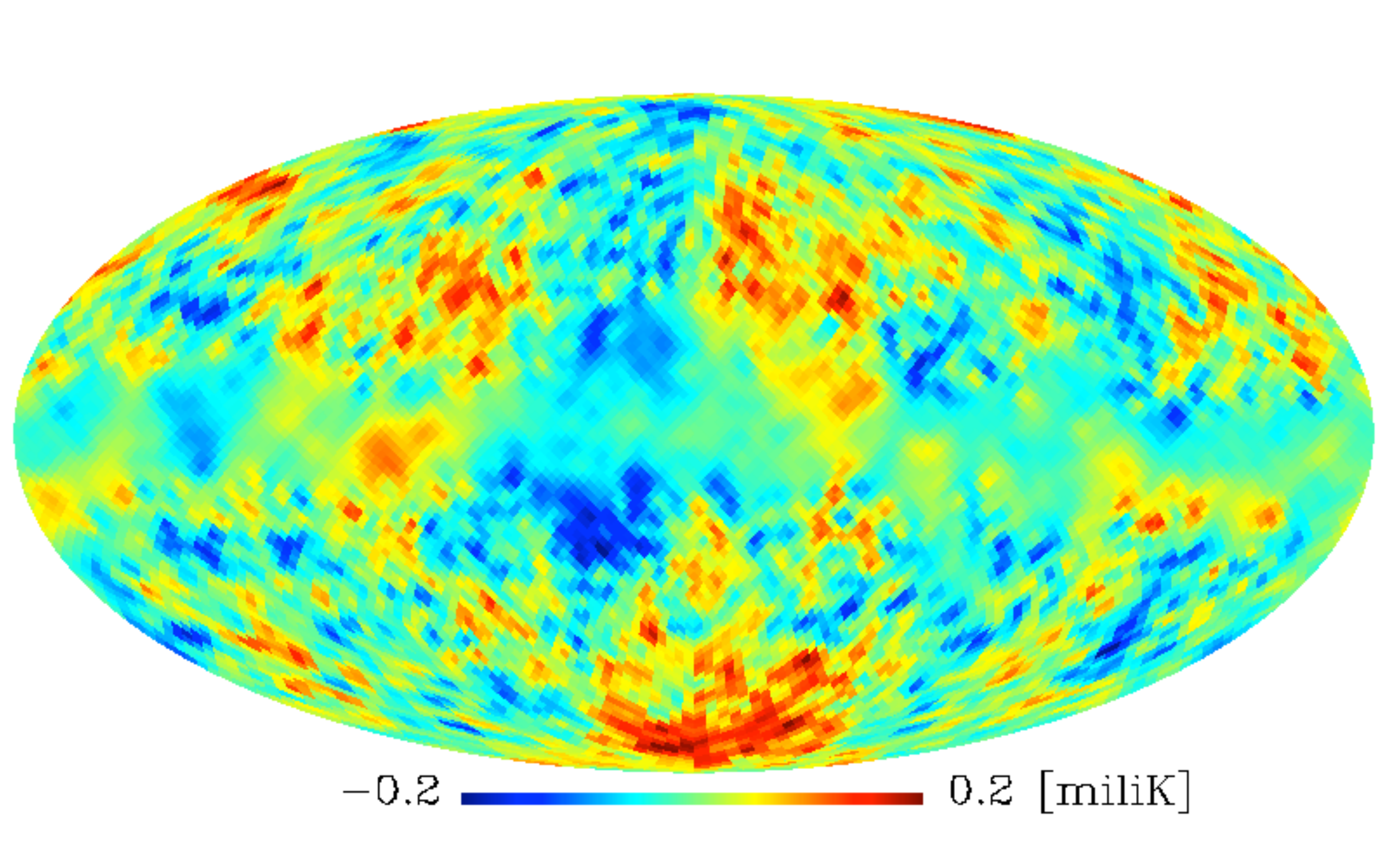}
}%
\subfigure[]{
  \includegraphics[width=.48\textwidth]{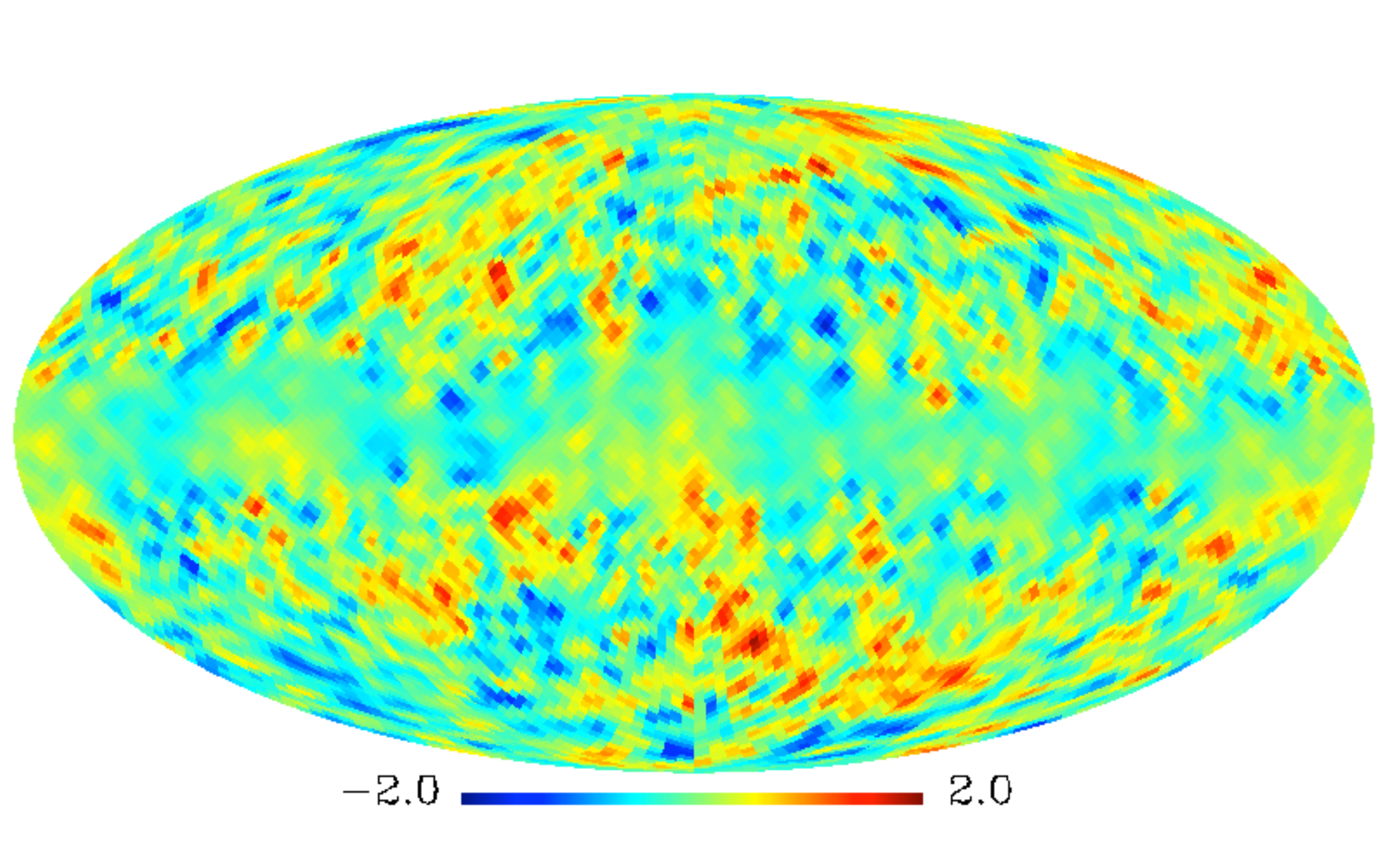}
} \\ % end of first row
\subfigure[]{
  \includegraphics[width=.48\textwidth]{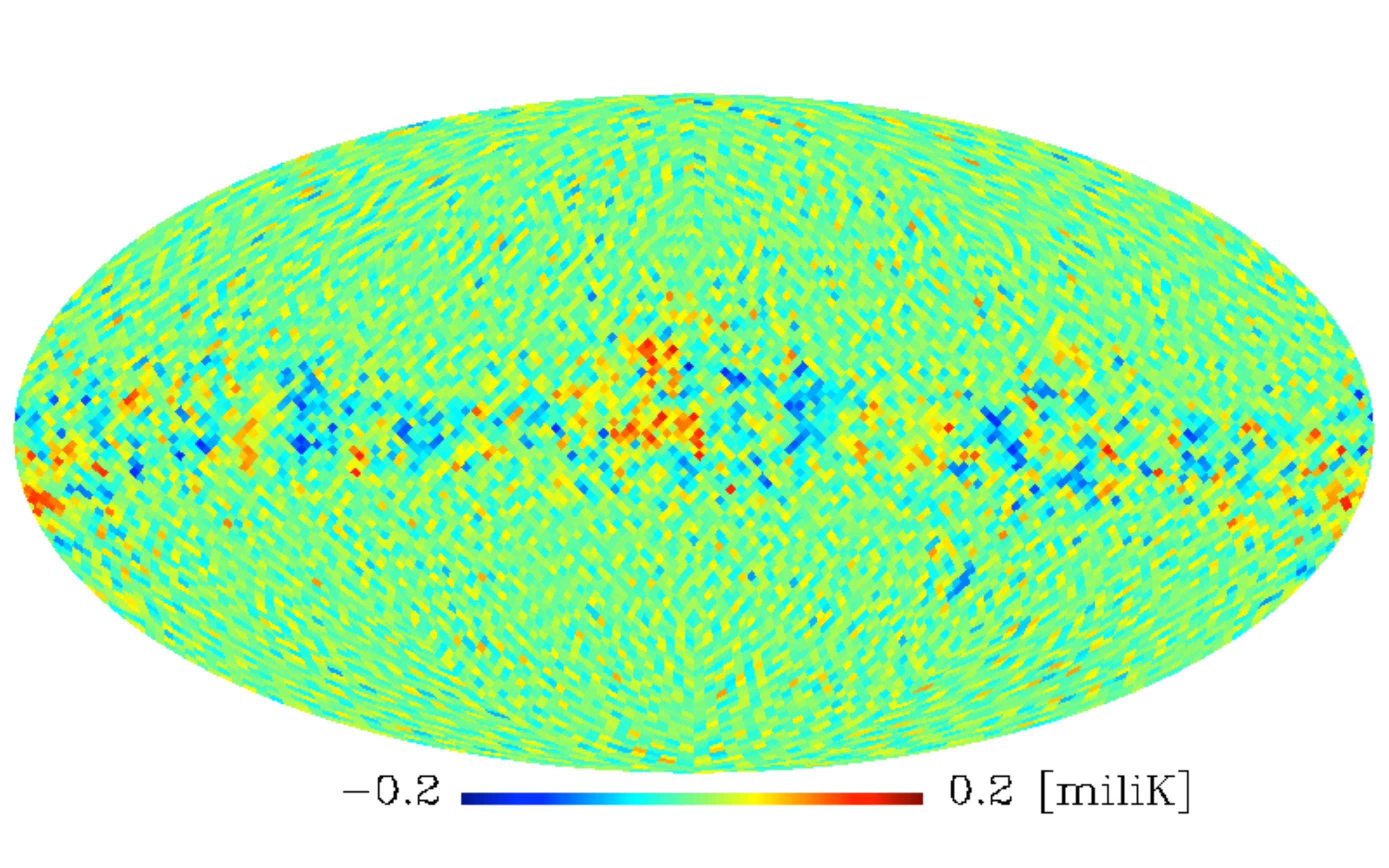}
}%
\subfigure[]{
  \includegraphics[width=.48\textwidth]{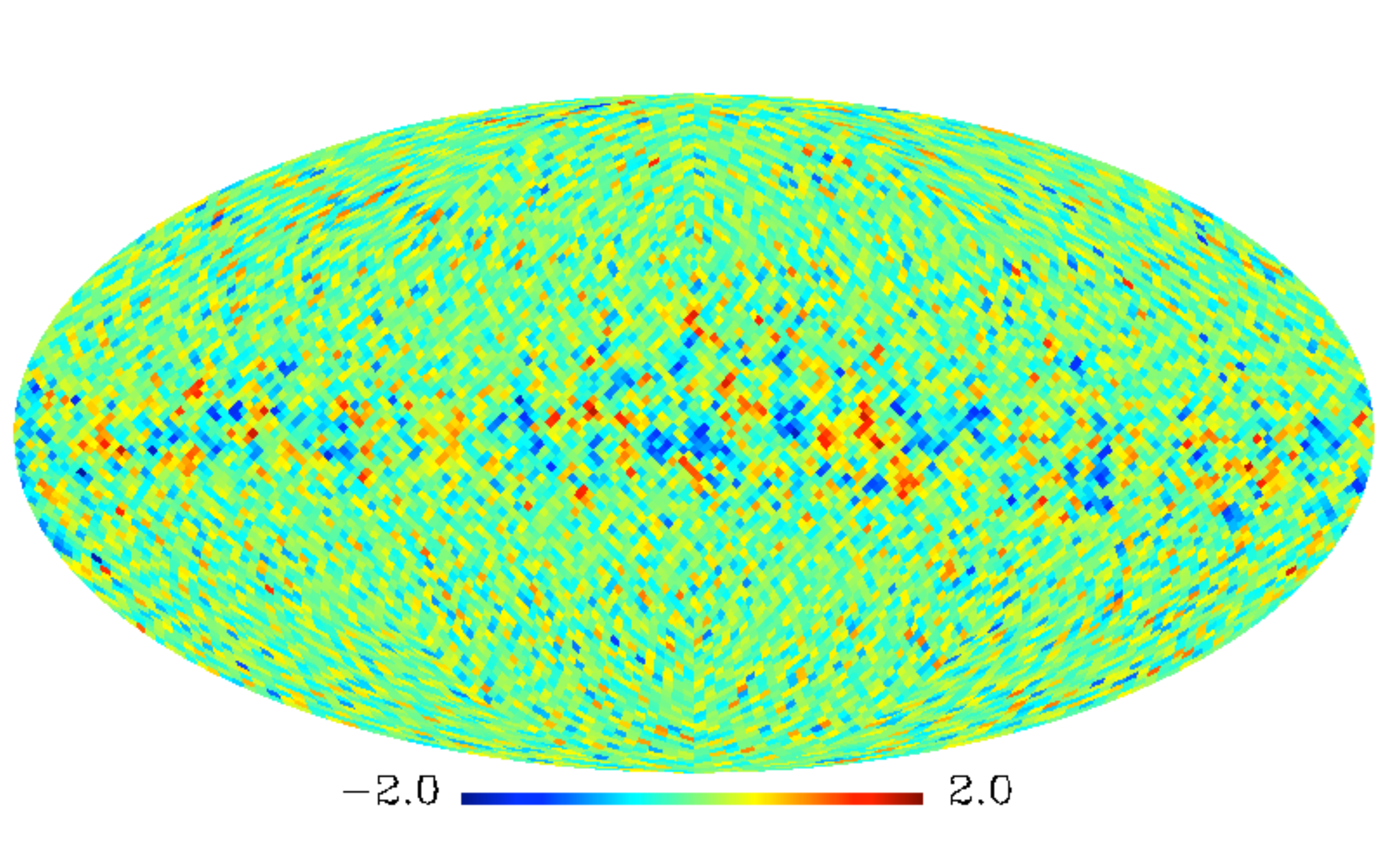}
} \\ % end of second row
\subfigure[]{
  \includegraphics[width=.48\textwidth]{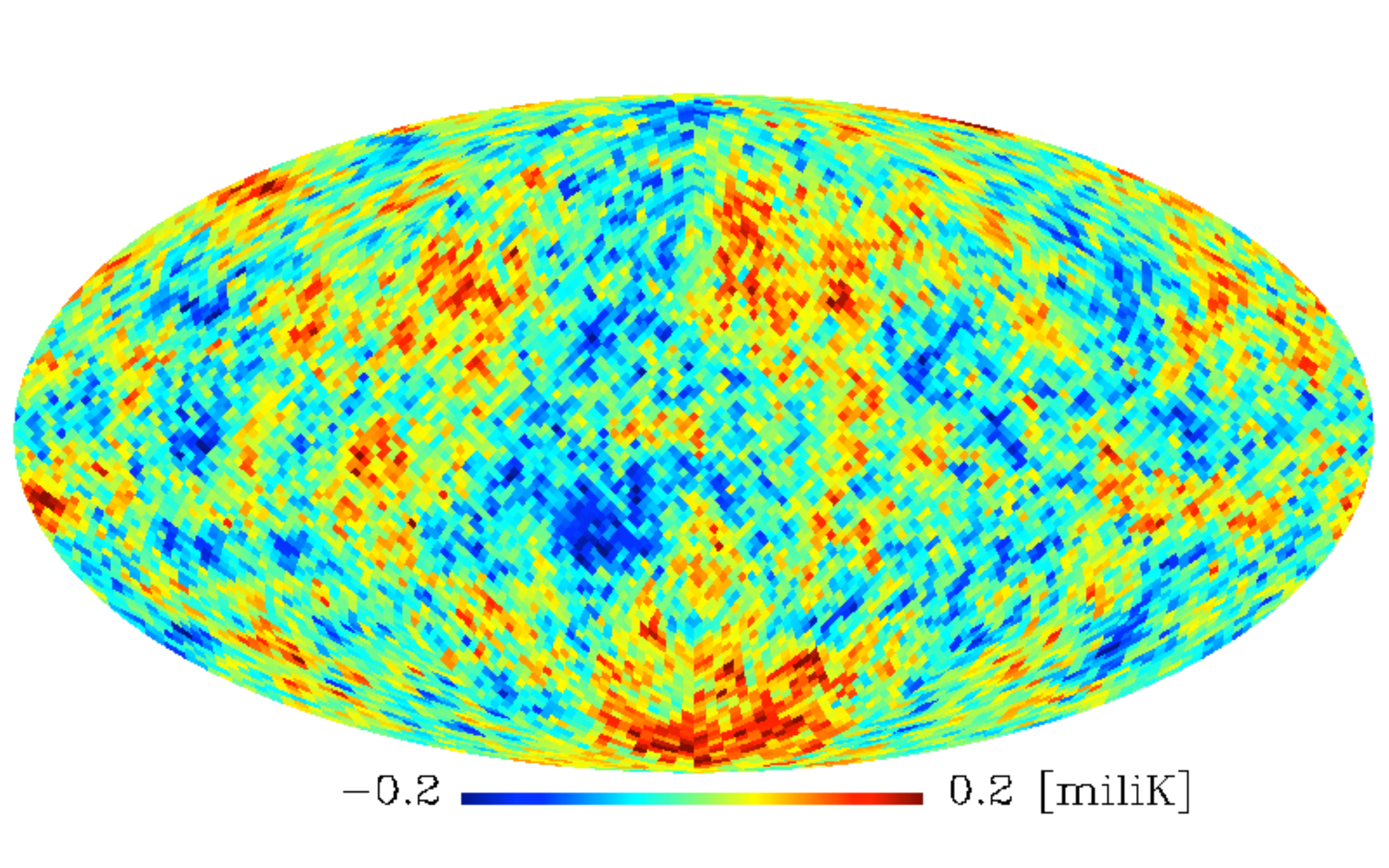}
}%
\subfigure[]{
  \includegraphics[width=.48\textwidth]{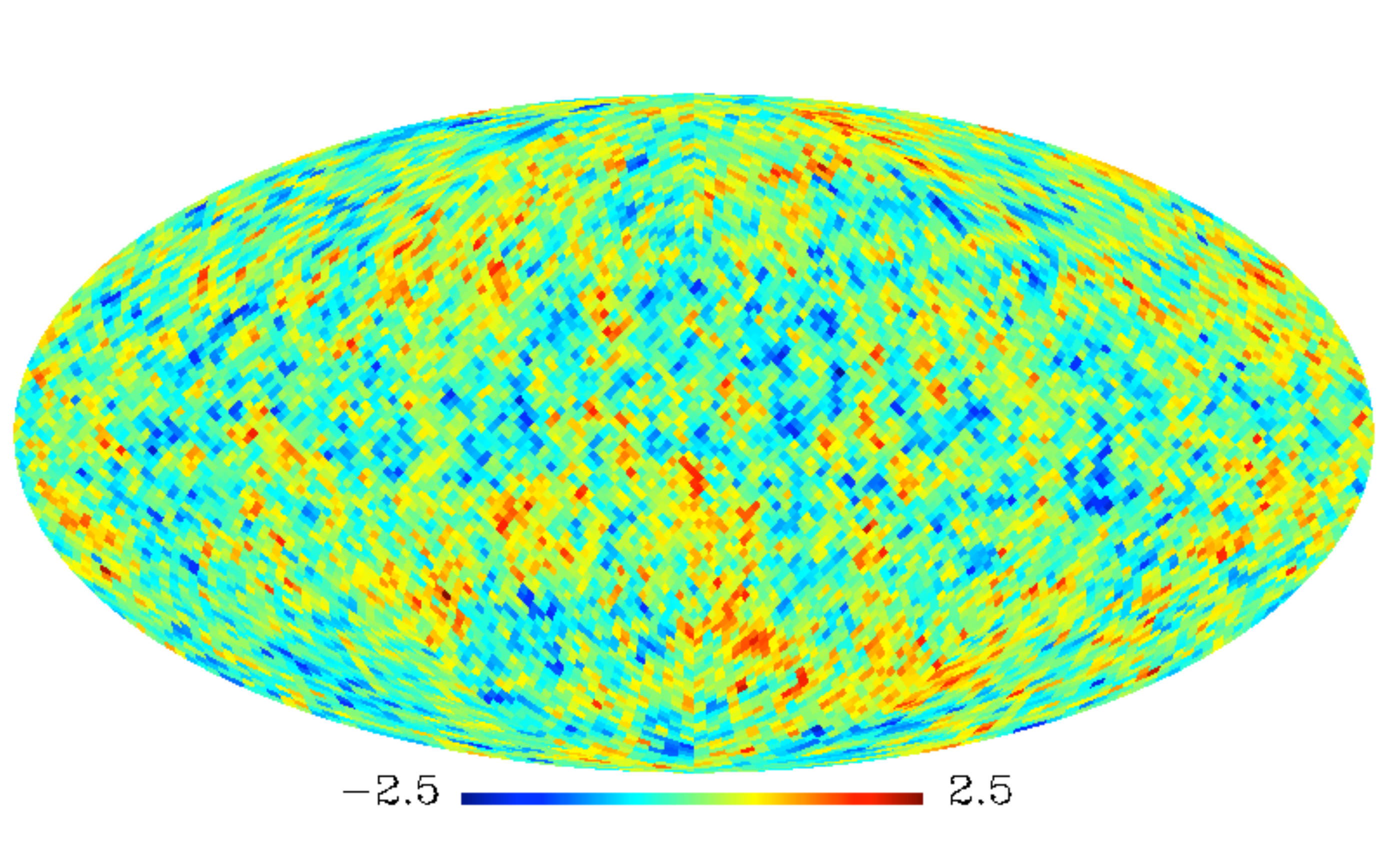}
} \\ % end of third row
\end{center}
%\vspace*{-0.5cm}
\caption{Sky maps associated with the Gibbs chain sampled fields. Left:
  CMB temperature field. Right: galaxy contrast field. Top: mean
  field $\boldsymbol{x}$. Middle: fluctuation field $\boldsymbol{y}$;
  Bottom: total field $\boldsymbol{x}+\boldsymbol{y}$. The
  corresponding 2MASS selection function is that of band 1.}
\label{fig:gibbs_samples_6maps_band1}
\end{figure*}

The four sets of combined CMB/contrast maps (one for each of the 2MASS selection functions) described above were then used to feed a Gibbs chain, whose
sampling equations were described in Section~\ref{sec:gibbs}. The mask
used was the combination of 2MASS and {\it WMAP}9 masks described in Section~\ref{sec:processing}, 
with a final sky fraction of $f_{\text{sky}}=0.70$. The initial
state of the chain was chosen to be the full covariance
matrix of the fiducial model itself. The monopole and dipole, both
null for the fiducial spectrum, were kept fixed at these values along
the whole chain. Multipoles were sampled up to $\ell_{max}=51$, and 
kept fixed for $51 < \ell \le 96$ at the fiducial model
spectra. For $\ell > 96$, the spectrum was taken to be zero. We also have used the Jeffreys scale-invariant prior, $\pi(\mathbf{S})= |\mathbf{S}|^{-1}$ \citep{Wandelt:2003uk}.

\begin{figure*}[ht]
%\captionsetup[subfigure]{aboveskip=-1pt,belowskip=-1pt}
\begin{center}
\subfigure[]{
  \includegraphics[width=.48\textwidth]{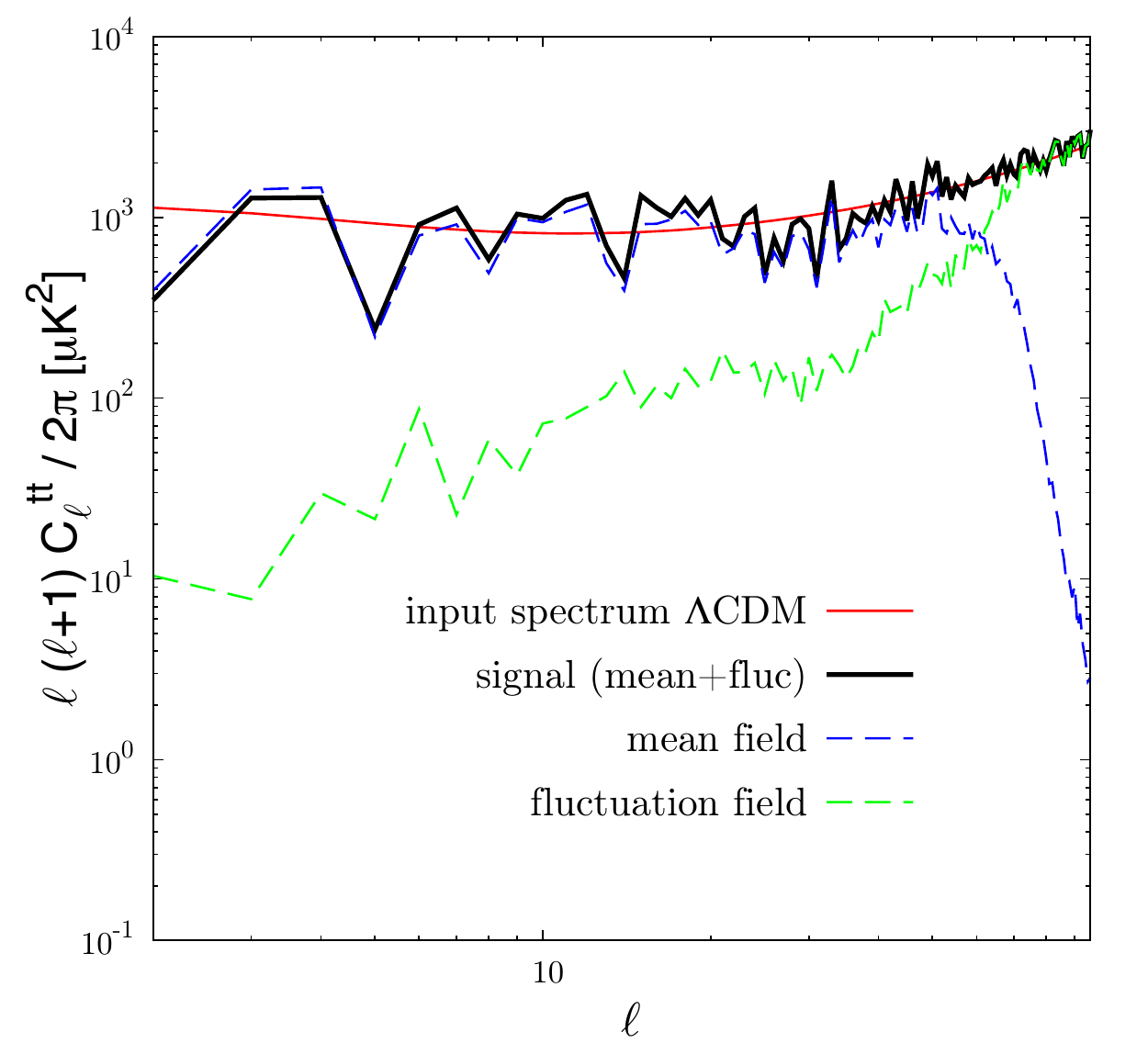}
}%
\subfigure[]{
  \includegraphics[width=.48\textwidth]{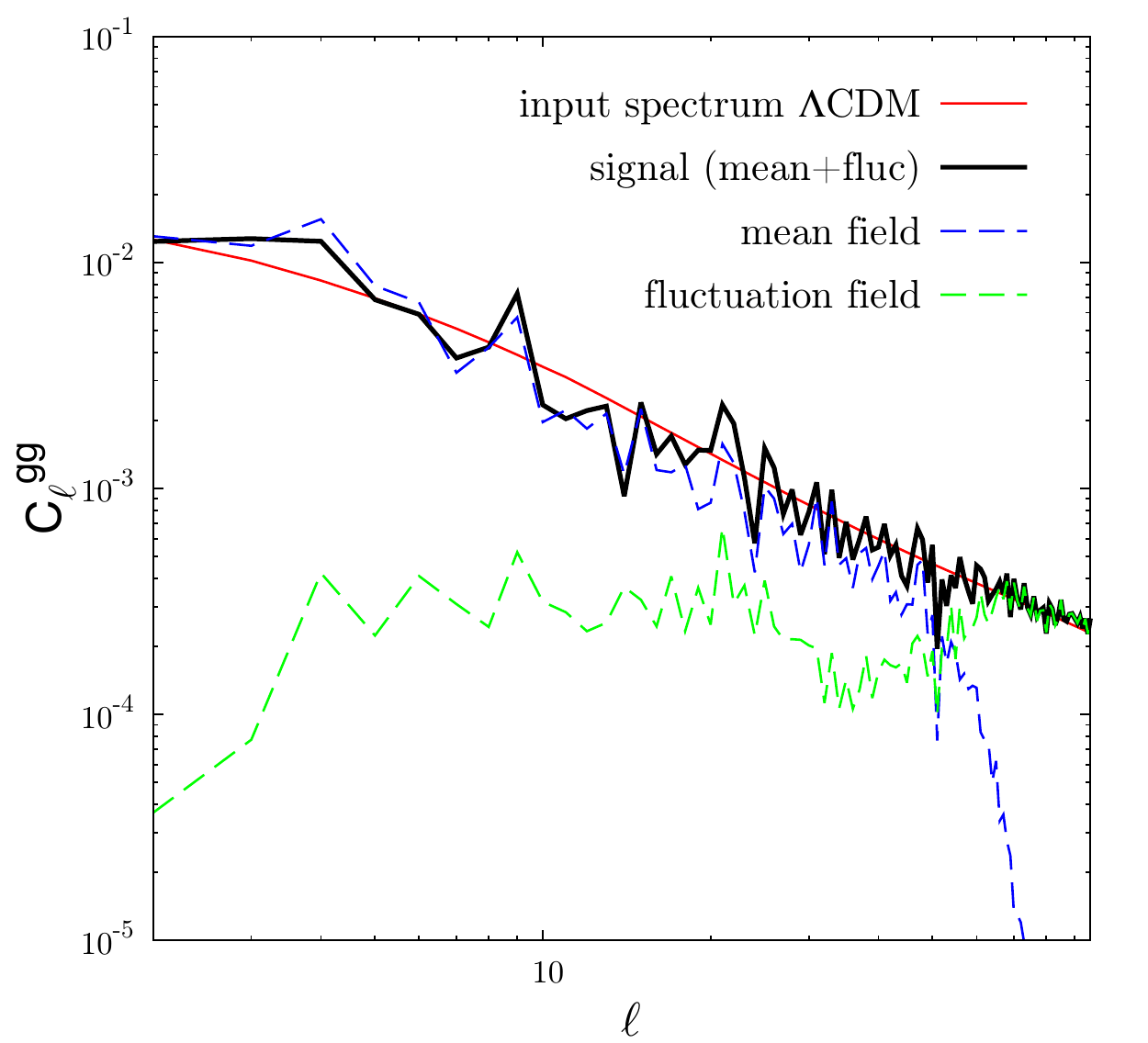}
} \\ % end of first row
\end{center}
%\vspace*{-1.0cm}
\caption{Autocorrelation power spectrum of the Gibbs chain fields of Figure
  \ref{fig:gibbs_samples_6maps_band1}. The total
  ($\boldsymbol{x}+\boldsymbol{y}$), mean ($\boldsymbol{x}$), and
  fluctuation ($\boldsymbol{y}$)
fields are shown separately.}
\label{fig:last_gibbs_ttgg_band1}
\end{figure*}

Even though the initial spectrum was already at its true state, a
burn-in phase of 100 samples was considered,\footnote{We have also
  tested the chain convergence with different initial spectra and
  verified that such burn-in phase size was enough to approach the
  true solution.} and all the samples drawn
after this initial phase were used for statistical analysis, that is,
we have ignored possible correlations between adjacent samples of the
chain. A total of 50,000 samples for each 2MASS selection function were generated using 50 independent
parallel chains. Figure~\ref{fig:gibbs_samples_6maps_band1} shows the
sky maps associated with the fields of Equation~\eqref{eq:linsys} (the mean
field $\boldsymbol{x}$, the fluctuation field $\boldsymbol{y}$ and the total field $\boldsymbol{x}+\boldsymbol{y}$) for one of the band 1 realizations. Some of the statistical properties of these fields can be seen
in Figure~\ref{fig:last_gibbs_ttgg_band1}, where the autocorrelation power spectrum of
each component is displayed separately. One sees that at large scales
($\ell \lsim 10$) the total field is dominated by its mean component $\boldsymbol{x}$,
whereas at small scales ($\ell \gsim 60$), where the added uncorrelated white noise is
larger than the primordial signal, the total field is dominated by
$\boldsymbol{y}$. Another interesting feature of these random fields is
the behavior of the autocorrelation at large scales for the
fluctuation $\boldsymbol{y}$. According to Equation~\eqref{eq:linsys},
the small-$\ell$ region is characterized by a large S/N ($S/N \gg 1 $), so the covariance of $\boldsymbol{y}$ in harmonic space is approximately
$\mathbf{B}^{\textrm{\tiny{-2}}}\mathbf{N}$:
\begin{equation}
\mathbf{C(\boldsymbol{y})}=E(\boldsymbol{y}\otimes\boldsymbol{y})
\stackrel{S/N\gg 1}{\simeq} \mathbf{B}^{-2}\mathbf{N}.
\end{equation}

Neglecting the effect of
the beam at these scales, the autocorrelations are quite 
different from those of the added full-sky noise (see Figures~\ref{fig:inputmapspowercmb} 
and \ref{fig:inputmapspowerlss}). This is
the direct effect of the mask whose mixing effect distorts also the
noise (see Equation~\eqref{eq:noisestruct}).  
\begin{figure*}[htb]
\begin{center}
%\vglue -0.3cm
\includegraphics[width=1.0\textwidth]{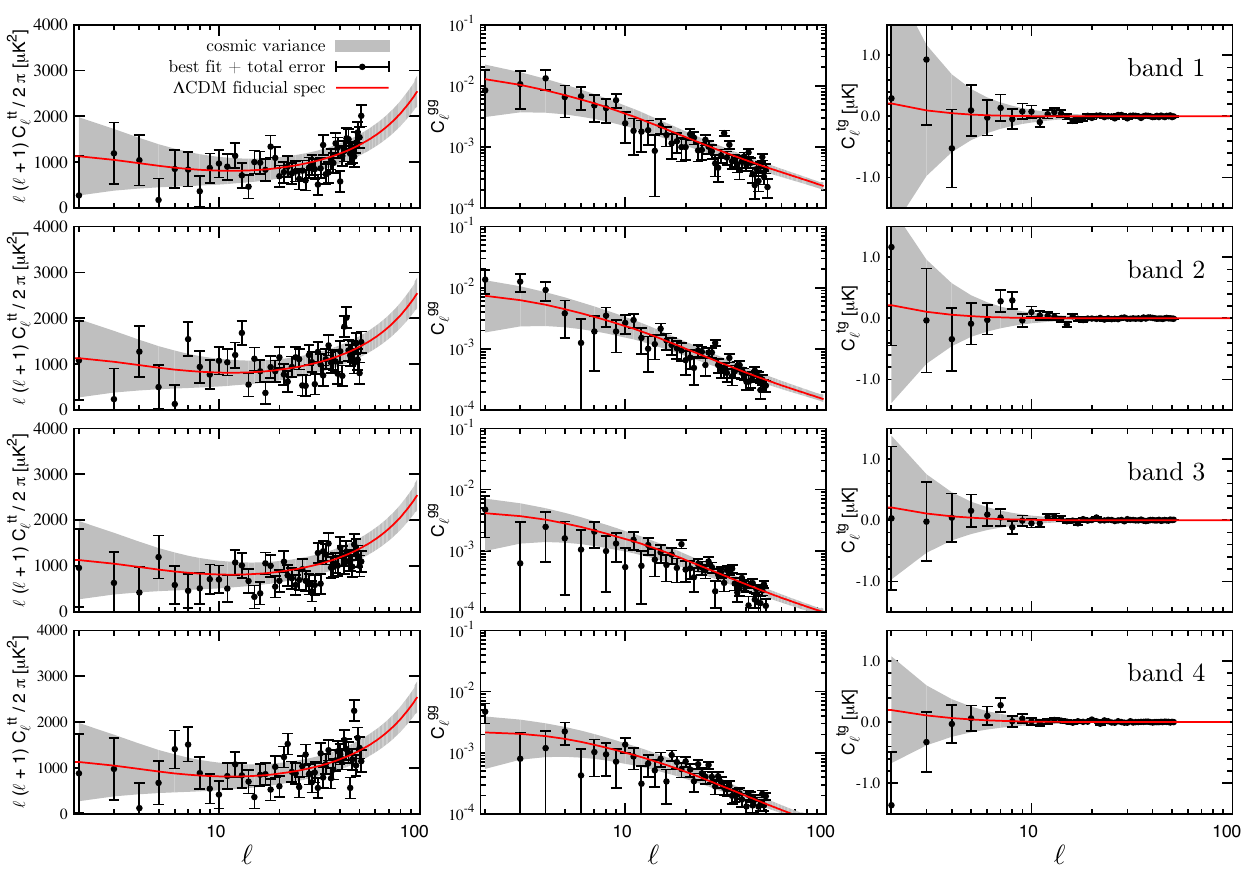}
\end{center}
%\vspace*{-1.2cm}
\caption{Monte Carlo. Best-fit values (dots with error bars) are shown for the full covariance matrix of a combined
  CMB--LSS experiment. Central values are the maximum of a
  log-likelihood Blackwell--Rao estimator, and the error bars contain
  contributions from experimental noise and cosmic variance (also
  shown separately as a gray band) around a
  fiducial $\Lambda$CDM (red line) model (see \citet{Greason:2012} for details).}
\label{fig:bestfit_mc}
\end{figure*}

With the 50,000 available samples, we have reconstructed a likelihood
for this combined CMB--LSS data set using a Blackwell--Rao estimator
\citep{Chu:2004zp,Dunkley:2008mk}. For a certain number of Gibbs
samples $n_G$, we were able to find an analytic solution for the
maximum of the log-likelihood (see Appendix~\ref{solution}), which
allowed us to construct the best-fitting
spectra shown in Figure~\ref{fig:bestfit_mc}. The error bars were
estimated assuming Gaussian errors, and the cosmic variance band was
calculated taking the $\Lambda$CDM fiducial model as reference (see
\citet{Greason:2012} for details). While
useful for many purposes, these central values with their associated
errors should only be taken as a first approximation to the full (and
non-Gaussian) likelihood function close to its maximum. 
\begin{table}[ht]
\caption{Summary of All parameters/maps/choices used to run the Gibbs chain}
\begin{center}
\begin{tabular}{lc}
\hline
\hline
Parameter & Choice \\
\hline
\hline
Input maps & ILC/$Q$/$V$/$W$ 9 years (r9) \\
Beaming & Effective $5^\circ$ (FWHM)\\
CMB noise level &2$\mu$K pixel$^{-1}$ (RMS)  \\
Galaxy contrast noise level & 5\% pixel$^{-1}$ (RMS)  \\
Mask & Combined KQ85y9 $\times$ 2MASS (r5)  \\
Final maps resolution & r5 ($nside=32$) \\
Input spectrum & $\Lambda$CDM best-fit to {\it WMAP}9 \\
Sampling mode & COMMANDER \\
Preconditioner & block-diagonal \\
Monopole/dipole & fixed (float. dipole) \\
Maximum multipole & $\ell_{max}=96$ \\
Start of fiducial spectrum & $\ell=52$ \\
Prior & Jeffrey (flat) \\
Burn-in phase & 100 (500) samples \\
Skipped samples & 0 (1) \\
CG convergence & $10^{-6}$\\
\hline 
\hline 
\end{tabular}
\end{center}
\label{table:fiducial}
\end{table}

%%%%%%%%%%%%%%%%%%%%%%%%%%%%%%%%%%%%%%%%%%%%%%%%%%%%%%%%%
\section{Application to the {\it WMAP}--2MASS combined data set}
\label{sec:results}
%%%%%%%%%%%%%%%%%%%%%%%%%%%%%%%%%%%%%%%%%%%%%%%%%%%%%%%%%

In Section~\ref{sec:valida} we used a fiducial power spectrum where $C_0^{tt} = C_1^{tt} = 0$. 
In order to apply this 
approach to real data, we had to follow a slightly different procedure. Although the 
monopole and dipole terms are removed in the input {\it WMAP}9 maps and their residual 
values are small in comparison to other multipoles at large scale, we verified that 
they could not be fixed at zero. Otherwise, the Gibbs chains converge to anomalous states 
at those scales. Thus, decomposing the ILC, {\it Q}, {\it V} and {\it W} maps with partial sky coverage, we obtained 
estimates for $C_0^{tt}$ and $C_1^{tt}$. We used these estimates to perform the analyses, 
keeping them fixed along the chain. With the block-diagonal
preconditioner of Equation~\eqref{eq:preconditioner}, the average number
of iterations required to solve the linear systems of Equations~\eqref{eq:linsys0} and \eqref{eq:linsys} was
about 700 for all 2MASS bands. This is still a reasonable number for the map resolutions
adopted here, but if a finer pixelization is required, one needs to
seek for better preconditioners, like those used by \citet{Eriksen:2004ss} and \citet{Oh:1998sr} for example.
\begin{figure*}[ht]
\begin{center}
    \includegraphics[width=0.95\textwidth]{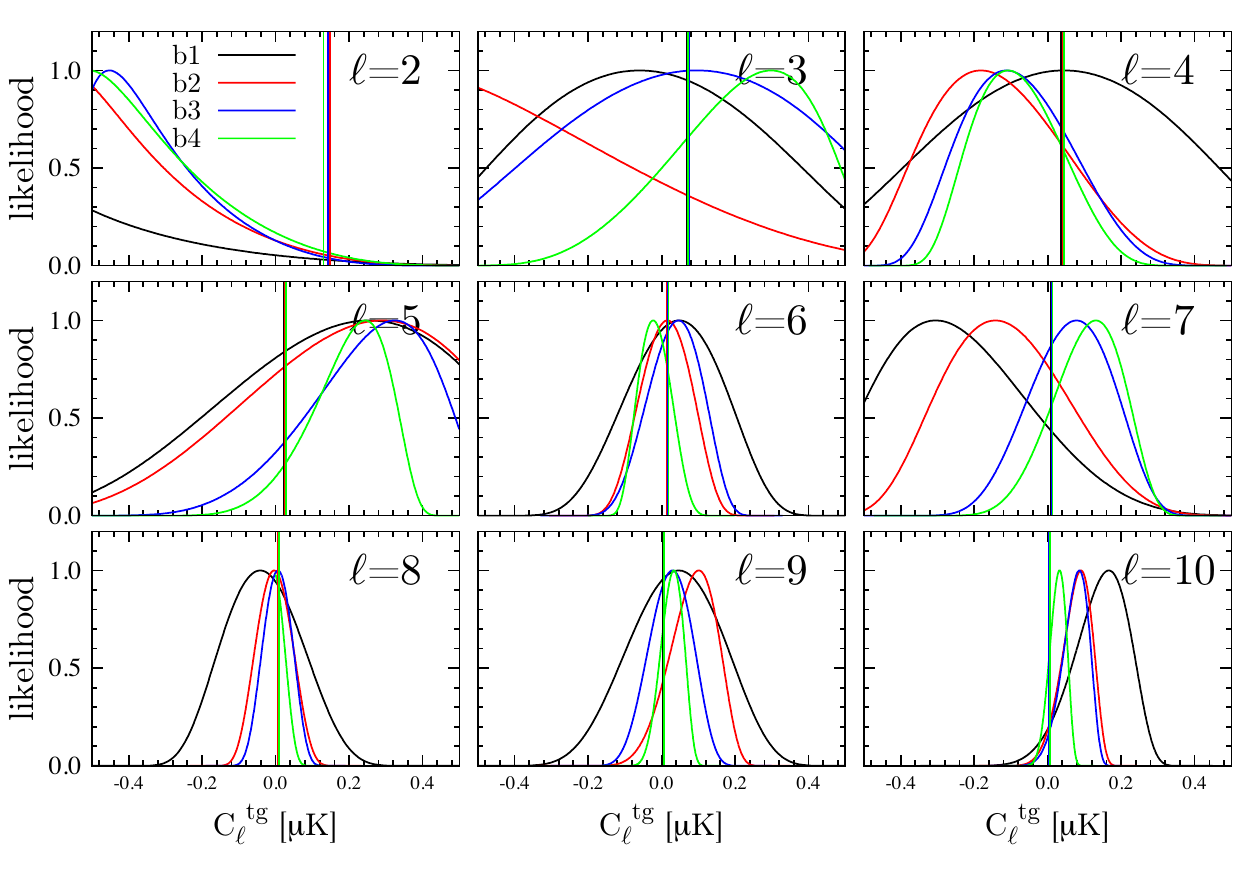}
%    \vspace*{-1.0cm}
\caption{One-dimensional slices of the Blackwell--Rao likelihood
  (normalized to 1 in its peak) built from 50,000 Gibbs samples. Nine slices are shown, corresponding to the first $C_{\ell}^{tg}$ multipoles, where all but the
  multipole being plotted are kept fixed at their maximum
  log-likelihood values. The different colors identify each of the
  four 2MASS selection functions (bands) considered in this
  work. Vertical lines show the expected values from a $\Lambda$CDM model.}
\label{fig:likeslices}
\end{center}
\end{figure*}

Figure~\ref{fig:likeslices} shows a few ($2\le \ell \le 10$) one-dimensional slices of the Blackwell--Rao
likelihood for the cross-correlation $C_{\ell}^{tg}$, where all but the multipole in the horizontal axis are
fixed at their maximum log-likelihood values. Similar to the results
with MC samples, one sees that for a shallow survey like
2MASS, the cross-correlation signal, if present, is always plunged
into a region of large cosmic variance. We stress here, however, that
this might not be the case if the cross-correlation signal is probed
with deeper selection function surveys. 
\begin{figure*}[htb]
  \begin{center}
    \includegraphics[width=1.\textwidth]{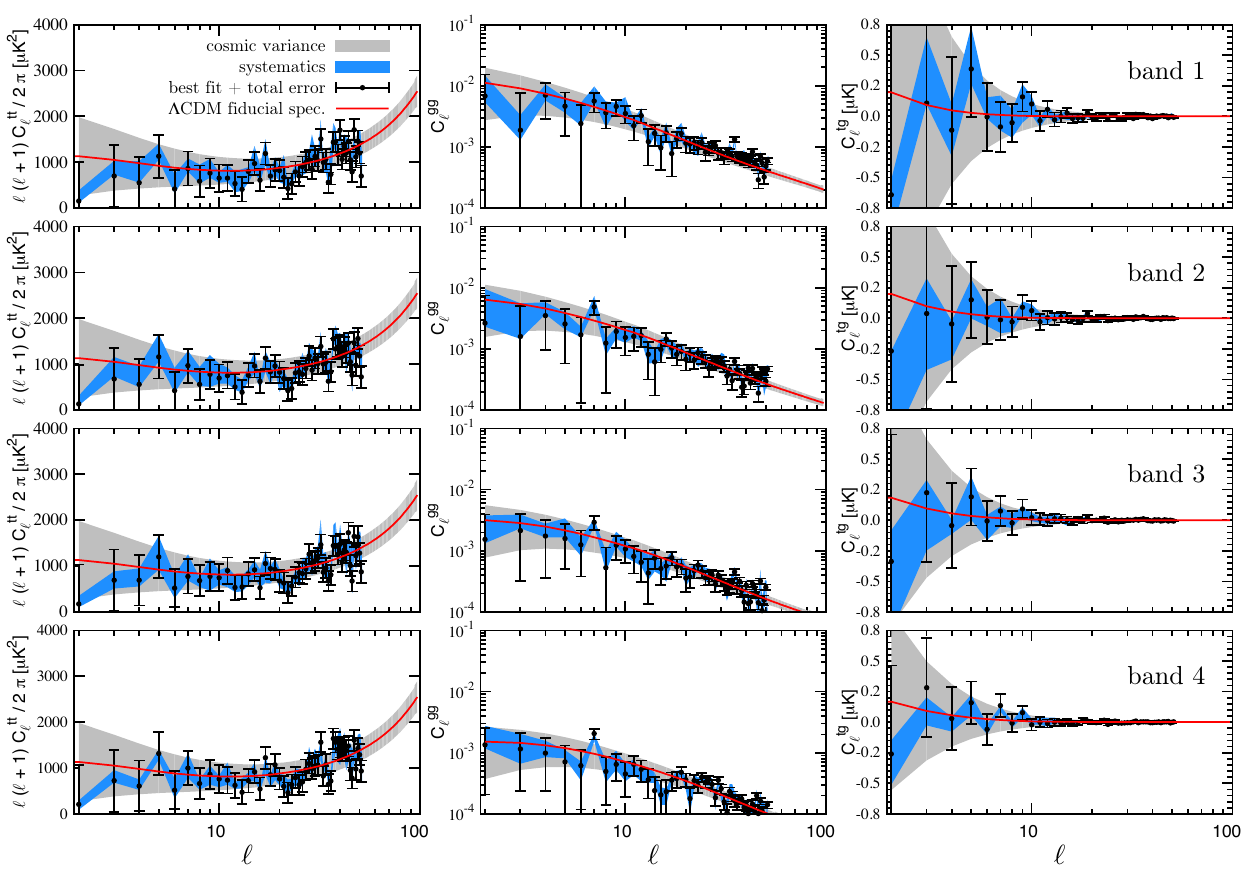}
  \end{center}
%  \vspace*{-1.0cm}
    \caption{The systematic blue band is bounded by the maximum and minimum
      values of best-fit multipole values obtained using different
      Gibbs chains fed with $W$ channel-2MASS and ILC-2MASS maps (see text of Section \ref{sec:results} for
      details). Cosmic variance around a $\Lambda$CDM fiducial model (red line)
      is also shown as a gray band. Data points with error bars are
      the weighted average for channels $Q$, $V$ and $W$ with weights given
      by the reciprocal of noise variances.}\label{fig:systematics_allbands}
\end{figure*}

In Figure~\ref{fig:systematics_allbands}, we show the result of the weighted
  average best-fit spectra of 2MASS $\times$ {\it Q}, {\it V}, {\it W} channels (using the 
  reciprocal of the noise variance as weight), together with the corresponding bands
of cosmic variance (gray) around the fiducial $\Lambda$CDM
model. As already largely discussed
in the literature, we also find a low value for the CMB quadrupole
$C_2^{tt}$ \footnote{A chain
run in temperature autocorrelation mode was also performed, leading to a similar
value of $C_2^{tt}$.}. The smallness of the quadrupole has already been
shown to be consistent with the level of cosmic variance
fluctuations at large angular scales
\citep{Efstathiou:2003wr,Efstathiou:2003hk,deOliveiraCosta:2003pu}.

In order to have an estimate of the systematic uncertainties coming from the 
Gibbs sampling algorithm, we have run the Gibbs chains slightly differently than in
the fiducial case shown in Table~\ref{table:fiducial} taking the {\it WMAP}9
$W$ channel and ILC as templates, namely:
\begin{enumerate}
\item a flat prior was used for the spectrum;

\item a larger burn-in phase was tested by throwing away the first 500
  samples of each chain;

\item possible correlations between adjacent samples were probed by
  skipping one of every two samples of the chain.
\end{enumerate}

The typical level of variation obtained in the final values of the maximum
of the log-likelihood is represented by the blue bands in
Figure~\ref{fig:systematics_allbands}. The use of the ILC map in
  building these bands provides a rough estimate on possible residual
  foregrounds at high galactic latitudes. One can see that at large scales,
where the signal from the ISW effect is expected to
be present in an accelerated universe, the most important source of uncertainty is still cosmic
variance. It is worth mentioning that the blue bands of Figure~\ref{fig:systematics_allbands}
cannot be taken as the full systematic uncertainties, since we only
investigated a few contributions coming from the Bayesian approach
adopted here and possible foregrounds. A more robust estimate of systematics affecting the temperature maps, such
as those introduced by residual foregrounds at high latitudes, for
example, will be obtained by applying the same methodology to the
independent temperature sky maps measured by the {\it Planck} satellite.

\section{Conclusions}
\label{sec:conclusions}

The ISW signal, a temperature anisotropy induced by time-varying
gravitational potentials in an accelerated universe, is a powerful
tool to probe the DE properties and break the degeneracy
among competing theoretical models. From the observational
point of view, the scientific community has been gifted by a number of outstanding 
results from different instruments, improving the understanding of the properties of 
both CMB and large-scale distribution of matter in our local universe.

In this work we explored the potential of the cross-correlation
between the CMB temperature fluctuations and the LSS as a probe for
the ISW effect, with data coming, respectively, from the {\it WMAP}9 data release and 
the 2MASS infrared galaxy catalog. The cross-correlation signal has been estimated by maximizing
the likelihood for observing the combined {\it WMAP}9--2MASS data sets, using
a Gibbs sampling technique in order to reconstruct the likelihood without a
direct (and computationally expensive) evaluation of it, in either
pixel or harmonic space. The four 2MASS
selection functions considered here lead to a shallow sampling of LSS
distribution and, in turn, to a cross-correlation signal severely
affected by cosmic variance. The role of such an intrinsic fluctuation
on the cross-spectrum was also studied by tracking the convergence of a
MASTER solution (as an ensemble average over a certain number of sky
realizations) toward its true value. The analysis pipeline, however, was successfully
validated using MC control samples and can be easily
applied to current deeper surveys, such as the SDSS \citep{sdss} or the Dark Energy Survey (DES; \citet{des}), as well as to future data coming from 
the Large Synoptic Survey Telescope (LSST; \citet{lsstSRD}). Ongoing efforts of the authors will compare the current results with those obtained using 
CMB maps produced by the {\it Planck} satellite. 

An initial estimate of the systematic uncertainty associated with the
sampling algorithm and possible foregrounds was also obtained and turned out to be  a
subdominant contribution to the total uncertainty when compared to
cosmic variance. We also tested the behavior of the method in the presence of a 
higher cross-correlation signal, more precisely, five times larger
than expected in the $\Lambda$CDM model. In such a scenario, we verified that, 
for a 2MASS-like survey, the statistical significance of the cross-correlation signal is still 
limited by cosmic variance. However, as can be seen from the analysis performed with MC
samples, when enough signal is present at smaller scales, where the noise, 
and not the irreducible cosmic variance, is the dominant source of
uncertainty, the Bayesian method is able to recover the
true signal. In \cite{Zhao:2009} and
\cite{Hojjati:2011}, for example, suitable general modifications of
the equations governing the growth of cosmological perturbations were
presented, allowing for the auto- and cross-correlation power spectra of CMB
and LSS to be calculated within a modified version of CAMB. The spectra for a few modified gravity models
diverge from that of general relativity from intermediate to small angular scales ($\ell\gsim 20$). Such
departures could be scrutinized with the methods presented in this
paper, as long as maps of higher resolution than the ones used here
and better preconditioners are fed into the Gibbs chains. Therefore,
we believe that the methodology described in this paper would be able to 
help solve existing issues in the field of ISW measurements, 
and we hope that it will become a complementary tool for the 
current studies of DE and modified gravity.

%%%%%%%%%%%%%%%%%%%%%%%%%%%%%%%%%%%%%%%%%%%%%%%%%%%%%%%%%
\section{Acknowledgements}
%%%%%%%%%%%%%%%%%%%%%%%%%%%%%%%%%%%%%%%%%%%%%%%%%%%%%%%%%

This publication makes use of data products from 2MASS, which is a joint project of the University of
Massachusetts and the Infrared Processing and Analysis
Center/California Institute of Technology, funded by the National
Aeronautics and Space Administration and the National Science
Foundation.

For the use of the 2MPZ catalog \citep{2mpzpaper} we acknowledge the Wide Field Astronomy Unit.

The authors acknowledge the use of the Legacy Archive for Microwave Background
Data Analysis (LAMBDA), part of the High Energy Astrophysics Science
Archive Center (HEASARC). HEASARC/LAMBDA is a service of the
Astrophysics Science Division at the NASA Goddard Space Flight Center.

Some of the results in this paper have been derived using the HEALPix package \citep{Gorski:2004by}.

E.M.-S. and M.P.-L. acknowledge the use of the University of S\~ao Paulo (USP) cloud
computing environment InterNuvem USP. F.C.C. was supported by
FAPERN/PRONEM and CNPq and acknowledges the use of the University of State
of Rio Grande do Norte computing cluster. M.P.-L. thanks CNPq
(PCI/MCTI/INPE program  and grant 202131/2014-9) for financial
support. C.P.N. thanks CAPES for financial support and the use of the
COAA/ON cluster. C.A.W. and E.M.-S. were partially supported by CNPq grants
313597/2014-6 and 309452/2012-0, respectively.

Finally, the authors would like to thank the referee for the
careful reading of the manuscript and valuable comments that
contributed to the improvement of the paper.

\appendix

\section{Analytical solution for the maximum of a Blackwell--Rao log-likelihood}
\label{solution}

We present here an analytic solution to the maximum of the logarithm of a
Blackwell--Rao estimator. 

For a block-diagonal matrix $\mathbf{S}$, whose $2 \times 2$
submatrices are $\mathbf{S}_{\ell}$ (given in Equation~\eqref{eq:matrices}), 
its posterior probability distribution can be written as product of
the prior $\pi(\mathbf{S})=|\mathbf{S}|^{-q}$ to the
$2\ell+1$ degrees of freedom inverse-Wishart distribution \citep{gupta1999matrix}:
\begin{equation}
P(\mathbf{S}|\mathbf{s}) \propto
\pi(\mathbf{S})\prod_{\ell}\frac{1}{2^{\frac{2\ell-2}{2}}\Gamma_2\left(\frac{2\ell-1}{2}\right)}\frac{|\bm{\sigma}_{\ell}|^{\frac{2\ell-2}{2}}}{|\mathbf{S}_{\ell}|^{\frac{2\ell+1}{2}}}
\exp\left[-\frac{1}{2}\textrm{Tr}\left(\bm{\sigma}_{\ell}\mathbf{S}_{\ell}^{-1}
  \right)\right].
%\frac{1}{\sqrt{|\mathbf{S}_{\ell}|^{2\ell+1}}}\textrm{exp}\left[-\frac{1}{2}\textrm{tr}\left(\bm{\sigma}_{\ell}\mathbf{S}_{\ell}^{-1}
%  \right)\right],
\end{equation}

Neglecting the terms that do not depend on $\mathbf{S}_{\ell}$, we can write the logarithm 
of the posterior as
\begin{equation}
\textrm{ln}P=-\frac{1}{2}\sum_{\ell}\left[(2\ell+1+2q)\textrm{ln}|\mathbf{S}_{\ell}|+\textrm{Tr}\left(\bm{\sigma}_{\ell}\mathbf{S}_{\ell}^{-1}
  \right)\right]
\end{equation}
and, for $n_G$ Gibbs samples represented by the total signal power
matrices $\bm{\sigma}_{\ell}^{(n)}$ ($n=1,...,n_G$), an estimate of
the Blackwell--Rao log-likelihood is given by
\begin{equation}
\textrm{ln}\mathscr{L}=\frac{1}{n_G}\sum_n\textrm{ln}P_n=-\frac{1}{2}\sum_{\ell}(2\ell+1+2q)\textrm{ln}|\mathbf{S}_{\ell}|
-\frac{1}{2}\frac{1}{n_G}\sum_{n}\sum_{\ell}\textrm{Tr}\left(\bm{\sigma}_{\ell}^{(n)}\mathbf{S}_{\ell}^{-1}
  \right),
\end{equation}
whose maximum,
\begin{equation}
\frac{\partial}{\partial C_{\ell}}\textrm{ln}\mathscr{L}=0 \quad (0\le
\ell \le \ell_{max}), \quad C_{\ell}= C_{\ell}^{tt},C_{\ell}^{gg},C_{\ell}^{tg},
\end{equation}
provides
\begin{equation}
\frac{\partial}{\partial C_{\ell}}\textrm{ln}\mathscr{L} =
-\frac{1}{2}\frac{\partial}{\partial
  C_{\ell}}\textrm{ln}|\mathbf{S}_{\ell}|
-\frac{1}{2}\frac{1}{n_G}\sum_n \frac{\partial}{\partial C_{\ell}}\textrm{Tr}\left(\bm{\sigma}_{\ell}^{(n)}\mathbf{S}_{\ell}^{-1}
  \right).
\end{equation}

Denoting the average over Gibbs samples as
$\overline{\alpha}=\sum_{n}\alpha_n$ and using the relation
\begin{equation}
\overline{\textrm{Tr}\left(\bm{\sigma}_{\ell}\mathbf{S}_{\ell}^{-1}
  \right)}=\textrm{Tr}\left(\overline{\bm{\sigma}}_{\ell}\mathbf{S}_{\ell}^{-1}
  \right),
\end{equation}
we can write the following set of three coupled equations:
\begin{eqnarray}
C_\ell^{tt}\left[ \textrm{Tr}\left(\overline{\bm{\sigma}}_{\ell}\mathbf{S}_{\ell}^{-1}
  \right) - (2\ell+1+2q)\right] &=& \overline{\sigma}_{11}\\
C_\ell^{gg}\left[ \textrm{Tr}\left(\overline{\bm{\sigma}}_{\ell}\mathbf{S}_{\ell}^{-1}
  \right) - (2\ell+1+2q)\right] &=& \overline{\sigma}_{22}\\
C_\ell^{tg}\left[ \textrm{Tr}\left(\overline{\bm{\sigma}}_{\ell}\mathbf{S}_{\ell}^{-1}
  \right) - (2\ell+1+2q)\right] &=& \overline{\sigma}_{12},
\end{eqnarray}
whose solution is given by
\begin{equation}
\mathbf{S}_{\ell}=\frac{1}{2\ell+1+2q}\overline{\bm{\sigma}}_{\ell}=\frac{1}{2\ell+1+2q}
\left( 
\begin{array}{cc}
 \overline{\sigma}_{11} &  \overline{\sigma}_{12} \\
 \overline{\sigma}_{12} &  \overline{\sigma}_{22}
\end{array}
\right).
\end{equation}

%%%%%%%%%%%%%%%%%%%%%%%%%%%%%%%%%%%%%%%%%%%%%%%%%%%%%%%%%
%\bibliography{biblio}

\begin{thebibliography}{dummy}

\bibitem[Afshordi et al.(2004)]{Afshordi:2003xu}   
Afshordi, N., Loh, Y.-S., \& Strauss, M. A. 2004, PRD, 69, 083524
%      eprint         = "astro-ph/0308260",

\bibitem[Abazajian et al.(2009)]{Abazajian:2008wr}   
Abazajian, K. N., Adelman-McCarthy, J. K., Agüeros, M. A., et al. 2009, ApJS, 182, 543
%      eprint         = "0812.0649",

\bibitem[Alam et al.(2015)]{sdss}   
Alam, S., Albareti, F.~D., Allende Prieto, C., et al. 2015, ApJS, 219, 12
%   eprint = {1501.00963},

\bibitem[Betoule et al.(2014)]{Betoule2014}   
Betoule, M., Kessler, R., Guy, et al. 2014, A\&A, 568, A22

\bibitem[Bennett et al.(2013)]{Bennett:2012zja}   
Bennett, C. L., Larson, D., Weiland, J.~L., et al. 2013, ApJS, 208, 20
%   eprint = {1212.5225},

\bibitem[Bilicki et al.(2014)]{2mpzpaper}
Bilicki et al. 2014, ApJS, 210, 9
% eprint = "1311.5246",

\bibitem[Bond(1998)]{Bond:1998zw}   
Bond, J. R., Jaffe, A. H., Knox, \& L. 1998, PRD, 57, 2117
%      eprint         = "astro-ph/9708203",

\bibitem[Boughn \& Crittenden(2002)]{Boughn:2001zs}   
Boughn, S. P., \& Crittenden, R. G. 2002, PRL, 88, 021302
%      eprint         = "astro-ph/0111281",

\bibitem[Boughn \& Crittenden(2004a)]{Boughn:2004ah}   
Boughn, S. P., \& Crittenden, R. G. 2004a, ApJ, 612, 647
%      eprint         = "astro-ph/0404348",

\bibitem[Boughn \& Crittenden(2004b)]{Boughn:2003yz}   
Boughn, S. P., \& Crittenden, R. G. 2004b, Nature, 427, 45
%      eprint         = "astro-ph/0305001",

\bibitem[Boughn \& Crittenden(2005a)]{Boughn:2004zm}   
Boughn, S. P., \& Crittenden, R. G. 2005a, NewAR, 49, 75
%      eprint         = "astro-ph/0404470",

\bibitem[Boughn \& Crittenden(2005b)]{Boughn:2004yy}   
Boughn, S. P., \& Crittenden, R. G. 2005b, MNRAS, 360, 1013
%      eprint         = "astro-ph/0408242",

\bibitem[Bull et al.(2015)]{Bull2015}   
Bull, P., Wehus, I.~K., Eriksen, H.~K., et al. 2015, ApJS, 219, 10
%   eprint = {1410.2544},

\bibitem[Cabre et al.(2006)]{Cabre:2006qm}   
Cabre, A., Gaztanaga, E., Manera, M., Fosalba, P., \& Castander, F. 2006, MNRAS, 372, L23
%      eprint         = "astro-ph/0603690",

\bibitem[Carvalho et al.(2006)]{Carvalho:2006fy}   
Carvalho, F. C., Alcaniz, J. S., Lima, J. A. S., \& Silva, R. 2006, PRL, 97, 081301
%      eprint         = "astro-ph/0608439",

\bibitem[Chu et al.(2005)]{Chu:2004zp}   
Chu, M., Eriksen, H. K., Knox, L., et al. 2005, PRD, 71, 103002
%     eprint         = "astro-ph/0411737",

\bibitem[Coles \& Jones (1991)]{lognormal2}
Coles, P, \& Jones, B. 1991, MNRAS, 248, 1

\bibitem[Copeland et al.(2006)]{Copeland:2006wr}   
Copeland, E. J., Sami, M., \& Tsujikawa, S. 2006, IJMPD, D15, 1753
%      eprint         = "hep-th/0603057",

\bibitem[Cooray(2002)]{Cooray:2001ab}   
Cooray, A. 2002, PRD, 65, 103510
%      eprint         = "astro-ph/0112408"

\bibitem[Crittenden \& Turok (1996)]{Crittenden:1995ak}   
Crittenden, R. G., \& Turok, N. 1996, PRL, 76, 575
%      eprint         = "astro-ph/9510072",

\bibitem[de Oliveira-Costa et al.(2004)]{deOliveiraCosta:2003pu}   
de Oliveira-Costa, A., Tegmark, M., Zaldarriaga, M., \& Hamilton, A. 2004, PRD, 69, 063516
%     eprint         = "astro-ph/0307282",

\bibitem[DES Collaboration (2005)]{des}
DES Collaboration 2005, The Dark Energy Survey, http://adsabs.harvard.edu/abs/2005astro.ph.10346T

\bibitem[Dunkley et al.(2009)]{Dunkley:2008mk}   
Dunkley, J., Spergel, D. N., Komatsu, E., et al. 2009, ApJ, 701, 1804
%      eprint         = "0811.4280",

\bibitem[Dup{\'e} et al.(2011)]{Dupe2011}   
Dup{\'e}, F.-X., Rassat, A., Starck, J.-L., \& Fadili, M.~J. 2011, A\&A, 534, A51
%   eprint = {1010.2192},

\bibitem[Efstathiou(2003b)]{Efstathiou:2003hk}   
Efstathiou, G. 2003b, MNRAS, 343, L95
%      eprint         = "astro-ph/0303127"

\bibitem[Efstathiou(2003a)]{Efstathiou:2003wr}   
Efstathiou, G. 2003a, MNRAS, 346, L26
%      eprint         = "astro-ph/0306431",

%\bibitem[Enander et al.(2015)]{Enander:2015}  
%Enander, J., Akrami, Y., M{\"o}rtsell, E., Renneby, M., Solomon, A.~R. 2015, PhRvD, 91, 084046
% Integrated Sachs-Wolfe e ect in massive bigravity, WMAP9+WISE
%      eprint         = "arXiv:1501.02140"

\bibitem[Eriksen et al.(2004)]{Eriksen:2004ss}   
Eriksen, H. K., O'Dwyer, I. J., Jewell, J. B., et al. 2004, ApJS, 155, 227
%      eprint         = "astro-ph/0407028",

\bibitem[Ferraro et al.(2015)]{Ferraro:2015}  
Ferraro, S., Sherwin, B.~D., Spergel, D.~N. 2015, PRD, 91, 083533
% A WISE measurement of the ISW e ect, WMAP9+WISE
%      eprint         = "arXiv:1401.1193",

\bibitem[Fosalba et al.(2003)]{Fosalba:2003ge}   
Fosalba, P., Gaztanaga, E., \& Castander, F. 2003, ApJ, 597, L89
%      eprint         = "astro-ph/0307249",

\bibitem[Fosalba \& Gaztanaga(2004)]{Fosalba:2003iy}   
Fosalba, P., \& Gaztanaga, E. 2004, MNRAS, 350, L37
%      eprint         = "astro-ph/0305468"

\bibitem[Gaztanaga et al.(2006)]{Gaztanaga:2004sk}   
Gaztanaga, E., Manera, M., \& Multamaki, T. 2006, MNRAS, 365, 171
%      eprint         = "astro-ph/0407022",

\bibitem[Giannantonio et al.(2006)]{Giannantonio:2006du}   
Giannantonio, T., Crittenden, R. G., Nichol, R. C., et al. 2006, PRD, 74, 063520
%      eprint         = "astro-ph/0607572",

\bibitem[Giannantonio et al.(2012)]{Giannantonio2012}   
Giannantonio, T., Crittenden, R., Nichol, R., \& Ross, A.~J. 2012, MNRAS, 426, 2581
% The significance of the integrated Sachs-Wolfe effect revisited
%   eprint = {1209.2125},

\bibitem[Giannantonio et al.(2014)]{Giannantonio:2013uqa}   
Giannantonio, T., Ross, A. J., Percival, W. J., et al. 2014, PRD, 89, 023511
%      eprint         = "1303.1349",

\bibitem[Gorski et al.(2005)]{Gorski:2004by}   
Gorski, K. M., Hivon, E., Banday, A. J., et al. 2005, ApJ, 622, 759
%      eprint         = "astro-ph/0409513",

\bibitem[Goto et al.(2012)]{Goto:2012}  
Goto, T., Szapudi, I., Granett, B. R. 2012, MNRAS, 422, L77
% Cross-correlation of WISE galaxies with the cosmic microwave background, WMAP7+WISE
%      eprint         = "arXiv:1202.5306",

\bibitem[Granett et al.(2008)]{Granett2008}
Granett, B.~R. and Neyrinck, M.~C. and Szapudi, I. 2008, ApJL, 683, L99
% An Imprint of Superstructures on the Microwave Background due to the Integrated Sachs-Wolfe Effect
% {0805.3695},
 
\bibitem[Granett et al.(2015)]{Granett2015}   
Granett, B.~R., Kov{\'a}cs, A., \& Hawken, A.~J. 2015, MNRAS, 454, 2804
%   eprint = {1507.03914},

\bibitem[Greason et al.(2012)]{Greason:2012}   
Greason, M. R., Limon, M., Wollack, E., et al. 2012, Nine-Year Wilkinson Microwave Anisotropy Probe ({\it WMAP}) Observations: Nine Year Explanatory Supplement, %(LAMBDA archive), 
http://lambda.gsfc.nasa.gov%/data/map/doc/MAP_supplement.pdf
%
%@unpublished{Greason:2005,
%      author         = "Greason, M. R. and others",
%      title          = "{Nine-Year Wilkinson Microwave Anisotropy Probe (WMAP)
%                        Observations: Nine Year Explanatory Supplement}",
%      collaboration  = "WMAP",
%      note           = "{available at LAMBDA archive}",
%      year           = "2005",
%      pages          = "189",
%}

\bibitem[Gupta \& Nagar(1999)]{gupta1999matrix}   
Gupta, A.K., \& Nagar, D.K. 1999, Matrix Variate Distributions (Chapman and Hall/CRC)
%@book{gupta1999matrix,
%  title={Matrix Variate Distributions},
%  author={Gupta, A.K. and Nagar, D.K.},
%  isbn={9780582369856},
%  series={PMS Series},
%  year="1999",
%  publisher={Addison-Wesley Longman, Limited}
%}

\bibitem[Hern{\'a}ndez-Monteagudo et al.(2014)]{Hernandez-Monteagudo:2013vwa}   
Hern{\'a}ndez-Monteagudo, C. Ross, A. J., Cuesta, A., et al. 2014, MNRAS, 438, 1724
%      eprint         = "1303.4302",

\bibitem[Hinshaw et al.(2013)]{Hinshaw:2012aka}   
Hinshaw, G., Larson, D., Komatsu, E., et al. 2013, ApJS, 208, 19
%      eprint         = "1212.5226",

\bibitem[Hivon et al.(2002)]{Hivon:2001jp}   
Hivon, E., Gorski, K. M., Netterfield, C. B., et al. 2002, ApJ, 567, 2
%      eprint         = "astro-ph/0105302",

\bibitem[Ho et al.(2008)]{Ho2008}   
Ho, S., Hirata, C., Padmanabhan, N., Seljak, U., \& Bahcall, N. 2008, PRD, 78, 043519
%  eprint = {0801.0642},

\bibitem[Hobson(2010)]{hobson2010bayesian}   
Hobson, M.P. 2010, Bayesian Methods in Cosmology (Cambridge: Cambridge University Press)
%@book{hobson2010bayesian,
%  title={Bayesian Methods in Cosmology},
%  author={Hobson, M.P.},
%  isbn={9780521887946},
%  lccn={2009035034},
%  series={Bayesian Methods in Cosmology},
%  url={https://books.google.com.br/books?id=RzehSUtgxvsC},
%  year="2010",
%  publisher={Cambridge University Press}
%}

\bibitem[Hojjati et al.(2011)]{Hojjati:2011}   
Hojjati, A., Pogosian, L. \& Zhao, G.-B. 2011, JCAP, 8, 005
%  eprint = {1106.4543},

\bibitem[Howlett et al.(2012)]{camb1}   
Howlett, C., Lewis, A., Hall, A., \& Challinor, A. 2012, JCAP, 1204, 027
%     eprint  = "1201.3654",

\bibitem[Hubble (1934)]{lognormal1}
Hubble, E. 1934, ApJ, 79, 8

\bibitem[Jewell et al.(2004)]{Jewell:2002dz}   
Jewell, J., Levin, S., \& Anderson, C. H. 2004, ApJ, 609, 1
%      eprint         = "astro-ph/0209560",

\bibitem[Joyce et al.(2015)]{Joyce2015}   
Joyce, A., Jain, B., Khoury, J., \& Trodden, M. 2015, PR, 568, 1

\bibitem[Kamionkowski(1996)]{Kamionkowski:1996ra}   
Kamionkowski, M. 1996, PRD, 54, 4169
%      eprint         = "astro-ph/9602150",

\bibitem[Kinkhabwala \& Kamionkowski(1999)]{Kinkhabwala:1998z}   
Kinkhabwala, A., \& Kamionkowski, M. 1999, PRL, 82, 4172
%      eprint         = "astro-ph/9808320",

\bibitem[Kinkhabwala \& Kamionkowski(1999)]{Kinkhabwala:1998zj}   
Kinkhabwala, A., \& Kamionkowski, M. 1999, PRL, 82, 4172
%      eprint         = "astro-ph/9808320",

\bibitem[Larson et al.(2007)]{Larson:2006ds}   
Larson, D. L., Eriksen, H. K., Wandelt, B. D., et al. 2007, ApJ, 656, 653
%      eprint         = "astro-ph/0608007",

\bibitem[Lewis(2013)]{camb4}   
Lewis, A. 2013, PRD, 87, 103529
%     eprint  = "1304.4473",

\bibitem[Lewis \& Bridle(2002)]{camb3}
Lewis, A., \& Bridle, S. 2002, PRD, 66, 103511
%     eprint  = "0205436",

\bibitem[Lewis et al.(2000)]{camb2}   
Lewis, A., Challinor, A., \& Lasenby, A. 2000, ApJ, 538, 473
%     eprint  = "9911177",

\bibitem[LSST Science Collaboration(2013)]{lsstSRD}   
{LSST Science Collaboration, Ivezi{\'c}, {\v{Z}.}, et al. 2013, LSST Science Requirements Document, http://ls.st/LPM-17}

\bibitem[McEwen et al.(2007)]{McEwen:2007ni}   
McEwen, J. D., Vielva, P., Wiaux, Y., et al. 2007, J. Fourier Anal. Appl., 13, 495
%      eprint         = "0704.3158",

\bibitem[Nadathur et al.(2012)]{Nadathur2012}  
Nadathur, S., Hotchkiss, S., Sarkar, S. 2012, JCAP, 2012, 042
% The integrated Sachs-Wolfe imprint of cosmic superstructures: a problem for ΛCDM
%      eprint         = "arXiv:1109.4126",

\bibitem[Nojiri \& Odintsov (2011)]{Nojiri2011}   
Nojiri, S., \& Odintsov, S.~D. 2011, PR, 505, 59
%   eprint = {1011.0544},

\bibitem[Nolta et al.(2004)]{Nolta:2003uy}   
Nolta, M. R., Wright, E. L., Page, L., et al. 2004, ApJ, 608, 10
%      SLACcitation   = "%%CITATION = ASTRO-PH/0305097;%%"

\bibitem[Oh et al.(1999)]{Oh:1998sr}   
Oh, S. P., Spergel, D. N., \& Hinshaw, G. 1999, ApJ, 510, 551
%      eprint         = "astro-ph/9805339",

\bibitem[Planck Collaboration(2014)]{Ade:2013dsi}   
Planck Collaboration, Ade, P. A. R., Aghanim, N., et al. 2014, A\&A, 571, A19
%      eprint         = "1303.5079",

\bibitem[Planck Collaboration(2015a)]{Ade:2015xua}   
Planck Collaboration, Ade, P. A. R., Aghanim, N., et al. 2015a, arXiv:1502.01589
%      eprint         = "1502.01589",

\bibitem[Planck Collaboration(2015b)]{2015arXiv150201590P}   
Planck Collaboration, Ade, P.~A.~R., Aghanim, N., et al. 2015b, arXiv:1502.01590

\bibitem[Planck Collaboration(2015c)]{2015arXiv150201595P}   
Planck Collaboration, Ade, P.~A.~R., Aghanim, N., et al. 2015c, arXiv:1502.01595

\bibitem[Padmanabhan et al.(2005)]{Padmanabhan:2004fy}   
Padmanabhan, N., Hirata, C. M., Seljak, U., et al. 2005, PRD, 72, 043525
%      eprint         = "astro-ph/0410360",

\bibitem[Rassat et al.(2007)]{Rassat:2006kq}   
Rassat, A., Land, K., Lahav, O., \& Abdalla, F. B. 2007, MNRAS, 377, 1085
%      eprint         = "astro-ph/0610911",

\bibitem[Schlegel et al.(1998)]{Schlegel:1997yv}   
Schlegel, D. J., Finkbeiner, D. P., \& Davis, M. 1998, ApJ, 500, 525
%      eprint         = "astro-ph/9710327",

\bibitem[Skrutskie et al.(2006)]{Skrutskie:2006wh}   
Skrutskie, M.~F., Cutri, R.~M., Stiening, R., et al. 2006, AJ, 131, 1163
%      doi            = "10.1086/498708",

\bibitem[Takahashi et al.(2012)]{halofit2012}   
Takahashi, R., Sato, M., Nishimichi, T., Taruya, A., \& Oguri, M. 2012, ApJ, 761, 152
%   eprint = {1208.2701},

\bibitem[Xia et al.(2011)]{Xia:2011hj}   
Xia, J.-Q., Baccigalupi, C., Matarrese, S., Verde, L., \& Viel, M. 2011, JCAP, 1108, 033
%      eprint         = "1104.5015",

\bibitem[Wandelt et al.(2004)]{Wandelt:2003uk}   
Wandelt, B. D., Larson, D. L., \& Lakshminarayanan, A. 2004, PRD, 70, 083511
%      eprint         = "astro-ph/0310080",

\bibitem[Zhao et al.(2009)]{Zhao:2009}   
Zhao, G.-B., Pogosian, L., Silvestri, A. \& Zylberberg, J. 2009, PRD, 79, 083513
%      eprint         = "0809.3791",

\end{thebibliography}

%%%%%%%%%%%%%%%%%%%%%%%%%%%%%%%%%%%%%%%%%%%%%%%%%%%%%%%%%

\end{document}